%% file: main.tex
\documentclass[runningheads]{llncs}

\usepackage[T1]{fontenc}
\usepackage[utf8]{inputenc}

\usepackage{svg}

\usepackage{xcolor}

\usepackage{graphicx}
\usepackage{float}

\usepackage{algorithm}
\usepackage{algpseudocode}

\usepackage{adjustbox}

\usepackage{booktabs}
\usepackage{tabularx}
\usepackage{caption}

\usepackage{amsmath, amssymb}

\usepackage{microtype}

\usepackage{tikz}
\usetikzlibrary{positioning, arrows.meta, calc, decorations.pathreplacing}

\usepackage{placeins}

\usepackage{pdflscape}
\usetikzlibrary{backgrounds}
\usepackage{subcaption}

\usepackage{braket}

\definecolor{lblue}{RGB}{0, 90, 255}
\definecolor{lgreen}{RGB}{0, 170, 60}
\definecolor{lblack}{RGB}{0, 0, 0}
\definecolor{lpink}{RGB}{255, 0, 200}
\definecolor{lorange}{RGB}{255, 110, 0}

\usepackage[
  colorlinks=true,
  linkcolor=lblack,
  citecolor=lblue,
  urlcolor=lpink,
  filecolor=lorange
]{hyperref}

\begin{document}

\title{Intrinsic Redundancy and Local Robustness in Finite $\beta$-Expansion Systems}
\titlerunning{Intrinsic Redundancy in $\beta$-Expansions}

\author{Adilbek Taizhanov\inst{1} 
\and Miras Seilkhan\inst{2}
}
\authorrunning{A. Taizhanov and M. Seilkhan}

\institute{
\email{adilbek300108@gmail.com}
\and
\email{seilkhan.miras6117@gmail.com}\\
}

\maketitle
\begin{abstract}
Redundancy in non-standard numeration systems is often associated with
robustness, but its operational value in finite digital arithmetic depends on
how representation, storage, and repair are defined. This paper studies
intrinsic redundancy in finite \(\beta\)-expansion systems through a
bounded-window model that separates ambient semantic non-uniqueness from
canonical codebook admissibility. The model distinguishes arithmetic
canonicalization from corruption repair and evaluates three outcomes:
structural detectability, value-preserving re-admissibilization of the observed
state, and semantic survival of the original value.

For the golden-ratio system and related multinacci bases, we formalize a
single-digit impossibility result: within a canonically injective finite
codebook, genuine single-digit corruptions cannot be semantically recovered by
exact repair without external information. Under exact structural repair,
semantic survival is possible only for localized multi-digit perturbations whose
error vector lies in the algebraic kernel of the evaluation map, as in local
rewrite identities such as \(100 \leftrightarrow 011\). Comparative experiments
across standard binary, signed-digit non-adjacent form (NAF), and multinacci
systems \((\varphi,T_3,T_4)\) quantify the trade-offs among codebook sparsity,
structural fault visibility, bounded-window canonicalization cost, residual
error, and boundary loss. The results identify intrinsic
\(\beta\)-redundancy as a constrained-language resource for structural digital
integrity, distinct from classical error-control redundancy.
\end{abstract}


\input{sections/1_introduction}

\input{sections/2_background}

\input{sections/3_models}

\input{sections/4_experiments}

\input{sections/5_discussion}

\input{sections/6_conclusion}

\newpage

\section*{Data and Code Availability}

All source code used for the experiments in this paper is open-source. The exact
release generating our results is archived on Zenodo
{\hypersetup{urlcolor=blue!55!black}%
\href{https://doi.org/10.5281/zenodo.21140212}{doi:10.5281/zenodo.21140212}}.
The maintained development version is available on GitHub:
{\hypersetup{urlcolor=blue!55!black}%
\href{https://github.com/The-Creator16/beta-redundancy-experiments}%
{The-Creator16/beta-redundancy-experiments}}.

\bibliographystyle{splncs04}
\bibliography{references}

\newpage

\input{sections/appendix}

\end{document}

%% file: sections/1_introduction.tex
{
\setlength{\parskip}{0.4em}

\section{Introduction}

Modern digital computation is built almost entirely on binary representation.
At the same time, positional numeration systems with a real base \(\beta>1\)
provide a broader design space in which the base, the digit alphabet, and the
admissible language can be chosen independently. This makes it possible to study
not only compactness of representation, but also the structural constraints that
a representation imposes on stored words and arithmetic intermediates.
Representations in non-integer bases, commonly referred to as
\(\beta\)-expansions, were introduced and systematically studied in the
literature on numeration systems and symbolic dynamics
\cite{Renyi1957,Parry1960,Sidorov2010,Hare2007}. A central feature of many
\(\beta\)-systems is semantic non-uniqueness: for bases in the interval
\(1<\beta<2\), one numerical value may admit several distinct expansions
\cite{Ward2008Robustness,Lai2009Thesis}.

This non-uniqueness has motivated several notions of robustness. In analog
\(\beta\)-encoders and golden-ratio encoders, redundant expansions can provide
stability against imperfect circuit components during analog-to-digital
conversion \cite{Ward2008Robustness,Daubechies2010GRE}. In non-standard
numeration systems, algebraic bases and extended digit alphabets can also support
structured arithmetic, including parallel addition
\cite{Frougny2011Parallel}. These results motivate a finite digital question
that is narrower but operationally sharper: when numbers are stored as finite
canonical codewords, what does intrinsic \(\beta\)-redundancy actually provide?

The key distinction is between ambient semantic non-uniqueness and redundancy
inside a stored canonical codebook. In the finite model studied here, the
admissible codebook selects canonical representatives from a larger ambient set
of digit strings. A perturbed string can often be re-admissibilized while preserving its observed value.
This operation is re-admissibilization of the observed state. It is distinct
from recovering the pre-fault value. The paper therefore studies three outcomes
separately: structural detectability, value-preserving normalization of the
observed state, and true semantic survival of the original value.

Our main question is:
\[
\textit{
What operational robustness, if any, is supplied by intrinsic
}
\]
\[
\textit{
\(\beta\)-redundancy in a finite canonical digital model?
}
\]

We answer this question through a bounded-window computational framework. The
arithmetic part of the framework is deliberately addition-only: digit-wise
addition is used as a controlled benchmark for upward carry propagation and
bounded-window canonicalization. Corruption repair is evaluated separately as a
digit-level process applied to already stored codewords. We compare standard binary, signed NAF as an alphabet-driven redundancy baseline, and selected multinacci systems, with emphasis on the golden-ratio base. In the
\(\varphi\)-system, the local identity
\[
100 \leftrightarrow 011
\]
induced by \(\varphi^2=\varphi+1\) makes the distinction between ambient
semantic redundancy and canonical admissibility especially transparent. The
strings \(100\) and \(011\) have the same value, but only \(100\) satisfies the
strict no-adjacent-ones convention used as the stored canonical language. Higher
multinacci systems, such as tribonacci and tetranacci, are included to test how
rewrite length influences detectability, semantic survival, and
canonicalization cost.

The study separates two operational pipelines. The first is addition
canonicalization: valid operands are combined by digit-wise addition, producing
non-canonical intermediate states that must be returned to the target codebook.
This addition-only benchmark isolates upward carry propagation and
bounded-window canonicalization cost; it is not intended to cover subtraction,
signed intermediate states, negative borrows, or mixed arithmetic workloads. The
second pipeline is corruption repair: an already stored codeword is perturbed,
possibly leaving the admissible language or moving into a different semantic
class. Treating these pipelines separately allows the experiments to distinguish
representation overhead, structural fault visibility, normalization effort,
boundary loss, and semantic preservation.

The primary contributions are as follows:
\begin{enumerate}
    \item \textbf{Operational separation of redundancy mechanisms.}
    We distinguish ambient semantic non-uniqueness from canonical codebook
    admissibility in bounded representation windows.

    \item \textbf{Single-digit impossibility result.}
    We formalize the limit that canonically injective finite codebooks cannot
    recover the original value after a genuine single-digit corruption without
    external information.

    \item \textbf{Algebraic-kernel condition for semantic survival.}
    We characterize the exceptional multi-digit perturbations that remain
    semantically neutral because their error vector lies in the kernel of the
    evaluation map.

    \item \textbf{Comparative finite-window evaluation.}
    We quantify codebook sparsity, structural detectability, bounded-window
    canonicalization cost, residual error, overflow, and truncation across
    Binary, Signed NAF, and multinacci systems \((\varphi,T_3,T_4)\).
\end{enumerate}

The resulting claim is deliberately precise: intrinsic \(\beta\)-redundancy is
not a general-purpose error-correcting mechanism, but it is a measurable
constrained-language resource for structural digital integrity. The scope and
limits of the finite-window protocol are collected explicitly in
Section~\ref{subsec:scope-limitations-finite-window}.

Finally, to support reproducible research and open science, the complete source
code, experimental configurations, and scripts used to generate all figures and
tables in this manuscript are publicly available. The repository has been
archived with a permanent DOI:
{\hypersetup{urlcolor=blue!55!black}%
\href{https://doi.org/10.5281/zenodo.21140212}{10.5281/zenodo.21140212}}.
}

%% file: sections/2_background.tex
\section{Background and Preliminaries on $\beta$-Representations}

We study representations of real numbers in non-integer bases, known as $\beta$-expansions, of the form
\[
x = \sum_i d_i \beta^i,
\]
where the digits $d_i$ belong to a finite alphabet. Unlike standard integer-base representations, such expansions are generally non-unique: the same numerical value may have multiple syntactic representatives. This motivates the study of normalization, admissibility, and local semantic equivalence in finite computational models.

We fix a real base $\beta > 1$ and a finite digit alphabet $D \subset \mathbb{Z}$. Our primary system is the golden-ratio base
\[
\beta = \varphi := \frac{1 + \sqrt{5}}{2}
\]
with $D = \{0,1\}$. The general $\beta$-framework is introduced only when needed.

\subsection{$\beta$-Representations, Greedy Expansions, and Admissibility}

We begin by separating three levels used throughout the paper: digit-string syntax, numerical semantics, and normalization procedures selecting valid or canonical representatives.

\phantomsection\label{def:beta-digit-string}
\textbf{Definition 1 ($\beta$-digit string and its value).}
A finite $\beta$-digit string over alphabet $D$ is a sequence
\[
d = (d_i)_{i \in \mathbb{Z}}, \qquad d_i \in D,
\]
with only finitely many nonzero digits. Its numerical value is
\[
\operatorname{val}_{\beta}(d) := \sum_{i \in \mathbb{Z}} d_i \beta^i \in \mathbb{R}.
\]

Two strings $d,e$ are semantically equivalent, written $d \sim_\beta e$, if
\[
\operatorname{val}_{\beta}(d) = \operatorname{val}_{\beta}(e).
\]
Thus, two digit strings may differ syntactically while representing the same value. For example, for $\beta=\varphi$, the strings $100$ and $011$ satisfy $\operatorname{val}_\varphi(100)=\varphi^2$ and $\operatorname{val}_\varphi(011)=\varphi+1$. Since $\varphi^2=\varphi+1$, we have $100 \sim_\varphi 011$, although $100 \neq 011$ as strings.

\phantomsection\label{def:beta-expansion}
\textbf{Definition 2 ($\beta$-expansion).}
Let $x \in \mathbb{R}$. A $\beta$-expansion of $x$ over $D$ is a sequence
$d=(d_i)_{i \in \mathbb{Z}}$ with $d_i \in D$ for which there exists an
integer $k$ such that
\[
d_i=0 \quad \text{for all } i>k,
\qquad
x=\sum_{i \in \mathbb{Z}} d_i\beta^i.
\]
Thus, the expansion has a highest nonzero position $k$: the integer part is
finite, while the fractional part may contain infinitely many nonzero digits.
The set of all such expansions for $x$ is denoted by
$\operatorname{Rep}_\beta(x)$.

This separates \hyperref[def:beta-digit-string]{Definition~1} from \hyperref[def:beta-expansion]{Definition~2}: hardware-level syntax uses finitely supported strings, whereas general $\beta$-expansions may require infinite fractional tails. In the computational model, such sequences are truncated to bounded hardware windows.

A number $x$ may therefore admit several distinct $\beta$-expansions, corresponding to different ways of distributing weight among powers of $\beta$.

The greedy $\beta$-expansion serves as a canonical representation. For our primary base $\beta=\varphi$, the digit alphabet is
\[
D=\{0,1\}.
\]
For $x \in [0,1)$, the greedy $\beta$-expansion is the lexicographically largest digit sequence over $D$ representing $x$. Equivalently, it is generated by the multiply-and-extract algorithm, formalized by the $\beta$-transformation
\[
T_\beta(x)=\beta x-\lfloor \beta x \rfloor.
\]

For \(x\ge 1\), one first extracts the integer-side digits down to \(d_0\)
and then applies the fractional procedure to the remaining part. For
\(x_0\in[0,1)\), the fractional greedy digits are generated by
\[
d^*_{-n}:=\lfloor \beta x_{n-1}\rfloor,
\qquad
x_n:=\beta x_{n-1}-d^*_{-n}=T_\beta(x_{n-1}).
\]
The process terminates if some remainder \(x_n\) becomes zero; otherwise it
continues as an infinite fractional expansion.

\textbf{Example 1 (Greedy expansion in the golden-ratio base).}

The display below shows the first five greedy steps for \(x_0=0.783\) in
base \(\varphi\). Each update is computed exactly in \(\mathbb{Q}(\varphi)\);
the displayed decimal values are rounded to five decimal places.
\input{figures/fig1_greedy_conversion}

The first generated digits are
\[
(d^*_{-1},d^*_{-2},d^*_{-3},d^*_{-4},d^*_{-5},\dots)
=
(1,0,0,1,0,\dots),
\]
so that
\[
0.783_{10}\approx 0.10010\ldots_\varphi.
\]

For the golden-ratio base, admissible greedy expansions follow a simple local rule: the block $11$ is forbidden. We now explain where this rule comes from.

In general, the greedy algorithm produces the fractional part of a
\(\beta\)-expansion as a right-infinite sequence
\[
x=\sum_{n=1}^{\infty} d^*_{-n}\beta^{-n}.
\]
Not every such sequence is greedy-admissible. Parry's condition characterizes
admissibility by comparing every proper suffix with the quasi-greedy expansion
of \(1\), the lexicographically largest infinite digit sequence representing the
unit value~\cite{Parry1960}.

For \(\beta=\varphi\), the boundary computation is especially simple:
\[
1=0.11_\varphi=\varphi^{-1}+\varphi^{-2}.
\]
Equivalently, the finite greedy expansion is
\[
d_\varphi(1)=110^\infty.
\]
Since this expansion terminates, the associated quasi-greedy expansion is
\[
d_\varphi^*(1)=(10)^\infty,
\]
so the Parry boundary sequence is
\[
\alpha_\varphi=(10)^\infty.
\]

It follows that any suffix beginning with \(11\) exceeds the boundary
\(10\ldots\) lexicographically and is therefore inadmissible. Conversely, if a
binary sequence contains no occurrence of \(11\), then every suffix either begins
with \(0\), or begins with \(10\), and cannot exceed \((10)^\infty\) without
creating a \(11\) block. Apart from the boundary sequence itself, the Parry
condition is therefore equivalent to forbidding \(11\). Since the computational
strings used below are finitely supported, the boundary equality case does not
occur, and admissibility reduces to
\[
s_i+s_{i+1}\le 1
\qquad\text{for all }i.
\]
Thus, admissible sequences must not contain the substring \(11\)~\cite{Parry1960,Blanchard1989}.

The same local obstruction appears algebraically in the normalization rule. If an
unnormalized digit string contains the sequence \(011\), then it evaluates
locally to the exact next higher power
\[
\varphi^{-2} + \varphi^{-3} = \varphi^{-1},
\]
which corresponds to the string $100$. The greedy representative therefore replaces the non-admissible block by the
higher-order digit:
\[
011 \to 100.
\]
This local carry operation is the normalization mechanism enforcing the no-\(11\)
rule.

Although the $\beta$-shift is classically defined on one-sided right-infinite sequences representing fractional parts, its admissibility constraints are translation-invariant. Since forbidden patterns do not depend on the radix-point position, we can lift this local constraint to finitely supported two-sided sequences. This justifies the following working definition for the computational model.

\phantomsection\label{def:admissible-phi}
\textbf{Definition 3 (Admissible string for $\beta=\varphi$).}
Using this lifted local constraint for finitely supported two-sided sequences, we say that a finite digit string
\[
d \in \{0,1\}^{(\mathbb{Z})}
\]
is admissible if it contains no two consecutive ones, that is,
\[
d_i + d_{i-1} \le 1 \qquad \text{for all } i \in \mathbb{Z}.
\]
Equivalently, the forbidden local pattern is the block $11$.

Admissibility therefore defines the valid code-space of the representation system. Any arithmetic operation, transient corruption, or storage fault that introduces the forbidden pattern produces a syntactically invalid codeword, even if its numerical value is well-defined. Recovery is therefore formulated first as re-admissibilization: mapping a perturbed string back into the admissible language while tracking the numerical effect.

\paragraph{Admissibility in higher-order multinacci bases.}
While \hyperref[def:admissible-phi]{Definition~3} gives the strict
golden-ratio constraint, forbidding $11$, the experiments also include
higher-order multinacci bases such as tribonacci ($T_3$) and tetranacci
($T_4$). Their Parry conditions forbid $111$ and $1111$, respectively,
as derived from the corresponding quasi-greedy expansions of $1$.
The formal constraints and finite-window codebooks are detailed in
\hyperref[app:multinacci-admissibility]{Appendix~C}.

With admissibility fixed as the syntactic target, we use three repair notions
throughout the paper. Here $d^\star$, $\widetilde d$, and $\widehat d$ denote
the original, corrupted, and repaired words, respectively.

\[
\widehat d \text{ is admissible and }
\operatorname{val}_\varphi(\widehat d)
=
\operatorname{val}_\varphi(\widetilde d).
\]
It is \emph{semantically recovering} if it restores the original pre-fault value,
\[
\operatorname{val}_\varphi(\widehat d)
=
\operatorname{val}_\varphi(d^\star).
\]
These are different criteria: exact re-admissibilization preserves the observed
post-fault value, whereas semantic recovery concerns the original pre-fault
value.
\phantomsection\label{rem:finite-window-codebook}
\textbf{Remark 1 (Finite-precision model and the admissible codebook).}
For computational purposes, we fix integers $L,R \ge 0$ and work on the digit window
\[
I_{L,R} := \{-R,\dots,L\}.
\]
A general finite-precision binary code is a vector
\[
d \in \{0,1\}^{I_{L,R}}
\]
with value
\[
\operatorname{val}_{\varphi}^{L,R}(d) := \sum_{i=-R}^{L} d_i \varphi^i,
\]
where the radix point is implicitly located between indices $0$ and $-1$. The full unconstrained ambient space is therefore \(\{0,1\}^{I_{L,R}}\).

Borrowing terminology from coding theory, we distinguish the full unconstrained space $\{0,1\}^{I_{L,R}}$ from the valid codebook. The codebook $C_\varphi^{L,R}$ is the subset of syntactically valid, admissible codewords:
\[
C_\varphi^{L,R} := \left\{d \in \{0,1\}^{I_{L,R}} : d_i + d_{i-1} \le 1 \text{ for all } i \in \{-R+1,\dots,L\}\right\}.
\]

We use \emph{codebook} in a strict syntactic sense: it denotes the stored admissible language, not the full semantic quotient of finite binary strings. Thus the ambient space $\{0,1\}^{I_{L,R}}$ may contain several strings with the same numerical value, whereas $C_\varphi^{L,R}$ is the canonical admissible subset selected by the no-$11$ convention.

All subsequent computational analyses use this finite-window model. We track when arithmetic or transient faults move a string outside $C_\varphi^{L,R}$ and how normalization maps it back, so that encoding cost, normalization depth, and post-repair numerical error remain well-defined and finite.

\phantomsection\label{lem:phi-canonical-injectivity}
\textbf{Lemma 1 (Canonical uniqueness in the finite $\varphi$-codebook).}
\label{lem:phi-codebook-injective}
For fixed $L,R\ge 0$, the restriction of the value map
\[
\operatorname{val}_{\varphi}^{L,R}: C_\varphi^{L,R}\to \mathbb{R}
\]
is injective. Equivalently, if $d,e\in C_\varphi^{L,R}$ and
\[
\operatorname{val}_{\varphi}^{L,R}(d)=\operatorname{val}_{\varphi}^{L,R}(e),
\]
then $d=e$.

\emph{Proof.}
Let \(N:=L+R\). Multiplying by \(\varphi^R\) shifts the window
\(I_{L,R}\) to the nonnegative index set \(\{0,\ldots,N\}\). It is therefore enough to prove injectivity for no-adjacent-one binary strings
\[
a,b\in\{0,1\}^{\{0,\ldots,N\}},
\qquad
a_j+a_{j-1}\le 1,\quad b_j+b_{j-1}\le 1,
\]
under the value map
\[
a\mapsto \sum_{j=0}^{N} a_j\varphi^j .
\]

Assume, toward a contradiction, that \(a\neq b\) and
\[
\sum_{j=0}^{N} a_j\varphi^j
=
\sum_{j=0}^{N} b_j\varphi^j .
\]
Let \(m\) be the largest index with \(a_m\neq b_m\). Interchanging \(a\)
and \(b\), if necessary, assume \(a_m=1\) and \(b_m=0\). Then
\[
0=
\sum_{j=0}^{N}(a_j-b_j)\varphi^j
=
\varphi^m+\sum_{j=0}^{m-1}(a_j-b_j)\varphi^j .
\]

We bound the largest possible negative contribution of the lower-order terms. Let \(M_m\) denote the maximum value of
\[
\sum_{j=0}^{m-1} u_j\varphi^j
\]
over all no-adjacent-one strings \(u\in\{0,1\}^{\{0,\ldots,m-1\}}\). We claim that
\[
M_m<\varphi^m
\qquad\text{for all }m\ge 0.
\]
For \(m = 0\), this is immediate since \(M_0 = 0\). For \(m = 1\), \(M_1 = 1 < \varphi\).
Now let \(m \ge 2\). Any maximizing no-adjacent-one string either has
\(u_{m-1}=0\), in which case its value is bounded by \(M_{m-1}\), or has
\(u_{m-1}=1\). In the latter case the no-adjacent-one constraint forces
\(u_{m-2}=0\), so the remaining contribution can only come from positions
\(0,\ldots,m-3\), and is therefore bounded by \(M_{m-2}\). Hence
\[
M_m \le \max\{M_{m-1}, \varphi^{m-1}+M_{m-2}\}.
\]
By induction,
\[
M_{m-1}<\varphi^{m-1}<\varphi^m
\]
and
\[
\varphi^{m-1}+M_{m-2}
<
\varphi^{m-1}+\varphi^{m-2}
=
\varphi^m,
\]
where the last equality uses \(\varphi^2=\varphi+1\). Therefore
\(M_m<\varphi^m\).

Since \(b_0,\ldots,b_{m-1}\) form a no-adjacent-one string, we have
\[
\sum_{j=0}^{m-1} b_j\varphi^j \le M_m < \varphi^m.
\]
Consequently,
\[
\sum_{j=0}^{N}(a_j-b_j)\varphi^j
\ge
\varphi^m-\sum_{j=0}^{m-1}b_j\varphi^j
>
0,
\]
contradicting equality. Hence \(a=b\). Undoing the shift by \(\varphi^R\)
gives \(d=e\) on the original window \(I_{L,R}\). Thus
\(\operatorname{val}_{\varphi}^{L,R}\) is injective on \(C_\varphi^{L,R}\).

\subsection{Non-Uniqueness and Redundancy in $\beta$-Expansions}

A central feature of non-integer numeration systems is that decoding on the \emph{ambient} digit space need not be injective: distinct digit strings may represent the same number. 
In this work, this non-uniqueness is not viewed as a defect, but as a structural feature before canonicalization. 
The admissible codebook used for storage is separate: it is the canonical target language selected from the larger ambient space.

By comparison, standard integer-base systems with the canonical alphabet are essentially non-redundant at finite precision, apart from familiar infinite-trailing ambiguities. 
To introduce redundancy in integer bases, standard arithmetic often expands the digit alphabet, for example using $\{-1,0,1\}$ in base $2$, to enable carry-free parallel addition~\cite{Avizienis1961}. 
We refer to this as artificial or alphabet-driven redundancy.

In contrast, for many non-integer bases $\beta \in (1,2)$, multiple expansions occur generically even with the minimal alphabet $D=\{0,1\}$. 
This is structural redundancy in the ambient representation space, not multiplicity inside the canonical admissible codebook.

\paragraph{The local rewrite identity.}
For the golden ratio, the algebraic relation
\[
\varphi^2 = \varphi + 1
\]
implies that for every $k \in \mathbb{Z}$,
\[
\varphi^{k+2} = \varphi^{k+1} + \varphi^k.
\]
At the digit level, this yields the value-preserving local equivalence
\begin{equation}
\label{eq:phi-local-rewrite}
100 \longleftrightarrow 011.
\end{equation}
Replacing the block $(d_{k+2},d_{k+1},d_k)=(1,0,0)$ by $(0,1,1)$, or conversely, preserves the value of the full digit string. For canonicalization, however, we use the directed form
\[
011 \longrightarrow 100,
\]
since the goal is to eliminate $11$ and move toward the admissible greedy representative. Each application affects at most three consecutive positions.

This local equivalence is the main source of ambient redundancy in the $\varphi$-system. 
As illustrated by the state-space graph in \hyperref[app:redundancy-graph]{Appendix~\ref*{app:redundancy-graph}}, distinct syntactic paths in the full binary ambient space may converge to the same semantic node as digit-string length increases. 
For finitely supported binary strings, the representation set
\[
\operatorname{Rep}_\varphi(x) := \{d \in \{0,1\}^{(\mathbb{Z})} : \operatorname{val}_\varphi(d)=x\}
\]
may contain several distinct strings. 
After imposing the no-$11$ convention, however, the finite codebook $C_\varphi^{L,R}$ selects at most one canonical representative of each value by \hyperref[lem:phi-canonical-injectivity]{Lemma~1}. 
Thus, semantic multiplicity here refers to the ambient or pre-normalized representation space, not to multiplicity inside the stored admissible codebook.

\paragraph{Why the interval $1<\beta<2$ matters.}
The interval $\beta \in (1,2)$ is especially relevant because the canonical digit alphabet remains binary in spirit while the base is already non-integer. This is the regime where non-trivial redundancy can arise without a large alphabet. In particular, for every $\beta \in (1,2)$ and Lebesgue-almost every $x \in [0,1]$, the set of $\beta$-expansions of $x$ is uncountable~\cite{Sidorov2003}.

These results justify non-uniqueness conceptually, but our practical focus is finite precision. 
In finite windows, we distinguish two operational notions. 
First, \emph{ambient semantic multiplicity} counts how many strings in the full space $\{0,1\}^{I_{L,R}}$ share the same numerical value. 
Second, \emph{canonical admissibility} restricts storage to the no-$11$ codebook $C_\varphi^{L,R}$, where each admissible word is treated as the canonical representative of its value. 
Later normalization and corruption experiments study the interaction between these two levels: ambient semantic collisions and canonical re-admissibilization.

\paragraph{Information capacity and sparsity of the admissible codebook.}
To quantify the storage cost of a strict canonical language, we evaluate the admissible codebook from an information-theoretic perspective. 
The admissible sequences in the golden-ratio base form a constrained shift space characterized by a single forbidden block, $11$. 
This space corresponds to the classic Fibonacci shift~\cite{LindMarcus1995}. 
The asymptotic information capacity $C$, or topological entropy, of this constrained discrete channel is governed by the largest eigenvalue of its transition matrix, namely $\varphi$~\cite{Shannon1948}. 
Consequently, the theoretical channel capacity per symbol is
\[
C = \log_2(\varphi) \approx 0.694 \text{ bits/symbol}.
\]

Since an unconstrained binary alphabet carries $1$ bit per symbol, the no-$11$ admissible language has asymptotic storage overhead
\[
R = 1 - \log_2(\varphi) \approx 0.306 \text{ bits/symbol}.
\]

This metric measures \emph{syntactic sparsity} of the canonical codebook $C_\varphi^{L,R}$ inside the full ambient space $\{0,1\}^{I_{L,R}}$, not semantic multiplicity inside $C_\varphi^{L,R}$.

This is an information-rate overhead, not the finite-window fraction of forbidden strings. 
For a window of width $W$, the actual admissible fraction is
\[
s_\varphi^{L,R}
=
\frac{|C_\varphi^{L,R}|}{2^W}
=
\frac{F_{W+2}}{2^W},
\]
which decays exponentially with $W$. 

This sparsity provides a structural signal only when a perturbation moves a word outside the admissible language. 
If a corruption remains inside $C_\varphi^{L,R}$, the no-$11$ constraint alone cannot detect it, even if the represented value has changed. 
Consequently, detectability is not guaranteed by redundancy alone; it depends on the fault distribution, sampled codeword ensemble, and bounded-window repair convention. 
The corresponding detection rates are therefore evaluated explicitly in the corruption experiments rather than assumed a priori.

\paragraph{Ambient redundancy as a structural resource.}
From a coding-theoretic viewpoint, admissible strings in a fixed window $I_{L,R}$ form a constrained canonical codebook, distinct from the full ambient representation space. 
We therefore define the finite-window ambient equivalence class of a string $d\in\{0,1\}^{I_{L,R}}$ by
\[
[d]_{\varphi}^{L,R}
:=
\left\{
e\in\{0,1\}^{I_{L,R}}:
\operatorname{val}_{\varphi}^{L,R}(e)
=
\operatorname{val}_{\varphi}^{L,R}(d)
\right\}.
\]
The quantity $|[d]_{\varphi}^{L,R}|$ measures ambient semantic multiplicity, whereas $C_\varphi^{L,R}$ provides the canonical admissible representatives used for storage. 
For the $\varphi$-system, the local equivalence $011\leftrightarrow100$ generates nontrivial ambient equivalence classes, while the directed rule $011\to100$ selects the admissible no-$11$ representative whenever it exists within the window. 
The exact combinatorial properties of bounded $\varphi$-representations have been studied by Dekking and van Loon~\cite{DekkingVanLoon2023}. 
In this work, ambient multiplicity is used only to interpret normalization paths and possible semantic collisions under perturbation; it is not treated as multiple stored admissible representatives of the same value.

\subsection{Normalization, Automata, and the Role of Pisot and Parry Bases}

We now move from representation to computation. Once arithmetic or perturbations produce a raw, possibly unnormalized digit string, a rigorous procedure is needed to restore admissibility.

\paragraph{Normalization as a computational task.}
Given a digit string $d$ over an extended working alphabet $E \supset D$, which accommodates temporary symbols such as ``2'', we seek a transformation that returns an admissible representative of the same value.

\phantomsection\label{def:normalization-map}
\textbf{Definition 4 (Normalization map).}
Assume that $\operatorname{val}_\beta$ extends linearly to sequences over an extended finite alphabet $E \supset D$. A normalization map is a function
\[
N : E^{(\mathbb{Z})} \to D^{(\mathbb{Z})}
\]
such that for every input string $d$ in its domain:
\begin{enumerate}
    \item \textbf{Value preservation:}
    \[
    \operatorname{val}_\beta(N(d)) = \operatorname{val}_\beta(d);
    \]
    \item \textbf{Admissibility:}
    \[
    N(d) \text{ is admissible over } D.
    \]
\end{enumerate}
If $N$ is implemented by repeated local value-preserving rewrite rules, we call it \emph{rewrite normalization}.

Requiring the image of the normalization map to lie in $D^{(\mathbb{Z})}$, i.e. to have finite support, is non-trivial for a general base $\beta$. For the golden ratio, this closure is guaranteed by the finiteness property (F) discussed below.

Because multiple ambient representatives may exist for one value before canonicalization, idempotence $N(N(d))=N(d)$ requires $N$ to target a chosen admissible normal form. 
For the golden-ratio finite-window codebook $C_\varphi^{L,R}$, this target is the no-$11$ canonical representative described in \hyperref[lem:phi-canonical-injectivity]{Lemma~1}. 
Once a string has been normalized into this target language, further applications of the map leave it invariant.

We use \emph{renormalization} for re-applying normalization after arithmetic or a fault violates admissibility. In the ideal infinite-precision setting, renormalization is exact: it preserves value and returns an admissible representative of the same semantic state.

In the finite-window setting, however, this ideal notion must be relaxed. A bounded-storage repair procedure may encounter edge effects, overflow, or truncation, and therefore need not preserve value exactly. We reserve \emph{exact normalization} for the ideal map $N$, and use \emph{bounded-window repair} for the finite-precision procedures evaluated below.

\paragraph{Finite-state realizability.}
Whether normalization can be implemented by a finite-state transducer is non-trivial. Results due to Frougny show that, for Pisot bases, normalization is finite-state computable under standard assumptions~\cite{Frougny1992,Frougny2002}.

The golden ratio
\[
\varphi = \frac{1+\sqrt5}{2}
\]
is a quadratic Pisot number: it is the positive root of $x^2-x-1=0$, and its conjugate satisfies $|\widehat\varphi|<1$. Accordingly, the $\varphi$-system belongs to a class where finite-state normalization is guaranteed and especially simple.

Recent automata-theoretic work further illustrates the link between
\(\varphi\), Zeckendorf representations, and finite-state computation:
Barnoff, Bright, and Shallit show that the \(n\)-th digit of the ordinary
base-\(b\) expansion of the golden ratio can be computed by a finite automaton
from the Zeckendorf representation of \(b^n\), and extend the automata-theoretic method to quadratic irrationals~\cite{barnoff2026finite}.

\paragraph{Directed normalization vs. algebraic equivalence.}
Although identities such as \eqref{eq:phi-local-rewrite} establish equivalence in both directions, a deterministic normalization map $N$ requires a directed strategy. In $(\varphi,\{0,1\})$, the rule
\[
011 \to 100
\]
is applied left-to-right, or iteratively, to eliminate $11$ by carrying its weight to higher-order digits. Conversely, the reverse rule
\[
100 \to 011
\]
exhibits semantic redundancy but breaks admissibility. Since \(\varphi\) is a Pisot base,
general finite-state normalization results ensure that value-preserving normalization can be realized by suitable finite-state procedures under standard assumptions~\cite{Frougny1992,FrougnySolomyak1992}. In the finite-window model below, directed rewrites are evaluated through a deterministic scheduler with explicit overflow, truncation, and rewrite-budget accounting.

\paragraph{The finiteness property and arithmetic closure.}
A related notion is the finiteness property (F)~\cite{FrougnySolomyak1992}, which ensures that many numbers in $\mathbb{Z}[\beta^{-1}]$ have finite greedy $\beta$-expansions. For arithmetic, this means that addition and subtraction of finite expansions yield results with finite greedy expansions, making $\varphi$ a clean finite-precision testbed.

\paragraph{Parallel addition and extended alphabets.}
Our strict canonicalization requirement contrasts with constant-time parallel addition. Building on signed-digit arithmetic introduced by Avizienis for fast parallel computation~\cite{Avizienis1961}, Frougny, Pelantov{\'a}, and Svobodov{\'a} showed that golden-ratio addition can be performed in parallel, in $O(1)$ time and without carry cascades, if one uses an extended symmetric alphabet such as $\{-1,0,1\}$~\cite{Frougny2011Parallel}. In that framework, the raw arithmetic state is mapped locally to another valid representation over the redundant alphabet.

However, a permanently enlarged alphabet changes the structural profile of the
storage language. Accepting \(\{-1,0,1\}\) as valid storage digits removes the
strict binary no-\(11\) constraint defining the canonical \(\varphi\)-codebook.
Our framework instead returns to the minimal alphabet \(\{0,1\}\), accepting the
corresponding carry cascades because the strict admissible language can expose
faults that leave the codebook. Faults that remain admissible may still change
the represented value silently, so detectability must be measured under the
chosen corruption model rather than assumed.

\paragraph{Parry numbers and admissibility checking.}
Beyond Pisot numbers, Parry numbers form a broader class characterized by eventual periodicity of the greedy expansion of $1$~\cite{Parry1960}. For Parry bases, the associated $\beta$-shift is sofic, so admissibility can be recognized by a finite automaton~\cite{Blanchard1989,Frougny2002,LindMarcus1995}. From an implementation perspective, validity checking therefore requires only bounded memory, a useful property for hardware-oriented models.

\paragraph{Locality and normalization behavior.}
The practical consequence of finite-state normalization is bounded locality: admissibility violations can be detected, and normalization can be implemented using finite neighbourhoods. In the golden-ratio case, the relevant neighbourhood is especially small because both the forbidden pattern and the basic rewrite rule have length three. Each rewrite step is local, but a single perturbation may still trigger a cascade. This interplay motivates the later metrics for canonicalization cost, propagation depth, structural detectability, and semantic survival under local corruption.

\subsection{Robust Encoders and the Digital Robustness Question}

Three neighboring literatures motivate the finite-window question studied here.
First, analog \(\beta\)-encoders and golden-ratio encoders exploit redundant
expansions to obtain robustness against threshold imperfections and quantization
noise during analog-to-digital conversion
\cite{Daubechies2006,Ward2008Robustness,Daubechies2010GRE}. In that setting,
redundancy acts during the encoding process itself: deviations at one threshold
crossing can be compensated by later quantization steps, keeping the reconstructed
analog value close to the target.

Second, classical digital reliability uses external check redundancy. In
systems-on-chip and critical digital applications, robustness is usually supplied
by information-theoretic or hardware mechanisms such as parity, Hamming-style
single-error-correcting codes, and Triple Modular Redundancy
\cite{LinCostello2004,Hamming1950,LyonsVanderkulk1962}. These mechanisms keep
the arithmetic representation and the protection structure conceptually
separate: a fault is detected or corrected through auxiliary check information,
not through the positional digit string alone.

Third, constrained numeration systems and Fibonacci-style representations show
that arithmetic languages can contain useful local structure. The engineering
potential of intrinsic redundancy in linearly recurrent bases was anticipated in
work on Fibonacci computers and ternary mirror-symmetrical arithmetic
\cite{Stakhov2002}. Those architectures mainly targeted explicit error-checking
through synthesized parity logic; the present model instead isolates what can be
obtained from the stored admissible language and its local normalization rules.

The no-adjacent-ones constraint also has a useful analogy to Fibonacci coding in
data compression. In a standard Fibonacci code, every valid codeword is
terminated by the block \(11\), while the internal sequence obeys the Zeckendorf
constraint forbidding adjacent ones
\cite{ApostolicoFraenkel1987,FraenkelKlein1996}. In that variable-length
setting, an unexpected occurrence of \(11\) before the terminal marker can
indicate a synchronization error or a transmission fault
\cite{ApostolicoFraenkel1987}. The fixed-window arithmetic model studied here is
different: there is no terminal delimiter, and admissibility is checked inside a
bounded positional register. The analogy is therefore syntactic rather than
coding-theoretic: in both settings, a forbidden local pattern signals that the
observed word has left the intended language.

There is also a separate line of explicit error-control codes based on
Fibonacci, Fibonacci-polynomial, and higher-order Fibonacci matrix structures.
For example, Fibonacci-polynomial coding methods define encoding and decoding
matrices whose algebraic relations yield explicit error-detection and
error-correction criteria~\cite{EsmaeiliEsmaeili2010}. More recent variants use
\(k\)-order Gaussian Fibonacci matrices and related higher-order Fibonacci
constructions
\cite{AydinyuzAsci2023,BasuDas2014NStep,BasuDas2014Tribonacci}. These works are
closer to classical block-code design than to the finite-window
\(\beta\)-normalization model studied here: they introduce explicit algebraic
coding maps and decoding criteria, whereas our experiments ask which
perturbations can be detected by, or remain semantically neutral under, the
intrinsic admissibility structure of a stored positional representation.

The present work therefore isolates a narrower digital object: an already
encoded finite digit string stored or processed in a bounded register. The main
question is what the intrinsic admissible language itself provides under local
corruption and bounded-window canonicalization. We measure which faults become
syntactically visible, which perturbations achieve semantic survival or remain value-neutral, and what
normalization cost is incurred after arithmetic or corruption. In this model,
normalization is repair of the observed representation, not an implicit
reconstruction of the pre-fault value.

%% file: figures/fig1_greedy_conversion.tex
\begin{figure}[h]
\centering

\providecommand{\digitmark}[1]{\textcolor{green!50!black}{\mathbf{#1}}}

\begin{tikzpicture}[
    node distance=5mm and 6mm,
    >=Latex,
    every node/.style={font=\footnotesize},
    gcell/.style={
        draw=gray!35,
        rounded corners=1pt,
        minimum height=8mm,
        inner xsep=6pt,
        inner ysep=3pt,
        align=center,
        fill=white
    },
    gstate/.style={
        gcell,
        minimum width=26mm,
        fill=green!6,
        draw=green!45!black
    },
    gmult/.style={
        gcell,
        minimum width=33mm
    },
    gdigit/.style={
        gcell,
        minimum width=42mm
    },
    gflow/.style={
        ->,
        draw=gray!60,
        line width=0.45pt
    }
]

\node[gstate] (x0) {$x_0=0.78300$};
\node[gmult, right=of x0] (b0) {$\beta x_0\approx 1.26692$};
\node[gdigit, right=of b0] (d1)
{$d^*_{-1}=\digitmark{1}$\\[-1pt] $x_1\approx 0.26692$};

\node[gstate, below=of x0] (x1) {$x_1\approx 0.26692$};
\node[gmult, right=of x1] (b1) {$\beta x_1\approx 0.43189$};
\node[gdigit, right=of b1] (d2)
{$d^*_{-2}=\digitmark{0}$\\[-1pt] $x_2\approx 0.43189$};

\node[gstate, below=of x1] (x2) {$x_2\approx 0.43189$};
\node[gmult, right=of x2] (b2) {$\beta x_2\approx 0.69881$};
\node[gdigit, right=of b2] (d3)
{$d^*_{-3}=\digitmark{0}$\\[-1pt] $x_3\approx 0.69881$};

\node[gstate, below=of x2] (x3) {$x_3\approx 0.69881$};
\node[gmult, right=of x3] (b3) {$\beta x_3\approx 1.13069$};
\node[gdigit, right=of b3] (d4)
{$d^*_{-4}=\digitmark{1}$\\[-1pt] $x_4\approx 0.13069$};

\node[gstate, below=of x3] (x4) {$x_4\approx 0.13069$};
\node[gmult, right=of x4] (b4) {$\beta x_4\approx 0.21147$};
\node[gdigit, right=of b4] (d5)
{$d^*_{-5}=\digitmark{0}$\\[-1pt] $x_5\approx 0.21147$};

\draw[gflow] (x0.east) -- (b0.west);
\draw[gflow] (b0.east) -- (d1.west);

\draw[gflow] (x1.east) -- (b1.west);
\draw[gflow] (b1.east) -- (d2.west);

\draw[gflow] (x2.east) -- (b2.west);
\draw[gflow] (b2.east) -- (d3.west);

\draw[gflow] (x3.east) -- (b3.west);
\draw[gflow] (b3.east) -- (d4.west);

\draw[gflow] (x4.east) -- (b4.west);
\draw[gflow] (b4.east) -- (d5.west);

\node[
    below=6mm of b4,
    font=\footnotesize,
    text=gray!75
] (res)
{$d^*_{-1}d^*_{-2}d^*_{-3}d^*_{-4}d^*_{-5}=10010$};

\end{tikzpicture}

\caption{First five greedy steps for $x_0=0.783$ in base $\varphi$.}
\label{fig:greedy-beta-example}

\end{figure}

%% file: sections/3_models.tex
\setlength{\parskip}{0.4em}

\section{Computational Model, Corruption Model, and Evaluation Framework}\label{sec:computational-model}

This section defines the finite-window protocol used in the experiments. The
central modeling choice is to treat arithmetic canonicalization and corruption
repair as separate operational pipelines. Arithmetic canonicalization starts from
valid operands and measures the cost and boundary behavior of returning raw
arithmetic states to the codebook. Corruption repair starts from an already
stored codeword and measures detectability, re-admissibilization, and semantic
survival after an external digit-level perturbation.

Throughout this section, the golden-ratio base
\[
\varphi=\frac{1+\sqrt5}{2}
\]
with digit set $D=\{0,1\}$ remains the primary system under study. However, the protocol is formulated comparatively so that the same evaluation structure can be applied to the baseline systems specified in \hyperref[subsec:comparison-systems-experimental-protocol]{Section~\ref*{subsec:comparison-systems-experimental-protocol}}.

\subsection{Finite-Window Computational Model}\label{subsec:finite-window-computational-model}

All experiments are carried out in the finite-window model introduced in \hyperref[rem:finite-window-codebook]{Remark~1}. 
For fixed integers $L,R\ge 0$, we work on the index set
\[
I_{L,R}:=\{-R,\dots,L\}, \qquad W:=L+R+1,
\]
with the radix point placed between positions $0$ and $-1$.

\input{figures/fig_finite_window_register}

For the golden-ratio system, a stored word is a vector
\[
d\in\{0,1\}^{I_{L,R}},
\qquad
\operatorname{val}_{\varphi}^{L,R}(d):=\sum_{i=-R}^{L} d_i \varphi^i,
\]
and the admissible codebook is
\[
C_{\varphi}^{L,R}
:=
\left\{
d\in\{0,1\}^{I_{L,R}} : d_i+d_{i-1}\le 1 \text{ for all } i\in\{-R+1,\dots,L\}
\right\}.
\]
Its cardinality is $|C_{\varphi}^{L,R}|=F_{W+2}$, where $F_n$ denotes the $n$-th Fibonacci number.

Throughout the comparison, all systems use the same window width $W$. 
This equal-width convention is hardware-oriented: it fixes the number of available storage positions rather than the represented numeric range.

For bounded-window normalization and canonicalization, we record two boundary
effects. \emph{Overflow} denotes writes beyond position $L$, while
\emph{truncation} denotes accesses below position $-R$. Such out-of-window
contributions are discarded and counted accordingly, so exact value preservation
may fail at the boundaries. For the golden-ratio base, the discarded lower tail
has the worst-case bound
\[
\sum_{j=1}^{\infty} \varphi^{-R-j} = \varphi^{-R+1}.
\]
Thus, boundary truncation can introduce numerical error, but that error remains analytically bounded.

The same finite-window viewpoint is used for the multinacci comparison systems, with the corresponding admissibility constraints and local canonicalization rules.

\subsection{Corruption Model and Canonicalization Procedures}\label{subsec:corruption-model-canonicalization}

We now formalize the perturbations and procedures studied in the experiments. The key distinction is the following:

\begin{itemize}
    \item a \emph{corruption} is an external disturbance applied to an already encoded word;
    \item a \emph{raw arithmetic state} is an internal intermediate object produced by a valid arithmetic operation before canonicalization.
\end{itemize}

Only the first is a fault model in the strict sense.

\paragraph{Corruption types.}
Let
\[
d^\star \in C_{\varphi}^{L,R}
\]
denote the original, uncorrupted admissible codeword. A corruption is a map
\[
F:\{0,1\}^{I_{L,R}}\to E^{I_{L,R}}
\]
that produces a possibly inadmissible word
\[
\widetilde d := F(d^\star)
\]
over an extended alphabet $E\supseteq\{0,1\}$.

\phantomsection\label{def:single-digit-corruption}\textbf{Definition 5 (Single-digit corruption).}
For a system with digit alphabet \(A\), a single-digit corruption at position \(j\in I_{L,R}\) is a map \(F_j\) such that
\[
F_j(d)_i :=
\begin{cases}
e \sim \mathcal{U}(A \setminus \{d_j\}), & i=j,\\
d_i, & i\neq j.
\end{cases}
\]

Here, $\mathcal{U}(S)$ denotes the uniform distribution over a finite set $S$. Thus,
\[
e \sim \mathcal{U}\bigl(A \setminus \{d_j\}\bigr)
\]
means that $e$ is selected uniformly at random from all digits in the alphabet $A$ except the original digit $d_j$. In particular, the corrupted digit is guaranteed to differ from the original one.

For the standard binary systems, this reduces to the deterministic flip $F_j(d)_j = 1-d_j$. For the signed-digit system, $e \in \{-1,0,1\} \setminus \{d_j\}$.

A single-digit corruption may or may not violate structural admissibility. For the golden-ratio system, if $d^\star_j=0$ and at least one adjacent digit is $1$, a flip to $1$ creates the forbidden block $11$. A flip to $0$ cannot create a forbidden pattern.

\input{figures/fig4_fault_injection_diagram}

\FloatBarrier

\phantomsection\label{def:burst-corruption}\textbf{Definition 6 (Burst corruption).}
Let \(b\geq 1\) denote the burst width, measured as the number of logical digit
positions affected by the burst window. For a finite window
\[
I_{L,R}=\{-R,\ldots,L\},
\]
we define the admissible set of burst starting positions by
\[
J_{L,R}(b):=\{-R,\ldots,L-b+1\},
\]
so that for every \(j\in J_{L,R}(b)\) the full burst interval
\[
B_{j,b}:=\{j,j+1,\ldots,j+b-1\}
\]
is contained in \(I_{L,R}\).

A burst corruption of width \(b\) starting at position \(j\in J_{L,R}(b)\) is a map
\[
F^\varepsilon_{j,b}
\]
that corrupts the contiguous interval \(B_{j,b}\) according to a nonzero binary mask
\[
\varepsilon=(\varepsilon_j,\ldots,\varepsilon_{j+b-1})
\in \{0,1\}^{b}\setminus\{0^b\}.
\]
Formally,
\[
F^\varepsilon_{j,b}(d)_i :=
\begin{cases}
F_i(d)_i, & i\in B_{j,b} \text{ and } \varepsilon_i=1,\\
d_i, & \text{otherwise}.
\end{cases}
\]
where \(F_i(d)_i\) is the single-digit corruption rule defined in
\hyperref[def:single-digit-corruption]{Definition~5}. Thus, the burst window
specifies the contiguous logical region exposed to a local disturbance, while the
mask determines which positions inside that window are actually corrupted. Since
the mask is nonzero and each active position is corrupted using the alphabet-aware
substitution rule, every sampled burst corruption changes at least one digit and
remains well-defined for binary, signed-digit, and non-integer-base systems.

\FloatBarrier

\input{figures/fig4_5_fault_burst_corruption}

Burst corruptions are used as a logically contiguous stress test for local
admissibility constraints and rewrite neighborhoods. This is an algorithmic
fault model: the physical mapping of a contiguous logical burst depends on the
memory layout, data path, register organization, and interleaving policy of a
particular architecture. The model is therefore most relevant to structures where
logical adjacency can remain operationally meaningful, such as local arithmetic
datapaths, non-interleaved registers, or densely packed sequential logic. This
choice is motivated by the increasing relevance of multi-bit upsets in scaled
technologies~\cite{Ibe2010MBU}, but the experiments do not claim to model a
specific DRAM, SRAM, or particle-strike layout.

\vspace{-1.0\baselineskip}

\paragraph{Raw arithmetic states.}
Arithmetic is treated separately from corruption.

\phantomsection\label{def:raw-arithmetic-state}\textbf{Definition 7 (Raw arithmetic state).}
Let
\[
a,b\in C_{\varphi}^{L,R}
\]
be admissible operands. Focusing strictly on the carry dynamics of addition, their digit-wise summation produces the raw arithmetic state
\[
c_i := a_i + b_i \in \{0,1,2\},
\qquad i\in I_{L,R}.
\]
Thus
\[
c\in \{0,1,2\}^{I_{L,R}}.
\]

This object is generally neither admissible nor canonical. It is a deterministic
intermediate arithmetic representation that must be canonicalized back into the
target codebook. Addition is used here as a controlled benchmark for upward
carry propagation over minimal canonical alphabets. This isolates the
canonicalization mechanism evaluated in the experiments; subtraction, signed
intermediates, negative borrows, and mixed arithmetic workloads are outside the
present protocol.

This restriction is methodological rather than conceptual. In Zeckendorf-style
arithmetic, addition and subtraction can be implemented by combinational logic
networks with \(\mathcal{O}(\log W)\) parallel depth and linear sequential
worst-case limits~\cite{Ahlbach2013Zeckendorf}. For higher-order multinacci
systems, digit-wise addition may also leave the target alphabet or violate the
admissibility constraints, so a bounded-window canonicalization procedure is
required~\cite{Frougny2002,Frougny2011Parallel}. The known trade-off between
digit-alphabet size and parallel evaluation depth~\cite{Frougny2014kblock}
therefore motivates comparing minimal canonical alphabets under a fixed
deterministic rewrite convention.

\paragraph{Exact normalization and bounded-window procedures.}
A normalization or canonicalization procedure takes a non-canonical word and attempts to return an admissible one.

\phantomsection\label{def:exact-normalization}\textbf{Definition 8 (Exact normalization).}
The exact normalization map
\[
N:E^{(\mathbb Z)}\to \{0,1\}^{(\mathbb Z)}
\]
is the ideal infinite-precision procedure satisfying:
\begin{enumerate}
    \item \textbf{Value preservation:}
    \[
    \operatorname{val}_{\varphi}(N(d))=\operatorname{val}_{\varphi}(d);
    \]
    \item \textbf{Admissibility:}
    \[
    N(d)\in C_{\varphi}^{\infty},
    \]
    where $C_{\varphi}^{\infty}$ denotes the set of all finitely supported admissible strings over $\mathbb Z$.
\end{enumerate}

\phantomsection\label{def:bounded-window-normalization}\textbf{Definition 9 (Bounded-window normalization / repair).}
A bounded-window procedure is a map
\[
R_{L,R}:E^{I_{L,R}}\to \{0,1\}^{I_{L,R}}
\]
that applies local rewrite rules within the finite window $I_{L,R}$ and attempts to restore admissibility.
If a rewrite would create or modify a digit outside $I_{L,R}$, the out-of-window contribution is discarded:
writes to indices $i>L$ are recorded as overflow, and writes to indices $i<-R$ are recorded as truncation.
Consequently, bounded-window repair need not preserve value exactly.

When the input to $R_{L,R}$ arises from a corruption, we refer to the procedure as \emph{repair}. When the input arises from a raw arithmetic state, we refer to it as \emph{canonicalization}. The underlying mechanism is the same; only the interpretation differs.

For the golden-ratio system, the principal local rewrite rules are:
\[
(0,1,1)\longrightarrow (1,0,0), \tag{2}\label{eq:phi-011-to-100}
\]
derived from $\varphi^2=\varphi+1$, and, more generally, when a temporary digit \(q\ge 2\) appears,
\[
d_k=q\ge 2
\quad\Longrightarrow\quad
\begin{cases}
d_k := q-2,\\
d_{k+1} := d_{k+1}+1,\\
d_{k-2} := d_{k-2}+1,
\end{cases}
\tag{3}\label{eq:phi-two-carry-rule}
\]
which follows from
\[
q\varphi^k
=
(q-2)\varphi^k+\varphi^{k+1}+\varphi^{k-2}.
\]
Equivalently, each application resolves two units of excess weight at position
\(k\) using the identity
\[
2\varphi^k = \varphi^{k+1}+\varphi^{k-2}.
\]
For \(q=2\), this reduces to the original temporary-digit rule. For \(q>2\),
the same rule may be applied repeatedly until the digit at position \(k\) lies
in the target alphabet \(\{0,1\}\), unless the finite rewrite budget is exhausted
first.

A single application of \eqref{eq:phi-011-to-100} or
\eqref{eq:phi-two-carry-rule} may create new nearby violations or may increase
another temporary digit. In particular, although raw digit-wise addition produces
digits only in \(\{0,1,2\}\), later carry propagation can accumulate at a
previously non-binary position and create values \(q\ge 3\). Such digits are
handled by the generalized form of \eqref{eq:phi-two-carry-rule}, which resolves
two units of excess weight per application. General finite-state normalization results for Pisot
numeration systems, including the golden-ratio base, ensure that
value-preserving normalization can be realized under suitable finite-state
procedures~\cite{Frougny1992,FrougnySolomyak1992}. The experiments below use a
more specific object: a deterministic bounded-window MSB-first scheduler with
explicit overflow, truncation, and rewrite-budget accounting. Thus the reported
quantities \(\kappa\), \(\pi\), overflow, and truncation are protocol metrics of
this scheduler, not order-independent invariants of the abstract rewrite
relation.

The local identities \eqref{eq:phi-011-to-100} and
\eqref{eq:phi-two-carry-rule} are value-preserving when their required digits
are available inside the computational window. Boundary loss is tracked
separately. In particular, lower-boundary truncation admits the guard-digit
bound below, which treats finite-window precision loss as a tunable storage
trade-off.

\phantomsection\label{lem:guard-digits}
\textbf{Lemma 2 (Guard-Digit Error Bound).}
\textit{Let a bounded window be extended by $g \ge 0$ lower-order guard digits, moving the truncation boundary to $-(R+g)$. Let $D_{\max} = \max \{|d| : d \in E\}$ be the maximum absolute value of a digit in the extended working alphabet $E$. The absolute numerical error $\varepsilon$ introduced by truncating all fractional contributions below the guard boundary satisfies the strict bound
\[
|\varepsilon| \le D_{\max} \frac{\beta^{-(R+g+1)}}{1 - \beta^{-1}}.
\]
}

Thus, when the only discarded contribution is the tail below the extended
lower boundary, guard digits give a tunable upper bound on that lower-boundary
tail component of the finite-window error. End-to-end arithmetic error may also
include overflow and accumulated rewrite effects, which are measured separately
in the experiments.

\phantomsection
\paragraph{Bounded-window MSB-first scheduler.}\label{alg:bounded-msb-scheduler}
Let \(X\in\{\Phi,T_3,T_4\}\) be a local-rewrite system. Let
\(I_{L,R}=\{-R,\ldots,L\}\) be the finite digit window and let
\(W=L+R+1\). Let \(\mathcal{R}_X\) denote the finite set of local rewrite
schemas specified for system \(X\), including the generalized temporary-digit
resolution rule for every visible value \(q\ge 2\).

Because the reported procedure is run with a finite rewrite budget
\(K_{\max}(W)\), the effective working alphabet can be chosen finite:
\[
E_X(W):=\{0,1,\ldots,2+K_{\max}(W)\}.
\]
Indeed, the initial arithmetic states have digits in \(\{0,1,2\}\), corruption
states have binary digits, and each rewrite step can increase any fixed digit by
at most one. Therefore, before the procedure stops after at most
\(K_{\max}(W)\) rewrite attempts, no digit can exceed \(2+K_{\max}(W)\). In the reported experiments we use a fixed finite rewrite budget
\(K_{\max}(W)\), specified in
Section~\ref{subsec:scope-limitations-finite-window}. The budget is part of the
experimental protocol rather than a mathematical upper bound on all possible
rewrite cascades.

For a current word \(z\), define the residual defect set
\[
\mathcal{D}_X(z)
\]
as the set of all currently visible local defects in \(I_{L,R}\). These defects include temporary digits outside the target alphabet
\(\{0,1\}\), which in the local non-integer systems means digits \(q\ge 2\),
forbidden admissibility blocks, and rule matches whose required write position
would fall outside \(I_{L,R}\). Each defect \(q\in\mathcal{D}_X(z)\) is assigned the anchor
\[
a(q):=\max\{i:\text{ position } i \text{ is read or written by the corresponding local rule}\}.
\]

The bounded-window scheduler is the following deterministic procedure.

\begin{quote}
\small
\noindent\textbf{Input:} a word \(z\in E^{I_{L,R}}\), the rule set
\(\mathcal{R}_X\), and the budget \(K_{\max}(W)\).

\noindent\textbf{Initialize:}
\[
\kappa:=0,\qquad \mathrm{overflow}:=0,\qquad \mathrm{truncation}:=0.
\]

\begin{enumerate}
    \item While \(\mathcal{D}_X(z)\neq\emptyset\) and
    \(\kappa<K_{\max}(W)\), do the following.

    \begin{enumerate}
        \item Choose a defect \(q^\star\in\mathcal{D}_X(z)\) with maximal
        anchor \(a(q^\star)\).

        \item If several defects have the same anchor, resolve temporary-digit
        defects before admissibility-block defects. Any remaining ties are
        resolved by the fixed local pattern order used in the implementation.

        \item Apply the local rewrite associated with \(q^\star\).

        \item If the rewrite attempts to write to an index \(i>L\), set
        \(\mathrm{overflow}:=1\) and discard that out-of-window write.

        \item If the rewrite attempts to write to an index \(i<-R\), set
        \(\mathrm{truncation}:=1\) and discard that out-of-window write.

        \item Set \(\kappa:=\kappa+1\).
    \end{enumerate}

    \item If \(\mathcal{D}_X(z)=\emptyset\), set \(H:=1\). Otherwise set
    \(H:=0\).
\end{enumerate}

\noindent\textbf{Return:}
\[
z,\qquad \kappa,\qquad \mathrm{overflow},\qquad \mathrm{truncation},\qquad H.
\]
\end{quote}

The flag \(H\) separates successful canonicalization from budget exhaustion.
Thus the bounded-window procedure is total as an experimental algorithm, since
it always returns after at most \(K_{\max}(W)\) rewrite attempts. It is counted
as successful only when no residual defect remains. This supplies a finite
termination boundary for the reported measurements without claiming that
MSB-first scheduling is mathematically necessary for the existence of
normalization in the underlying \(\beta\)-system. Consequently, the reported
\(\kappa\) and \(\pi\) values are properties of this specified bounded-window
scheduler, not order-independent invariants of the abstract rewrite relation.

For tribonacci and tetranacci, the corresponding admissibility-restoring
canonicalization procedures are based on the local multinacci identities
\[
1000 \leftrightarrow 0111
\qquad\text{and}\qquad
10000 \leftrightarrow 01111,
\]
together with the temporary-digit resolution rules, deterministic MSB-first
priority convention, and bounded-window overflow/truncation policy specified in
\hyperref[app:multinacci-admissibility]{Appendix~C}.

\paragraph{Remark 2 (Analog-like perturbations).}\phantomsection\label{rem:analog-like-perturbations}
In applications closer to $\beta$-encoders, one may also consider perturbations that do not necessarily violate admissibility but shift the represented value slightly near the least significant positions~\cite{Daubechies2006,Ward2008Robustness,Daubechies2010GRE}. Such perturbations are natural in the analog domain. The present work, however, focuses on discrete digital corruption and bounded-window canonicalization.

\subsection{Correctness Criteria and Evaluation Metrics}\label{subsec:correctness-criteria-metrics}

We now define the correctness notions and quantitative metrics used in the
reported experiments. Since corruption trials and arithmetic trials address different questions, we distinguish their evaluation criteria explicitly.

\input{figures/fig5_correctness_hierarchy_figure}

\paragraph{Corruption trials.}
Let
\[
d^\star\in C_{\varphi}^{L,R}
\]
be the original codeword, let
\[
\widetilde d = F(d^\star)
\]
be the corrupted word, and let
\[
\widehat d = R_{L,R}(\widetilde d)
\]
be the repaired output.

\phantomsection\label{def:structural-detectability}\textbf{Definition 10 (Structural detectability).}
A corruption is structurally detectable if the corrupted word is not admissible:
\[
\widetilde d \notin C_{\varphi}^{L,R}.
\]
This notion captures whether the codebook itself exposes the corruption before any repair is attempted.

\phantomsection\label{def:structural-correctness}\textbf{Definition 11 (Structural correctness).}
The repair is structurally correct if
\[
\widehat d \in C_{\varphi}^{L,R}.
\]
This is a purely syntactic requirement.

\phantomsection\label{def:structural-exactness}\textbf{Definition 12 (Structural exactness / value preservation).}
The repair is structurally exact if it is structurally correct and also preserves the value of the corrupted word:
\[
\operatorname{val}_{\varphi}^{L,R}(\widehat d)
=
\operatorname{val}_{\varphi}^{L,R}(\widetilde d).
\]

\phantomsection\label{def:semantic-survival}\textbf{Definition 13 (Semantic survival).}
The repair achieves semantic survival of the original value if
\[
\operatorname{val}_{\varphi}^{L,R}(\widehat d)
=
\operatorname{val}_{\varphi}^{L,R}(d^\star).
\]

This is the strongest corruption-oriented criterion. In general, structural exactness and semantic survival coincide only if the corruption itself did not alter the represented value.

The distinction between structural exactness and semantic survival gives a
simple but important limit on what intrinsic redundancy can reconstruct. The
following statement is formulated for canonically injective finite-window
codebooks. Canonical injectivity ensures that semantic recovery of the original
value is equivalent to recovery of the original canonical codeword within the
stored codebook. This assumption holds for all comparison codebooks used in the
experiments: for \(\Phi\), by
\hyperref[lem:phi-codebook-injective]{Lemma~1}, and for Binary, Signed NAF,
\(T_3\), and \(T_4\), by
\hyperref[app:canonical-injectivity-comparison]{Appendix~\ref*{app:canonical-injectivity-comparison}}.

\phantomsection\label{thm:no-free-lunch}
\textbf{Proposition 1 (Single-Digit Impossibility for Semantic Recovery).}
\textit{Let \(C_{\beta}^{L,R}\) be a canonically injective finite-window codebook,
and let \(d^\star \in C_{\beta}^{L,R}\) be an original admissible codeword. Let
\(\widetilde d\) be the result of a genuine single-digit corruption at position
\(j\), so that \(\widetilde d_j\neq d^\star_j\). Under exact structural repair
of the observed corrupted word \(\widetilde d\), semantic recovery of
\(d^\star\) is impossible.}

\textit{Proof.}
Let the represented value of the original codeword be
\[
v^\star=\operatorname{val}_{\beta}^{L,R}(d^\star).
\]
The single-digit corruption introduces the numerical delta
\[
\Delta=(\widetilde d_j-d^\star_j)\beta^j.
\]
Since \(\beta>1\), we have \(\beta^j\neq 0\), and since the corruption is
genuine, \(\widetilde d_j\neq d^\star_j\). Hence \(\Delta\neq 0\), and therefore
\[
\operatorname{val}_{\beta}^{L,R}(\widetilde d)\neq v^\star.
\]
By \hyperref[def:structural-exactness]{Definition~12}, exact structural repair
produces an admissible output \(\widehat d\) satisfying
\[
\operatorname{val}_{\beta}^{L,R}(\widehat d)
=
\operatorname{val}_{\beta}^{L,R}(\widetilde d).
\]
Consequently,
\[
\operatorname{val}_{\beta}^{L,R}(\widehat d)\neq v^\star,
\]
so the repaired output cannot semantically recover the original value.
\qed

Consequently, for canonical initial states, the probability of single-digit
semantic survival under exact repair is zero. In bounded-window experiments,
any apparent deviation from this ideal limit must be traced to boundary loss,
truncation, overflow, or another explicitly recorded finite-window artifact.

\phantomsection\label{thm:algebraic-kernel}
\textbf{Theorem 2 (Algebraic-Kernel Condition for Burst Semantic Survival).}
\textit{Let \(d^\star \in C_{\beta}^{L,R}\) undergo a localized multi-digit burst
corruption resulting in \(\widetilde d\). Let
\[
\Delta = \widetilde d - d^\star
\]
be the digit-wise error vector. Under exact structural repair of the observed
corrupted word \(\widetilde d\), semantic survival
\[
\operatorname{val}_{\beta}^{L,R}(\widehat d)
=
\operatorname{val}_{\beta}^{L,R}(d^\star)
\]
occurs if and only if
\[
\operatorname{val}_{\beta}(\Delta)=0.
\]}

\textit{Proof.}
Under exact structural repair,
\[
\operatorname{val}_{\beta}^{L,R}(\widehat d)
=
\operatorname{val}_{\beta}^{L,R}(\widetilde d)
=
\operatorname{val}_{\beta}^{L,R}(d^\star)
+
\operatorname{val}_{\beta}(\Delta).
\]
Semantic survival requires
\[
\operatorname{val}_{\beta}^{L,R}(\widehat d)
=
\operatorname{val}_{\beta}^{L,R}(d^\star),
\]
which holds if and only if
\[
\operatorname{val}_{\beta}(\Delta)=0.
\]
Thus, the perturbation vector \(\Delta\) must lie in the kernel of the evaluation
map. \qed

For the golden-ratio system, this kernel condition is realized by local
value-preserving substitutions such as \(100 \leftrightarrow 011\). Thus, burst
semantic survival is not a generic correction effect; it occurs only when the
error vector matches a value-preserving algebraic identity of the base.

\paragraph{Arithmetic trials.}
Let
\[
a,b\in C_{\varphi}^{L,R}
\]
be admissible operands. Their raw arithmetic state is
\[
c_i := a_i+b_i,
\qquad i\in I_{L,R},
\]
so that
\[
c=(c_i)_{i\in I_{L,R}}\in\{0,1,2\}^{I_{L,R}}.
\]
Let
\[
\overline d = R_{L,R}(c)
\]
be the canonicalized output.

\phantomsection\label{def:arithmetic-exactness}\textbf{Definition 14 (Arithmetic exactness).}
The arithmetic canonicalization is exact if
\[
\operatorname{val}_{\varphi}^{L,R}(\overline d)
=
\operatorname{val}_{\varphi}^{L,R}(a)
+
\operatorname{val}_{\varphi}^{L,R}(b).
\]

This criterion concerns arithmetic correctness, not fault recovery.

\paragraph{Quantitative metrics.}
Across the experiment families, we use the following metrics.

For comparative metrics, let
\[
X\in\{B,S,\Phi,T_3,T_4\}
\]
denote the representation system under consideration. We write \(\beta_X\) for
its base, \(A_X\) for its digit alphabet, \(C_X^{L,R}\) for its finite-window
canonical codebook, and
\[
\operatorname{val}_X^{L,R}(d)
:=
\sum_{i=-R}^{L} d_i\beta_X^i
\]
for the corresponding finite-window value map. For the standard binary and
signed-digit radix-\(2\) systems, \(\beta_B=\beta_S=2\); for the golden-ratio
system, \(\beta_\Phi=\varphi\); and for the tribonacci and tetranacci systems,
\(\beta_{T_3}\) and \(\beta_{T_4}\) denote the corresponding real multinacci
roots.

All residual error metrics in corruption trials compare the repaired output
\(\widehat d\) with the original pre-fault word \(d^\star\). By contrast,
structural exactness / value preservation compares \(\widehat d\) with the
corrupted word \(\widetilde d\).

\emph{Detectability rate.}

If \(T\) corruption trials are performed for system \(X\), the structural
detectability rate is
\[
\delta_{\mathrm{det},X}
:=
\frac{1}{T}
\sum_{t=1}^{T}
\mathbf{1}
\big[
\widetilde d^{(t)} \notin C_X^{L,R}
\big].
\]

\emph{Recovery / survival rates.}

For \(Y\in\{\mathrm{structural},\mathrm{exact},\mathrm{semantic}\}\), where
\(\mathrm{structural}\) denotes structural correctness, \(\mathrm{exact}\)
denotes structural exactness, and \(\mathrm{semantic}\) denotes semantic
survival, define
\[
\rho_{Y,X}
:=
\frac{1}{T}
\sum_{t=1}^{T}
\mathbf{1}
\big[
\text{trial } t \text{ in system } X \text{ satisfies criterion } Y
\big].
\]

\emph{Arithmetic exactness rate.}

For arithmetic trials in system \(X\),
\[
\rho_{\mathrm{arith},X}
:=
\frac{1}{T}
\sum_{t=1}^{T}
\mathbf{1}
\big[
\operatorname{val}_X^{L,R}(\overline d^{(t)})
=
\operatorname{val}_X^{L,R}(a^{(t)})
+
\operatorname{val}_X^{L,R}(b^{(t)})
\big].
\]

\emph{Mean absolute error (MAE) and worst-case error.}

For corruption trials in system \(X\), the raw mean absolute error is
\[
\mathrm{MAE}_X
:=
\frac{1}{T}
\sum_{t=1}^{T}
\left|
\operatorname{val}_X^{L,R}(\widehat d^{(t)})
-
\operatorname{val}_X^{L,R}(d^{\star(t)})
\right|.
\]
The raw worst-case error is
\[
\Delta_{\max,X}
:=
\max_{1\le t\le T}
\left|
\operatorname{val}_X^{L,R}(\widehat d^{(t)})
-
\operatorname{val}_X^{L,R}(d^{\star(t)})
\right|.
\]

\emph{Normalized mean absolute error (NMAE).}

Because the representation systems use different bases, digit alphabets, and
finite-window dynamic ranges, raw absolute errors are not directly comparable
across systems. To obtain a dimensionless cross-system measure, we define the
maximum representable value in the canonical finite-window codebook:
\[
V_{\max,X}^{L,R}
:=
\max_{d\in C_X^{L,R}}
\operatorname{val}_X^{L,R}(d).
\]
We then report the normalized mean absolute error and normalized worst-case
error:
\[
\mathrm{NMAE}_X
:=
\frac{\mathrm{MAE}_X}{V_{\max,X}^{L,R}},
\qquad
\mathrm{N}\Delta_{\max,X}
:=
\frac{\Delta_{\max,X}}{V_{\max,X}^{L,R}}.
\]
The normalized worst-case error \( \mathrm{N}\Delta_{\max,X} \) may exceed
\(1\) when the residual difference between the repaired word and the original
word is larger than the positive maximum canonical value. It is therefore a
dimensionless severity ratio rather than a probability.

\emph{Normalization / canonicalization success probability.}

For either corruption or arithmetic trials, let \(z^{(t)}\) denote the final
word produced in trial \(t\), and let \(H^{(t)}\in\{0,1\}\) be the halt flag,
where \(H^{(t)}=1\) means that the procedure terminated within the prescribed
rewrite budget. Let \(A_{\mathrm{tar}}\) be the target digit alphabet, let
\(\mathcal P_{\mathrm{bad}}\) be the set of forbidden admissibility patterns
for the system under consideration, and let \(\operatorname{Blk}(I_{L,R})\)
denote the set of contiguous index blocks contained in \(I_{L,R}\). Define
the residual defect set
\[
\mathcal Q_{\mathrm{res}}^{(t)}
:=
\left\{i\in I_{L,R}: z_i^{(t)}\notin A_{\mathrm{tar}}\right\}
\cup
\left\{J\in \operatorname{Blk}(I_{L,R}):
\left.z^{(t)}\right|_J\in \mathcal P_{\mathrm{bad}}\right\}.
\]
Then
\[
p_{\mathrm{norm}}
:=
\frac{1}{T}
\sum_{t=1}^{T}
\mathbf{1}\!\left\{
H^{(t)}=1
\ \text{and}\
\mathcal Q_{\mathrm{res}}^{(t)}=\varnothing
\right\}.
\]

\emph{Propagation depth.}

For one trial, let \(\pi(t)\) be the number of distinct anchor positions at which a
rewrite rule is applied. Then
\[
\overline{\pi}
:=
\frac{1}{T}
\sum_{t=1}^{T}
\pi(t),
\qquad
\pi_{\max}
:=
\max_{1\le t\le T}\pi(t).
\]
The quantity \(\pi\) measures the spatial extent of the cascade under the
deterministic sequential evaluation convention used in the experiments. It is a
structural propagation metric. As discussed in
\hyperref[app:parallel-canonicalization]{Appendix~B}, related canonicalization
procedures may admit parallel-prefix implementations with \(O(\log W)\)
critical-path depth~\cite{Ahlbach2013Zeckendorf}. Circuit-level quantities such
as switching activity, area, energy, and latency require a separate synthesized
implementation and are not estimated by \(\pi\).

\emph{Normalization cost.}

Let \(\kappa(t)\) be the total number of elementary rewrite, carry-resolution, or
recoding operations in a trial. Then
\[
\overline{\kappa}
:=
\frac{1}{T}
\sum_{t=1}^{T}
\kappa(t),
\qquad
\kappa_{\max}
:=
\max_{1\le t\le T}\kappa(t).
\]

\emph{Cost accounting across systems.}

The unit of \(\kappa\) is native to the deterministic procedure assigned to each
system. For \hyperref[sys:standard-binary]{System~B}, arithmetic
canonicalization after digit-wise addition is ordinary radix-\(2\) carry
propagation. The raw additive state has digits
\(c_i=a_i+b_i\in\{0,1,2\}\). The procedure scans from the least significant
position to the most significant position, resolves the current binary digit, and
propagates a carry when needed. One unit of cost is counted for each visited
digit position at which the carry state or raw digit must be resolved. A carry
leaving the most significant boundary \(i>L\) is recorded as overflow and
discarded under the bounded-window convention. No structural normalization is
required after corruption in System~B, because every binary word is
syntactically admissible.

For \hyperref[sys:signed-digit-radix-2]{System~S}, the reported cost is the
number of iterations or emitted digit positions of the deterministic
bounded-window NAF recoding procedure described below. This makes System~S a
global integer-recoding baseline.

For \hyperref[sys:golden-ratio-base]{System~\(\Phi\)} and the multinacci systems
\(T_3,T_4\), one unit of cost is one application of the specified local rewrite
rule or temporary-digit resolution rule under the deterministic MSB-first
priority convention. Full rescans used to locate the next defect are not counted;
\(\kappa\) measures canonicalization activity, while \(\pi\) records the spatial
extent of propagation.

Thus, \(\kappa\) is directly comparable within the local-rewrite family
\(\Phi,T_3,T_4\). Across Binary, Signed NAF, and the non-integer-base systems,
\(\kappa\) is a native deterministic procedure cost used for reproducibility and
baseline comparison.

\emph{Representation overhead and sparsity.}
For a system using a window of width $W$, we record its code length and codebook sparsity
\[
s
:=
\frac{|C|}{|\text{full ambient digit space}|}.
\]
For the golden-ratio system this becomes
\[
s_{\varphi}^{L,R}
=
\frac{|C_{\varphi}^{L,R}|}{2^W}
=
\frac{F_{W+2}}{2^W},
\]
which decays exponentially with $W$.

This sparsity metric is syntactic rather than semantic. 
It measures how small the canonical admissible codebook is relative to the full ambient digit space. 
Semantic multiplicity, when considered, is measured separately by equivalence classes in the ambient space under $\operatorname{val}_{\varphi}^{L,R}$, not by counting multiple representatives inside $C_\varphi^{L,R}$.

\emph{Average nonzero density.}
To compare representational concentration, we also record the relative Hamming weight
\[
\omega(d):=\frac{1}{W}\left|\{i\in I_{L,R}: d_i\neq 0\}\right|,
\]
and average it over sampled words.

\emph{Round-trip exactness.}
For clean representation experiments on exactly representable inputs, we record the encode--decode round-trip rate
\[
\rho_{\mathrm{rt}}
:=
\frac{1}{T}
\sum_{t=1}^{T}
\mathbf{1}
\big[
\operatorname{decode}(\operatorname{encode}(x^{(t)})) = x^{(t)}
\big].
\]
Approximation error for arbitrary real inputs is outside the present finite-window protocol.

\subsection{Comparison Systems and Experimental Protocol}\label{subsec:comparison-systems-experimental-protocol}

To assess what intrinsic redundancy actually provides, the golden-ratio system must be compared against systems representing different sources of redundancy: alphabet-driven redundancy, syntactic sparsity of a canonical codebook, and ambient semantic non-uniqueness before canonicalization.

\paragraph{Comparison systems.}

\phantomsection\label{sys:standard-binary}\textbf{System B: Standard binary.}
This system uses base $2$ and alphabet $\{0,1\}$. It is the non-redundant baseline. Every valid word is syntactically admissible, so structural detectability is absent, and no normalization is required after corruption. Arithmetic trials are different: digit-wise addition may produce temporary digits in \(\{0,1,2\}\), which are canonicalized by ordinary radix-\(2\) carry propagation under the cost-accounting convention defined above.

\phantomsection\label{sys:signed-digit-radix-2}\textbf{System S: Signed-digit radix-\(2\).}

This system uses base \(2\) and the alphabet \(\{-1,0,1\}\). 
It is the explicit redundancy baseline. 
Redundancy here is alphabet-driven rather than algebraic. 
Specifically, we strictly enforce the Non-Adjacent Form (NAF) as the canonical stored codebook, meaning that no two nonzero digits are adjacent. 
As in the \(\varphi\)-system, this canonical stored codebook should be distinguished from the larger ambient digit space from which non-canonical intermediate states may arise. 
This structural constraint provides a theoretical counterpart to the strict no-\(11\) admissibility rule of the golden-ratio system, allowing for a comparison of syntactic sparsity under different alphabets.

For reproducibility, restoration and arithmetic canonicalization in System~S are
performed by deterministic bounded-window NAF recoding. Given a finite-window
signed-digit word \(z\), we first interpret it as the scaled integer
\[
N(z):=\sum_{i=-R}^{L} z_i\,2^{i+R}.
\]
We then apply the standard Reitwiesner NAF recoding procedure~\cite{Reitwiesner1960}: while the current
integer \(N\neq 0\), if \(N\) is odd choose
\[
u = 2-(N \bmod 4)\in\{-1,1\},
\]
replace \(N\) by \((N-u)/2\), and emit digit \(u\); if \(N\) is even, emit digit
\(0\) and replace \(N\) by \(N/2\). The emitted digits are mapped back to the
window positions \(i=-R,\ldots,L\). Any emitted nonzero digit beyond position
\(L\) is recorded as overflow and discarded, while the finite-window scaling
prevents the creation of positions below \(-R\). This convention makes the NAF
baseline deterministic and removes any ambiguity between left-to-right and
right-to-left local restoration schedules.

The cost reported for System~S is therefore a NAF recoding cost: one unit of
\(\kappa\) is counted for each iteration of the integer recoding loop,
equivalently for each emitted digit position considered by the bounded-window
procedure. This choice makes the signed-digit baseline deterministic, but it
should not be identified with the local algebraic rewrite count used for
\(\Phi\), \(T_3\), and \(T_4\). We therefore interpret System~S as a global
recoding baseline rather than as a local-rewrite implementation of signed-digit
arithmetic.

\phantomsection\label{sys:golden-ratio-base}\textbf{System $\Phi$: Golden-ratio base.}
This is the primary system of the paper, with base $\varphi$ and alphabet $\{0,1\}$. 
Its ambient semantic redundancy is generated by the local identity
\[
100 \leftrightarrow 011.
\]
The stored canonical codebook, however, is the no-$11$ admissible subset.

\phantomsection\label{sys:tribonacci-base}\textbf{System T$_3$: Tribonacci base.}
This system uses the tribonacci base $\beta_3$, the real root of
\[
x^3 = x^2 + x + 1,
\]
with alphabet $\{0,1\}$. Its shortest local semantic equivalence is
\[
1000 \leftrightarrow 0111.
\]
It is included to test whether the observed effects in the golden-ratio system are specific to $\varphi$ or instead related more generally to rewrite locality within the multinacci family~\cite{Hare2007,Frougny2011Parallel}.

\phantomsection\label{sys:tetranacci-base}\textbf{System T$_4$: Tetranacci base.}
This system uses the tetranacci base $\beta_4$, the real root of
\[
x^4 = x^3 + x^2 + x + 1,
\]
again with alphabet $\{0,1\}$. Its shortest local semantic equivalence is
\[
10000 \leftrightarrow 01111.
\]
It extends the multinacci comparison by testing whether increasing rewrite length weakens local semantic survival and increases canonicalization cost.

The multinacci systems are included not because they are equally mature implementation candidates, but because they provide a controlled way to test whether local rewrite length influences robustness and canonicalization behavior.

\paragraph{Experimental families.}
The experimental study is organized into seven
benchmarks.

\phantomsection\label{exp:clean-representation}\textbf{Experiment 1: Clean representation benchmark.}
No corruption is applied. We compare:
\begin{itemize}
    \item code length / representational overhead,
    \item codebook sparsity,
    \item average nonzero density,
    \item round-trip exactness for exactly representable inputs.
\end{itemize}
Approximation error for arbitrary real inputs is outside the present
finite-window protocol.

\phantomsection\label{exp:arithmetic-canonicalization}\textbf{Experiment 2: Arithmetic canonicalization benchmark.}
Pairs of admissible words are added digit-wise to produce raw arithmetic states,
which are then canonicalized. We compare:
\begin{itemize}
    \item arithmetic exactness rate $\rho_{\mathrm{arith}}$,
    \item canonicalization success probability $p_{\mathrm{norm}}$,
    \item mean and maximum propagation depth,
    \item mean and maximum normalization cost,
    \item overflow / truncation frequency,
    \item distribution of cascade lengths.
\end{itemize}

\phantomsection\label{exp:single-digit-corruption}\textbf{Experiment 3: Single-digit corruption benchmark.}
A single uniformly random position is flipped in each sampled codeword. We compare:
\begin{itemize}
    \item structural detectability rate,
    \item structural correctness rate,
    \item structural exactness rate,
    \item semantic survival rate,
    \item MAE and worst-case error,
    \item normalization success probability,
    \item propagation depth and normalization cost.
\end{itemize}

\phantomsection\label{exp:local-burst-corruption}\textbf{Experiment 4: Local burst corruption benchmark.}

A contiguous burst window of width
\[
b\in\{2,3,4,5\}
\]
is injected using the burst-corruption model of
\hyperref[def:burst-corruption]{Definition~6}. The starting position is sampled
uniformly from \(J_{L,R}(b)\), and a uniformly random nonzero mask is sampled on
the corresponding interval \(B_{j,b}\). We report the same metrics as in the
single-digit benchmark, with particular emphasis on semantic survival and
residual error magnitude.

\phantomsection\label{exp:exhaustive-verification}\textbf{Experiment 5: Exhaustive small-window verification.}
To validate the single-digit impossibility result
\hyperref[thm:no-free-lunch]{Proposition~1} and the algebraic-kernel condition
\hyperref[thm:algebraic-kernel]{Theorem~2} without Monte Carlo sampling
artifacts, we perform an exhaustive combinatorial evaluation for small window
widths ($W \le 16$). For each system, we iterate over:
\begin{itemize}
    \item all admissible words $d^\star \in C_X^{L,R}$,
    \item all possible single-digit corruptions,
    \item all valid burst masks of width $b \in \{2,3,4,5\}$.
\end{itemize}
The goal is to confirm that the empirical probability of single-digit semantic survival is exactly zero (excluding boundary artifacts) and to establish exact baseline frequencies for structural detectability.

\phantomsection\label{exp:algebraic-burst}\textbf{Experiment 6: Algebraic burst injection benchmark.}
To empirically test the algebraic-kernel condition
(\hyperref[thm:algebraic-kernel]{Theorem~2}), we inject specific algebraic
perturbation patterns rather than uniform random noise. We apply targeted
multi-digit substitutions derived directly from the fundamental identities of the
bases (e.g., forcing \(011\to100\) and \(100\to011\) in \(\Phi\), or
\(0111\to1000\) in \(T_3\)). We measure whether the resulting semantic survival
rate strictly aligns with the theoretical kernel-membership prediction, directly
linking algebraic equivalence to fault resilience.

\phantomsection\label{exp:guard-digits}\textbf{Experiment 7: Guard-digit truncation analysis.}
To evaluate the practical bounds of \hyperref[lem:guard-digits]{Lemma~2} and demonstrate that bounded-window failure is a controllable hardware trade-off, we run arithmetic canonicalization with an extended fractional boundary using $g$ guard digits, where $g \in \{0,2,4,8,12,16\}$. We compare:
\begin{itemize}
    \item arithmetic exactness and residual semantic error,
    \item truncation rate and maximum propagation depth,
    \item normalization cost ($\kappa$) as $g$ increases.
\end{itemize}

The multinacci comparison is embedded across \hyperref[exp:clean-representation]{Experiments~1}, \hyperref[exp:arithmetic-canonicalization]{2}, \hyperref[exp:single-digit-corruption]{3}, \hyperref[exp:local-burst-corruption]{4}, \hyperref[exp:exhaustive-verification]{5}, \hyperref[exp:algebraic-burst]{6}, and \hyperref[exp:guard-digits]{7}, ensuring a consistent evaluation of variable-length rewrite locality.

\paragraph{Sampling protocol.}
For each system and each window width
\[
W\in\{8,12,16,20,24\},
\]
we sample admissible codewords uniformly from the corresponding finite-window codebook. For \hyperref[sys:standard-binary]{System~B}, this is simply a uniform draw from $\{0,1\}^W$. For \hyperref[sys:signed-digit-radix-2]{System~S}, we sample from the canonical signed-digit subset. For \hyperref[sys:golden-ratio-base]{System~$\Phi$},
\hyperref[sys:tribonacci-base]{T$_3$}, and
\hyperref[sys:tetranacci-base]{T$_4$}, direct rejection sampling becomes
computationally prohibitive due to the exponentially decaying density of valid
strings. Instead, admissible codewords are generated by dynamic-programming
unranking of the corresponding constrained regular language. The mathematical
uniqueness of the canonical representations is grounded in the Zeckendorf and
generalized numeration framework~\cite{Zeckendorf1972,Akiyama1998,Frougny2002},
while the algorithmic use of ranking and unranking follows standard
combinatorial generation methodology for constrained languages and numeration
systems~\cite{AlloucheShallit2003,KreherStinson1999}. Concretely, we precompute
the number of admissible continuations for each remaining suffix length and
state, draw an integer uniformly from \([0,|C|-1]\), and then select successive
digits by subtracting the appropriate continuation counts. This maps uniformly
drawn integers to unique canonical codewords without rejection, giving exact
uniform sampling of the constrained finite-window codebook in polynomial time.

We set
\[
L=\lfloor W/2\rfloor,
\qquad
R=W-L-1,
\]
so that integer and fractional positions remain balanced.

\paragraph{Fault injection protocol.}

In the single-digit benchmark, each trial chooses one position uniformly at random from
\[
I_{L,R}
\]
and corrupts that position according to the alphabet-aware single-digit corruption
rule of \hyperref[def:single-digit-corruption]{Definition~5}.

In the burst benchmark, each trial first chooses a burst width
\[
b\in\{2,3,4,5\},
\]
then chooses a starting position uniformly at random from
\[
J_{L,R}(b)=\{-R,\ldots,L-b+1\},
\]
so that the burst window
\[
B_{j,b}=\{j,j+1,\ldots,j+b-1\}
\]
is fully contained in \(I_{L,R}\). Finally, the trial samples a uniformly random
nonzero mask
\[
\varepsilon\in\{0,1\}^{b}\setminus\{0^b\}
\]
on \(B_{j,b}\). Each active mask position is corrupted according to the
alphabet-aware single-digit corruption rule of
\hyperref[def:single-digit-corruption]{Definition~5}.

In the arithmetic benchmark, pairs of admissible codewords are sampled
independently and added digit-wise to produce raw arithmetic states, which are
then canonicalized by the bounded-window procedure.

\paragraph{Number of trials.}

Each combination of system, window width, and experiment family is evaluated over
\[
T=10{,}000
\]
independent trials. The reported tables and figures give empirical point
estimates for the corresponding rates, residual errors, and repair-cost
statistics under the fixed random seed and bounded-window conventions specified
above.

\subsection{Scope and Limitations of the Finite-Window Protocol}
\label{subsec:scope-limitations-finite-window}

To isolate the effect of representation structure under a fixed storage budget, all systems are evaluated under:
\begin{enumerate}
    \item equal window width $W$,
    \item equal corruption-location distribution,
    \item equal burst-width distribution,
    \item deterministic bounded-window canonicalization or repair,
    \item native cost accounting for the specified procedure of each system.
\end{enumerate}

The maximal rewrite budget is set to
$$\kappa_{\max}^{\mathrm{budget}}=3W.$$
This serves as an experimental reproducibility cutoff rather than a proven upper bound on all finite-window cascades. Trials exceeding this budget are recorded as failures, but are not classified as true nontermination or unresolvable propagation cycles without separate mathematical proofs.

The equal-$W$ protocol fixes stored digit positions, not numerical dynamic range, as different bases and languages induce varying value distributions. For residual errors, we report NMAE normalized by $V_{\max,X}^{L,R}$—a scaling correction rather than a full matched-range experiment. A robustness check aligning maximum representable values ($V_{\max,X}^{L,R}$) is provided in \hyperref[app:matched_range]{Appendix~\ref*{app:matched_range}}, while the equal-$W$ tables remain the primary fixed-storage comparison.

Reported rates (detection, survival, residual error, repair cost) reflect the specified sampling ensemble. Unless stated otherwise, admissible words, corruption locations, and independent operands are sampled uniformly, with burst masks following the aforementioned distributions. These represent controlled finite-window benchmarks, not workload-independent probabilities for arbitrary programs or physical fault environments.

The addition-only arithmetic benchmark isolates upward carry propagation and bounded-window canonicalization over minimal alphabets. Subtractions, signed states, negative borrows, and mixed workloads remain outside this protocol's scope.

The logically contiguous burst model stresses local admissibility constraints and rewrite neighborhoods. While not a complete physical model (e.g., for interleaved DRAM, SRAM, or particle strikes), more detailed multi-bit upset models could be evaluated within this same framework.

Finally, propagation and cost metrics ($\pi$ and $\kappa$) evaluate the selected deterministic procedures. They act as structural indicators of normalization activity under the chosen scheduler, not as synthesized hardware metrics (latency, energy, area) or implementation-independent algebraic invariants.

%% file: figures/fig_finite_window_register.tex
\begin{figure}[htbp]
\centering

\begin{tikzpicture}[
    x=1.08cm,
    y=0.98cm,
    >=Stealth,
    line cap=round,
    line join=round,
    every node/.style={font=\small}
]
 
\def\Y{0}
\def\H{1.0}

\foreach \x in {0,1,2,3,4}{
    \fill[blue!8] (\x,\Y) rectangle (\x+1,\Y+\H);
}
\foreach \x in {5,6,7,8,9}{
    \fill[teal!8] (\x,\Y) rectangle (\x+1,\Y+\H);
}
\foreach \x in {0,1,...,9}{
    \draw[thick] (\x,\Y) rectangle (\x+1,\Y+\H);
}

\draw[very thick, violet!70!black] (5,\Y-0.14) -- (5,\Y+\H+0.14);

\node at (2.5,0.5) {\(\cdots\)};
\node at (7.5,0.5) {\(\cdots\)};

\draw[dashed, gray!70] (-0.08,\Y-0.28) -- (-0.08,\Y+\H+0.42);
\draw[dashed, gray!70] (10.08,\Y-0.28) -- (10.08,\Y+\H+0.42);

\node[font=\scriptsize, align=center] at (0,1.95)
{most significant\\boundary};

\node[font=\scriptsize, align=center] at (10,1.95)
{least significant\\boundary};

\node[font=\small\bfseries, text=blue!50!black] at (2.5,1.52)
{integer positions};

\node[font=\small\bfseries, text=teal!50!black] at (7.5,1.52)
{fractional positions};

\node[font=\scriptsize, align=center, text=violet!70!black] at (5,1.80)
{radix\\point};

\node[font=\scriptsize] at (0.5,-0.30) {\(L\)};
\node[font=\scriptsize] at (1.5,-0.30) {\(L-1\)};
\node[font=\scriptsize] at (3.5,-0.30) {\(1\)};
\node[font=\scriptsize] at (4.5,-0.30) {\(0\)};
\node[font=\scriptsize] at (5.5,-0.30) {\(-1\)};
\node[font=\scriptsize] at (6.5,-0.30) {\(-2\)};
\node[font=\scriptsize] at (8.5,-0.30) {\(-R+1\)};
\node[font=\scriptsize] at (9.5,-0.30) {\(-R\)};

\draw[<->, thick] (0,-1.00) -- (10,-1.00);
\node[font=\small] at (5,-1.34)
{hardware window \(I_{L,R}=\{-R,\dots,L\}\), \quad \(W=L+R+1\)};

\node[
    draw=red!60!black,
    fill=red!8,
    rounded corners=2pt,
    inner sep=4pt,
    align=center,
    font=\scriptsize
] (ovf) at (2.15,-2.65)
{\textbf{overflow}\\ write to \(i>L\) is discarded};

\draw[->, thick, red!60!black]
    (0,0.50) -- (0,-1.90) -- (2.15,-1.90) -- (ovf.north);

\node[
    draw=orange!70!black,
    fill=orange!10,
    rounded corners=2pt,
    inner sep=4pt,
    align=center,
    font=\scriptsize
] (trc) at (7.85,-2.65)
{\textbf{truncation}\\ write to \(i<-R\) is discarded};

\draw[->, thick, orange!70!black]
    (10,0.50) -- (10,-1.90) -- (7.85,-1.90) -- (trc.north);

\end{tikzpicture}

\caption{Finite hardware window \(I_{L,R}\) in the bounded-window model. The
radix point lies between positions \(0\) and \(-1\); writes beyond \(i>L\) are
counted as overflow, and writes beyond \(i<-R\) as truncation.}
\label{fig:finite-window-register}
\end{figure}
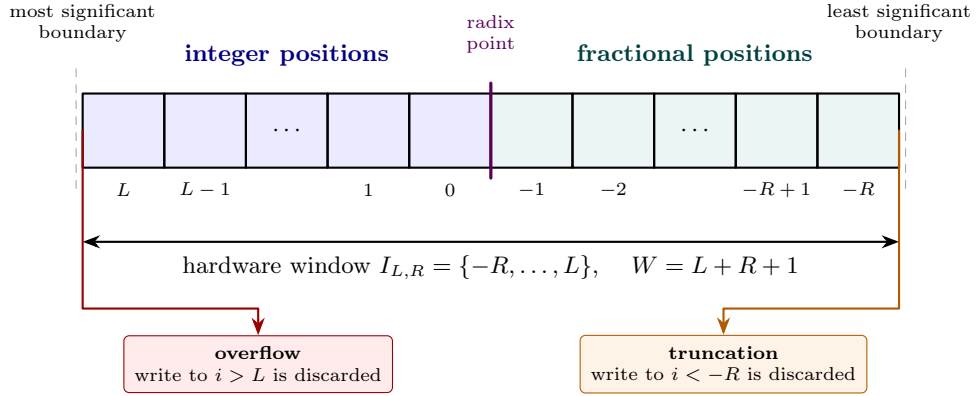

%% file: figures/fig4_fault_injection_diagram.tex
\begin{figure}[h]
\centering
\begin{tikzpicture}[
    >=Stealth,
    cell/.style={minimum width=0.78cm, minimum height=0.70cm,
                 draw=black!35, inner sep=0pt, font=\ttfamily\small},
    cellok/.style={cell, fill=green!10, draw=green!35!black},
    cellfault/.style={cell, fill=red!15, draw=red!55!black, thick},
    cellchanged/.style={cell, fill=yellow!18, draw=orange!55!black},
    celldone/.style={cell, fill=green!10, draw=green!35!black},
    rowlabel/.style={font=\small\bfseries, anchor=east},
    sidelabel/.style={font=\small\itshape, text=black!55, anchor=west},
]

\def\cw{0.78}  
\def\nc{8}      

\foreach \p/\l in {0/$d_7$, 1/$d_6$, 2/$d_5$, 3/$d_4$, 4/$d_3$, 5/$d_2$, 6/$d_1$, 7/$d_0$} {
    \node[font=\scriptsize, text=black!40] at (\p*\cw + \cw/2, 0.58) {\l};
}

\def\yone{0}
\node[rowlabel] at (-0.65, \yone) {$\mathbf{d}^{\star}$};
\foreach \p/\v in {0/0, 1/1, 2/0, 3/1, 4/0, 5/0, 6/1, 7/0} {
    \node[cellok] at (\p*\cw + \cw/2, \yone) {\v};
}
\node[sidelabel] at (\nc*\cw + 0.20, \yone + 0.12)  {admissible};
\node[font=\scriptsize, text=black!45, anchor=west]
    at (\nc*\cw + 0.20, \yone - 0.18)
    {$\mathrm{val}_{\varphi}\!=\!\varphi^{6}\!+\!\varphi^{4}\!+\!\varphi$};

\draw[->, thick, red!60!black]
    (4*\cw + \cw/2, \yone - 0.42) -- (4*\cw + \cw/2, \yone - 1.18)
    node[midway, right=3pt, font=\scriptsize, text=red!60!black]
    {flip $d_3$: $0 \!\to\! 1$};

\def\ytwo{-1.60}
\node[rowlabel] at (-0.65, \ytwo) {$\tilde{\mathbf{d}}$};
\foreach \p/\v/\s in {0/0/cell, 1/1/cell, 2/0/cell, 3/1/cellfault, 4/1/cellfault, 5/0/cell, 6/1/cell, 7/0/cell} {
    \node[\s] at (\p*\cw + \cw/2, \ytwo) {\v};
}
\node[sidelabel] at (\nc*\cw + 0.20, \ytwo + 0.12)  {inadmissible};
\node[font=\scriptsize, text=red!50!black, anchor=west]
    at (\nc*\cw + 0.20, \ytwo - 0.18)
    {$\mathrm{val}_{\varphi}\!=\!\varphi^{6}\!+\!\varphi^{4}\!+\!\varphi^{3}\!+\!\varphi$};

\draw[decorate, decoration={brace, amplitude=3.5pt, raise=2pt}, red!55!black]
    (3*\cw + 0.04, \ytwo + 0.39) -- (4*\cw + \cw - 0.04, \ytwo + 0.39)
    node[midway, above=6pt, font=\scriptsize, text=red!55!black] {\texttt{11}};

\draw[->, thick, orange!65!black]
    (3.5*\cw, \ytwo - 0.42) -- (3.5*\cw, \ytwo - 1.18)
    node[midway, right=3pt, font=\scriptsize, text=orange!65!black]
    {$0\underline{11} \to 100$};

\def\ythree{-3.20}
\node[rowlabel] at (-0.65, \ythree) {};
\foreach \p/\v/\s in {0/0/cell, 1/1/cellfault, 2/1/cellfault, 3/0/cellchanged, 4/0/cellchanged, 5/0/cell, 6/1/cell, 7/0/cell} {
    \node[\s] at (\p*\cw + \cw/2, \ythree) {\v};
}
\node[sidelabel] at (\nc*\cw + 0.20, \ythree) {cascade};

\draw[decorate, decoration={brace, amplitude=3.5pt, raise=2pt}, red!55!black]
    (1*\cw + 0.04, \ythree + 0.39) -- (2*\cw + \cw - 0.04, \ythree + 0.39)
    node[midway, above=6pt, font=\scriptsize, text=red!55!black] {new \texttt{11}};

\draw[->, thick, orange!65!black]
    (1.5*\cw, \ythree - 0.42) -- (1.5*\cw, \ythree - 1.18)
    node[midway, right=3pt, font=\scriptsize, text=orange!65!black]
    {$0\underline{11} \to 100$};

\def\yfour{-4.80}
\node[rowlabel] at (-0.65, \yfour) {$\hat{\mathbf{d}}$};
\foreach \p/\v/\s in {0/1/celldone, 1/0/celldone, 2/0/celldone, 3/0/celldone, 4/0/celldone, 5/0/celldone, 6/1/celldone, 7/0/celldone} {
    \node[\s] at (\p*\cw + \cw/2, \yfour) {\v};
}
\node[sidelabel] at (\nc*\cw + 0.20, \yfour + 0.12)  {admissible};
\node[font=\scriptsize, text=green!30!black, anchor=west]
    at (\nc*\cw + 0.20, \yfour - 0.18)
    {$\mathrm{val}_{\varphi}\!=\!\varphi^{7}\!+\!\varphi
      \;=\; \mathrm{val}_{\varphi}(\tilde{\mathbf{d}})$};

\draw[->, thick, blue!35!black, densely dashed]
    (-0.65 - 0.50, \ytwo + 0.25) -- (-0.65 - 0.50, \yfour - 0.25)
    node[midway, rotate=90, anchor=south, font=\scriptsize, text=blue!35!black]
    {carry propagation};

\end{tikzpicture}
\caption{Fault injection and carry-cascade repair in the $\varphi$-system.
An admissible codeword $\mathbf{d}^{\star}$ is corrupted by a single-digit
flip at position~$d_3$, introducing the forbidden pattern \texttt{11}.
The directed rewrite rule $011 \to 100$ eliminates the violation but propagates
a carry to higher positions, creating a new \texttt{11} at positions $d_6, d_5$.
A second application of the same rule yields the admissible output
$\hat{\mathbf{d}}$. The repair is \emph{structurally exact}:
$\mathrm{val}_{\varphi}(\hat{\mathbf{d}}) =
 \mathrm{val}_{\varphi}(\tilde{\mathbf{d}})$,
but does not achieve semantic recovery, since 
$\mathrm{val}_{\varphi}(\hat{\mathbf{d}}) \neq
 \mathrm{val}_{\varphi}(\mathbf{d}^{\star})$,
illustrating the distinction between value preservation and semantic recovery.}
\label{fig:fault-pipeline}
\end{figure}
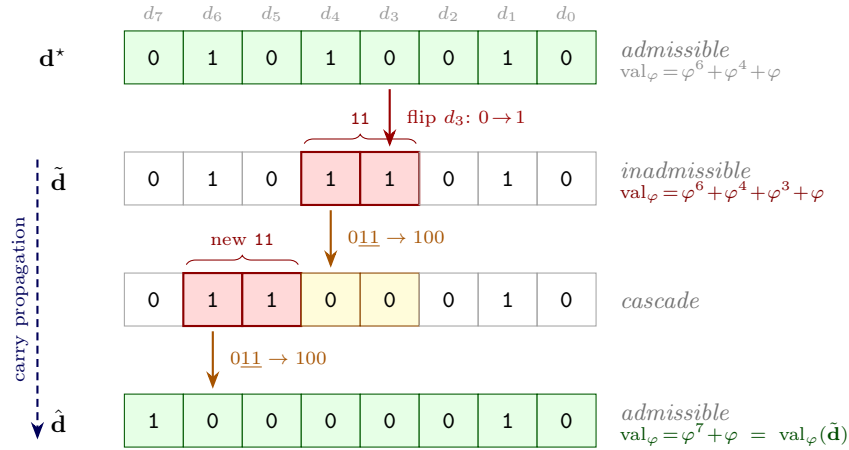

%% file: figures/fig4_5_fault_burst_corruption.tex
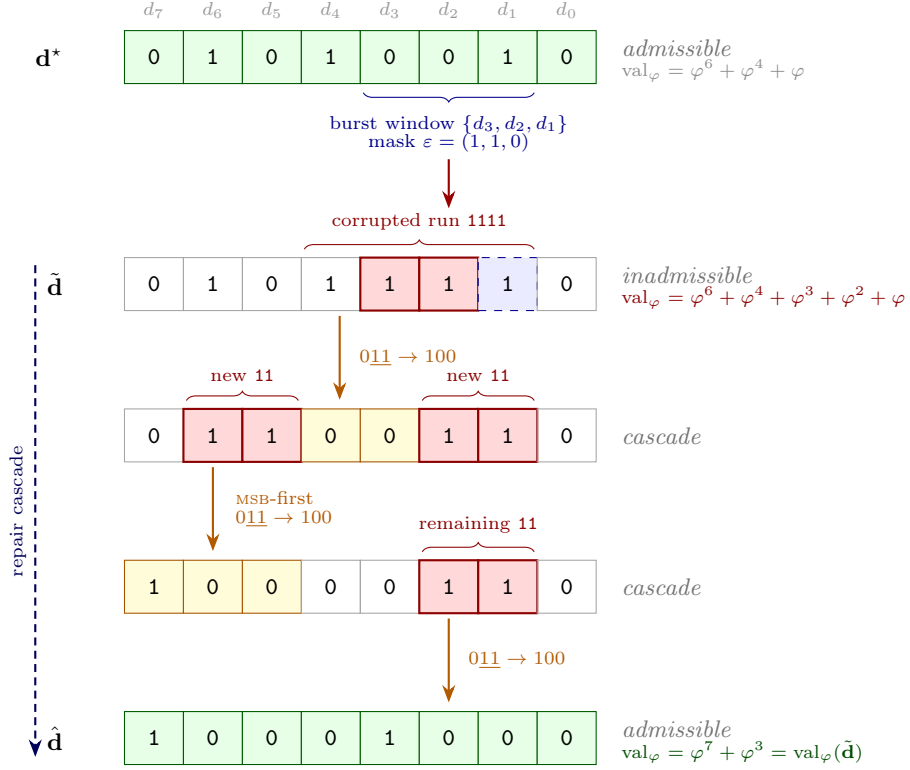
\begin{figure}[h]
\centering

\begin{tikzpicture}[
    >=Stealth,
    cell/.style={
        minimum width=0.78cm,
        minimum height=0.70cm,
        draw=black!35,
        inner sep=0pt,
        font=\ttfamily\small
    },
    cellok/.style={
        cell,
        fill=green!10,
        draw=green!35!black
    },
    cellfault/.style={
        cell,
        fill=red!15,
        draw=red!55!black,
        thick
    },
    cellmask/.style={
        cell,
        fill=blue!8,
        draw=blue!55!black,
        dashed
    },
    cellchanged/.style={
        cell,
        fill=yellow!18,
        draw=orange!65!black
    },
    celldone/.style={
        cell,
        fill=green!10,
        draw=green!35!black
    },
    rowlabel/.style={
        font=\small\bfseries,
        anchor=east
    },
    sidelabel/.style={
        font=\small\itshape,
        text=black!55,
        anchor=west
    },
    annotation/.style={
        font=\scriptsize,
        align=center
    }
]

\def\cw{0.78}

\def\nc{8}

\def\yone{0}
\def\ytwo{-3.00}
\def\ythree{-5.00}
\def\yfour{-7.00}
\def\yfive{-9.00}

\foreach \p/\l in {
    0/$d_7$,
    1/$d_6$,
    2/$d_5$,
    3/$d_4$,
    4/$d_3$,
    5/$d_2$,
    6/$d_1$,
    7/$d_0$
} {
    \node[
        font=\scriptsize,
        text=black!40
    ]
    at (\p*\cw+\cw/2,\yone+0.62)
    {\l};
}

\node[rowlabel]
    at (-0.70,\yone)
    {$\mathbf{d}^{\star}$};

\foreach \p/\v in {
    0/0,
    1/1,
    2/0,
    3/1,
    4/0,
    5/0,
    6/1,
    7/0
} {
    \node[cellok]
        at (\p*\cw+\cw/2,\yone)
        {\v};
}

\node[sidelabel]
    at (\nc*\cw+0.22,\yone+0.12)
    {admissible};

\node[
    font=\scriptsize,
    text=black!45,
    anchor=west
]
    at (\nc*\cw+0.22,\yone-0.18)
    {$\mathrm{val}_{\varphi}
      =\varphi^{6}+\varphi^{4}+\varphi$};

\draw[
    decorate,
    decoration={
        brace,
        mirror,
        amplitude=3.5pt,
        raise=2pt
    },
    blue!55!black
]
    (4*\cw+0.04,\yone-0.40)
    --
    (6*\cw+\cw-0.04,\yone-0.40)
    node[
        midway,
        below=7pt,
        annotation,
        text=blue!55!black
    ]
    {burst window $\{d_3,d_2,d_1\}$\\[-1pt]
     mask $\varepsilon=(1,1,0)$};

\draw[
    ->,
    thick,
    red!60!black
]
    (5*\cw+\cw/2,\yone-1.35)
    --
    (5*\cw+\cw/2,\ytwo+1.0);

\node[rowlabel]
    at (-0.70,\ytwo)
    {$\tilde{\mathbf{d}}$};

\node[cell]
    at (0*\cw+\cw/2,\ytwo)
    {0};

\node[cell]
    at (1*\cw+\cw/2,\ytwo)
    {1};

\node[cell]
    at (2*\cw+\cw/2,\ytwo)
    {0};

\node[cell]
    at (3*\cw+\cw/2,\ytwo)
    {1};

\node[cellfault]
    at (4*\cw+\cw/2,\ytwo)
    {1};

\node[cellfault]
    at (5*\cw+\cw/2,\ytwo)
    {1};

\node[cellmask]
    at (6*\cw+\cw/2,\ytwo)
    {1};

\node[cell]
    at (7*\cw+\cw/2,\ytwo)
    {0};

\node[sidelabel]
    at (\nc*\cw+0.22,\ytwo+0.12)
    {inadmissible};

\node[
    font=\scriptsize,
    text=red!55!black,
    anchor=west
]
    at (\nc*\cw+0.22,\ytwo-0.18)
    {$\mathrm{val}_{\varphi}
      =\varphi^{6}+\varphi^{4}
       +\varphi^{3}+\varphi^{2}+\varphi$};

\draw[
    decorate,
    decoration={
        brace,
        amplitude=3.5pt,
        raise=2pt
    },
    red!55!black
]
    (3*\cw+0.04,\ytwo+0.39)
    --
    (6*\cw+\cw-0.04,\ytwo+0.39)
    node[
        midway,
        above=6pt,
        annotation,
        text=red!55!black
    ]
    {corrupted run \texttt{1111}};

\draw[
    ->,
    thick,
    orange!70!black
]
    (3.65*\cw,\ytwo-0.42)
    --
    (3.65*\cw,\ythree+0.47)
    node[
        midway,
        right=4pt,
        annotation,
        text=orange!70!black
    ]
    {$0\underline{11}\to100$};

\node[rowlabel]
    at (-0.70,\ythree)
    {};

\node[cell]
    at (0*\cw+\cw/2,\ythree)
    {0};

\node[cellfault]
    at (1*\cw+\cw/2,\ythree)
    {1};

\node[cellfault]
    at (2*\cw+\cw/2,\ythree)
    {1};

\node[cellchanged]
    at (3*\cw+\cw/2,\ythree)
    {0};

\node[cellchanged]
    at (4*\cw+\cw/2,\ythree)
    {0};

\node[cellfault]
    at (5*\cw+\cw/2,\ythree)
    {1};

\node[cellfault]
    at (6*\cw+\cw/2,\ythree)
    {1};

\node[cell]
    at (7*\cw+\cw/2,\ythree)
    {0};

\node[sidelabel]
    at (\nc*\cw+0.22,\ythree)
    {cascade};

\draw[
    decorate,
    decoration={
        brace,
        amplitude=3.5pt,
        raise=2pt
    },
    red!55!black
]
    (1*\cw+0.04,\ythree+0.39)
    --
    (2*\cw+\cw-0.04,\ythree+0.39)
    node[
        midway,
        above=6pt,
        annotation,
        text=red!55!black
    ]
    {new \texttt{11}};

\draw[
    decorate,
    decoration={
        brace,
        amplitude=3.5pt,
        raise=2pt
    },
    red!55!black
]
    (5*\cw+0.04,\ythree+0.39)
    --
    (6*\cw+\cw-0.04,\ythree+0.39)
    node[
        midway,
        above=6pt,
        annotation,
        text=red!55!black
    ]
    {new \texttt{11}};

\draw[
    ->,
    thick,
    orange!70!black
]
    (1.50*\cw,\ythree-0.42)
    --
    (1.50*\cw,\yfour+0.47)
    node[
        midway,
        right=5pt,
        annotation,
        align=left,
        text=orange!70!black
    ]
    {\textsc{msb}-first\\[-1pt]
     $0\underline{11}\to100$};

\node[rowlabel]
    at (-0.70,\yfour)
    {};

\node[cellchanged]
    at (0*\cw+\cw/2,\yfour)
    {1};

\node[cellchanged]
    at (1*\cw+\cw/2,\yfour)
    {0};

\node[cellchanged]
    at (2*\cw+\cw/2,\yfour)
    {0};

\node[cell]
    at (3*\cw+\cw/2,\yfour)
    {0};

\node[cell]
    at (4*\cw+\cw/2,\yfour)
    {0};

\node[cellfault]
    at (5*\cw+\cw/2,\yfour)
    {1};

\node[cellfault]
    at (6*\cw+\cw/2,\yfour)
    {1};

\node[cell]
    at (7*\cw+\cw/2,\yfour)
    {0};

\node[sidelabel]
    at (\nc*\cw+0.22,\yfour)
    {cascade};

\draw[
    decorate,
    decoration={
        brace,
        amplitude=3.5pt,
        raise=2pt
    },
    red!55!black
]
    (5*\cw+0.04,\yfour+0.39)
    --
    (6*\cw+\cw-0.04,\yfour+0.39)
    node[
        midway,
        above=6pt,
        annotation,
        text=red!55!black
    ]
    {remaining \texttt{11}};

\draw[
    ->,
    thick,
    orange!70!black
]
    (5.50*\cw,\yfour-0.42)
    --
    (5.50*\cw,\yfive+0.47)
    node[
        midway,
        right=4pt,
        annotation,
        text=orange!70!black
    ]
    {$0\underline{11}\to100$};

\node[rowlabel]
    at (-0.70,\yfive)
    {$\hat{\mathbf{d}}$};

\foreach \p/\v in {
    0/1,
    1/0,
    2/0,
    3/0,
    4/1,
    5/0,
    6/0,
    7/0
} {
    \node[celldone]
        at (\p*\cw+\cw/2,\yfive)
        {\v};
}

\node[sidelabel]
    at (\nc*\cw+0.22,\yfive+0.12)
    {admissible};

\node[
    font=\scriptsize,
    text=green!30!black,
    anchor=west
]
    at (\nc*\cw+0.22,\yfive-0.18)
    {$\mathrm{val}_{\varphi}
      =\varphi^{7}+\varphi^{3}
      =\mathrm{val}_{\varphi}
        (\tilde{\mathbf{d}})$};

\draw[
    ->,
    thick,
    blue!35!black,
    densely dashed
]
    (-1.18,\ytwo+0.25)
    --
    (-1.18,\yfive-0.25)
    node[
        midway,
        rotate=90,
        anchor=south,
        annotation,
        text=blue!35!black
    ]
    {repair cascade};

\end{tikzpicture}

\caption{
Burst corruption and multi-step repair in the $\varphi$-system.
The burst mask $\varepsilon=(1,1,0)$ applied to the window
$\{d_3,d_2,d_1\}$ creates several forbidden \texttt{11} patterns.
Repeated MSB-first applications of $011\to100$ produce the admissible
output $\hat{\mathbf d}$. The repair preserves the corrupted value,
$\mathrm{val}_{\varphi}(\hat{\mathbf d})
=\mathrm{val}_{\varphi}(\tilde{\mathbf d})$,
but does not recover the original value.
}
\label{fig:burst-pipeline}

\end{figure}

%% file: figures/fig5_correctness_hierarchy_figure.tex
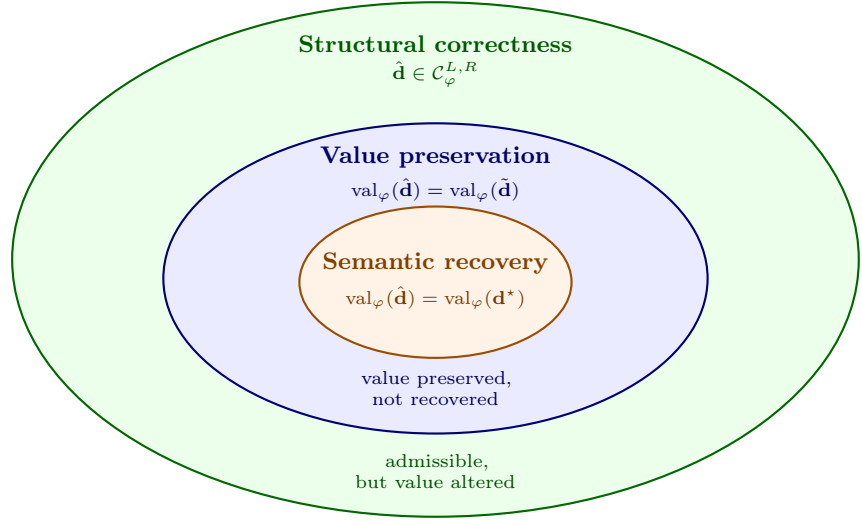
\begin{figure}[h]
\centering
\begin{tikzpicture}[
    every node/.style={font=\small},
]

\fill[green!8] (0,0) ellipse (5.6cm and 3.4cm);
\draw[green!40!black, thick] (0,0) ellipse (5.6cm and 3.4cm);

\node[font=\small\bfseries, text=green!35!black]
    at (0, 2.85) {Structural correctness};
\node[font=\scriptsize, text=green!30!black]
    at (0, 2.45) {$\hat{\mathbf{d}} \in \mathcal{C}_{\varphi}^{L,R}$};

\fill[blue!8] (0, -0.25) ellipse (3.6cm and 2.05cm);
\draw[blue!45!black, thick] (0, -0.25) ellipse (3.6cm and 2.05cm);

\node[font=\small\bfseries, text=blue!40!black]
    at (0, 1.35) {Value preservation};
\node[font=\scriptsize, text=blue!35!black]
    at (0, 0.92) {$\mathrm{val}_{\varphi}(\hat{\mathbf{d}})
                   = \mathrm{val}_{\varphi}(\tilde{\mathbf{d}})$};

\fill[orange!10] (0, -0.3) ellipse (1.8cm and 1.0cm);
\draw[orange!60!black, thick] (0, -0.3) ellipse (1.8cm and 1.0cm);

\node[font=\small\bfseries, text=orange!50!black]
    at (0, -0.05) {Semantic recovery};
\node[font=\scriptsize, text=orange!45!black]
    at (0, -0.50) {$\mathrm{val}_{\varphi}(\hat{\mathbf{d}})
                    = \mathrm{val}_{\varphi}(\mathbf{d}^{\star})$};

\node[
    font=\scriptsize,
    text=green!30!black,
    align=center,
    text width=2.5cm
]
at (0, -2.8)
{admissible,\\but value altered};

\node[
    font=\scriptsize,
    text=blue!35!black,
    align=center,
    text width=2.5cm
]
at (0, -1.7)
{value preserved,\\not recovered};

\end{tikzpicture}

\caption{Correctness hierarchy for bounded-window repair. The three nested
regions correspond to increasingly strict requirements: structural correctness
demands only that the repaired string is admissible; value preservation
additionally requires that the numerical value of the faulty string is
maintained; semantic recovery requires that the original pre-fault value is
restored. The converse inclusions are strict in general.}

\label{fig:correctness-hierarchy}
\end{figure}

%% file: sections/4_experiments.tex
\section{Experimental Results}
\label{sec:experimental-results}

This section evaluates the finite-window model developed in
\hyperref[sec:computational-model]{Section~\ref*{sec:computational-model}}.
The experiments are organized around three increasingly operational questions.
First, we measure the static information-theoretic and structural properties of
the admissible codebooks themselves. Second, we evaluate the bounded-window
canonicalization cost incurred after valid arithmetic operations. Third, we test
how the same admissibility structure responds to external digit-level corruption.

The clean representation benchmark is reported in
\hyperref[subsec:exp1]{Experiment~1}; arithmetic canonicalization is reported in
\hyperref[subsec:exp2]{Experiment~2}; single-digit corruption is reported in
\hyperref[subsec:exp3]{Experiment~3}; local burst corruption is reported in
\hyperref[subsec:exp4_local_burst_corruption]{Experiment~4}; exhaustive
single-digit and burst verification is reported in
\hyperref[subsec:exp5_exhaustive]{Experiment~5}; algebraic burst verification
is reported in \hyperref[subsec:exp6_algebraic_burst]{Experiment~6}; and
guard-digit arithmetic is reported in
\hyperref[subsec:exp7_guard_digits]{Experiment~7}.

\subsection{Experiment 1: Clean Representation Benchmark}
\label{subsec:exp1}

The first experiment establishes the static codebook baseline before any
arithmetic or corruption is applied. For each system, we sampled admissible words
across bounded hardware windows \(W \in \{8, 12, 16, 20, 24\}\) and measured
codebook size, ambient space cardinality, structural sparsity, capacity bits per
digit, normalized deficit, and average nonzero density. The main comparison at
the largest evaluated window, \(W=24\), is reported in
\hyperref[tab:clean_benchmark]{Table~\ref*{tab:clean_benchmark}}. The full
window-width sweep is provided in
\hyperref[app:full-clean-benchmark]{Appendix~\ref*{app:full-clean-benchmark}}, \hyperref[tab:clean_benchmark_full]{Table~\ref*{tab:clean_benchmark_full}}.

The capacity column reports
\[
C_X = \frac{\log_2 |C_X^{L,R}|}{W}
\]
in bits per digit. Since the compared systems do not all use digit alphabets of
the same cardinality, the deficit column is normalized by the ambient digit
alphabet:
\[
D_X = 1 - \frac{C_X}{\log_2 |A_X|},
\]
where \(A_X\) is the digit alphabet of system \(X\). Thus, for the binary-alphabet
systems \(D_X = 1-C_X\), while for Signed NAF the normalization is taken
relative to the ternary alphabet \(\{-1,0,1\}\).

\begin{table*}[!htbp]
\centering
\caption{Clean representation benchmark at the maximum evaluated window width
\(W=24\), with \(L=12\) and \(R=11\). Exact integer values for the full
window-width sweep are reported in
\hyperref[tab:clean_benchmark_full]{Table~\ref*{tab:clean_benchmark_full}}.}
\label{tab:clean_benchmark}

\begingroup
\footnotesize
\renewcommand{\arraystretch}{1.22}
\setlength{\tabcolsep}{4pt}

\begin{tabularx}{\textwidth}{@{}l *{6}{>{\centering\arraybackslash}X}@{}}
\toprule
\textbf{System}
&
\begin{tabular}[c]{@{}c@{}}
\textbf{Codebook}\\[-1pt]
\(\boldsymbol{|C|}\)
\end{tabular}
&
\begin{tabular}[c]{@{}c@{}}
\textbf{Ambient}\\[-1pt]
\(\boldsymbol{|\Omega|}\)
\end{tabular}
&
\begin{tabular}[c]{@{}c@{}}
\textbf{Sparsity}\\[-1pt]
\(\boldsymbol{s}\)
\end{tabular}
&
\begin{tabular}[c]{@{}c@{}}
\textbf{Capacity}\\[-1pt]
\(\boldsymbol{C_X}\)
\end{tabular}
&
\begin{tabular}[c]{@{}c@{}}
\textbf{Norm.}\\[-1pt]
\textbf{deficit}\\[-1pt]
\(\boldsymbol{D_X}\)
\end{tabular}
&
\begin{tabular}[c]{@{}c@{}}
\textbf{Density}\\[-1pt]
\(\boldsymbol{\bar{\omega}}\)
\end{tabular}
\\
\midrule

Binary
& \(1.68{\times}10^{7}\)
& \(1.68{\times}10^{7}\)
& \(1.0000\)
& \(1.0000\)
& \(0.0000\)
& \(0.5003\) \\

Signed NAF
& \(2.24{\times}10^{7}\)
& \(2.82{\times}10^{11}\)
& \(7.92{\times}10^{-5}\)
& \(1.0173\)
& \(0.3582\)
& \(0.3427\) \\

\(\Phi\)
& \(1.21{\times}10^{5}\)
& \(1.68{\times}10^{7}\)
& \(0.0072\)
& \(0.7037\)
& \(0.2963\)
& \(0.2823\) \\

\(T_3\)
& \(2.56{\times}10^{6}\)
& \(1.68{\times}10^{7}\)
& \(0.1523\)
& \(0.8869\)
& \(0.1131\)
& \(0.3883\) \\

\(T_4\)
& \(7.56{\times}10^{6}\)
& \(1.68{\times}10^{7}\)
& \(0.4504\)
& \(0.9520\)
& \(0.0480\)
& \(0.4407\) \\

\bottomrule
\end{tabularx}

\endgroup
\end{table*}

Because this benchmark uses only valid codewords, it does not test repair,
normalization, or semantic recovery. Its purpose is narrower: to quantify how
much syntactic structure each representation builds into the stored language
before any fault occurs.

Standard binary provides the reference point. It uses the entire ambient binary
space, so its sparsity is \(s=1\), its capacity is exactly \(1\) bit per digit,
and its normalized deficit is zero.

Signed NAF moves in the opposite direction. Measured inside its ternary ambient
space, it is sparse relative to the full ternary language, with sparsity
\(7.92\times10^{-5}\) at \(W=24\). This sparsity is achieved through an enlarged
signed alphabet rather than through a binary non-integer base.

The binary-alphabet non-integer bases show a cleaner comparison. Their
admissible languages become less restrictive as the multinacci order increases:
\(\Phi\) has the smallest codebook, \(T_3\) is intermediate, and \(T_4\) is the
least restrictive. At \(W=24\), the corresponding codebook sizes are
\(121{,}393\), \(2{,}555{,}757\), and \(7{,}555{,}935\). The same ordering
appears in the nonzero density: \(\Phi < T_3 < T_4\).

\subsection{Experiment 2: Arithmetic Canonicalization Benchmark}
\label{subsec:exp2}

The second experiment measures the bounded-window canonicalization overhead
incurred after valid arithmetic. Raw arithmetic states
(\hyperref[def:raw-arithmetic-state]{Definition~7}) were generated by digit-wise
addition of uniformly sampled admissible pairs, and the deterministic
canonicalization procedure for each system was then applied. For \(\Phi\),
\(T_3\), and \(T_4\), this is the bounded-window MSB-first scheduler defined in
\hyperref[alg:bounded-msb-scheduler]{Section~\ref*{subsec:corruption-model-canonicalization}}.

\begin{figure*}[p]
    \centering

    \begin{subfigure}[t]{0.495\textwidth}
        \centering
        \includegraphics[width=\linewidth]{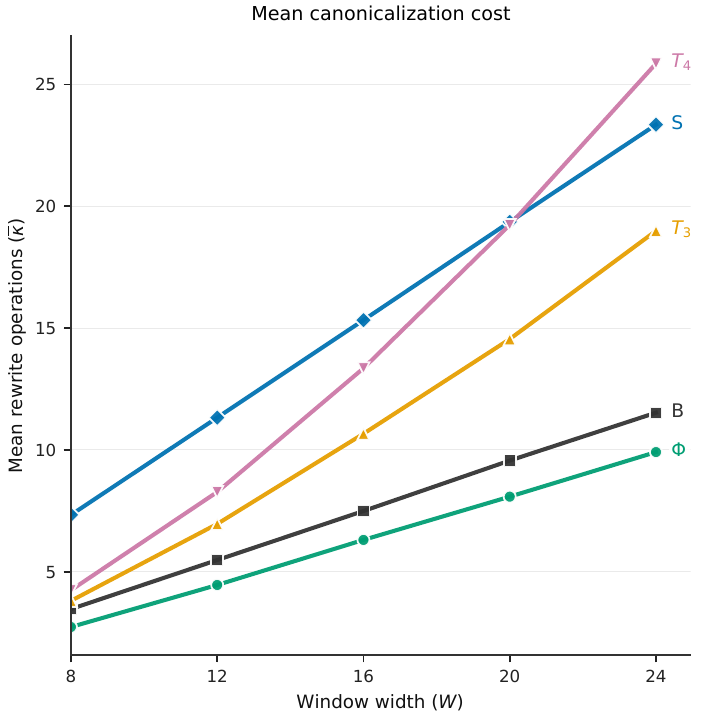}
        \caption{Mean rewrite cost.}
        \label{fig:exp2_sub_mean}
    \end{subfigure}
    \hfill
    \begin{subfigure}[t]{0.495\textwidth}
        \centering
        \includegraphics[width=\linewidth]{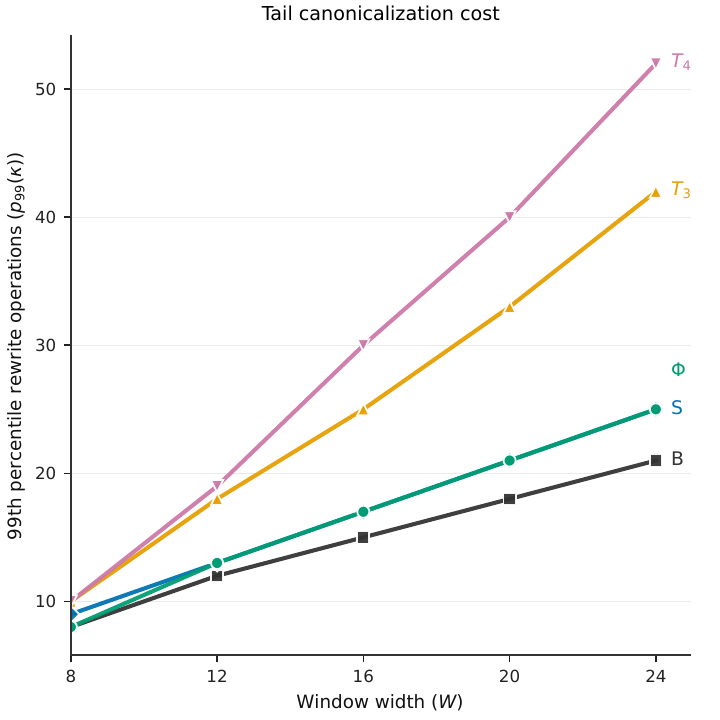}
        \caption{Tail rewrite cost.}
        \label{fig:exp2_sub_p99}
    \end{subfigure}

    \vspace{1em}

    \begin{subfigure}[t]{0.495\textwidth}
        \centering
        \includegraphics[width=\linewidth]{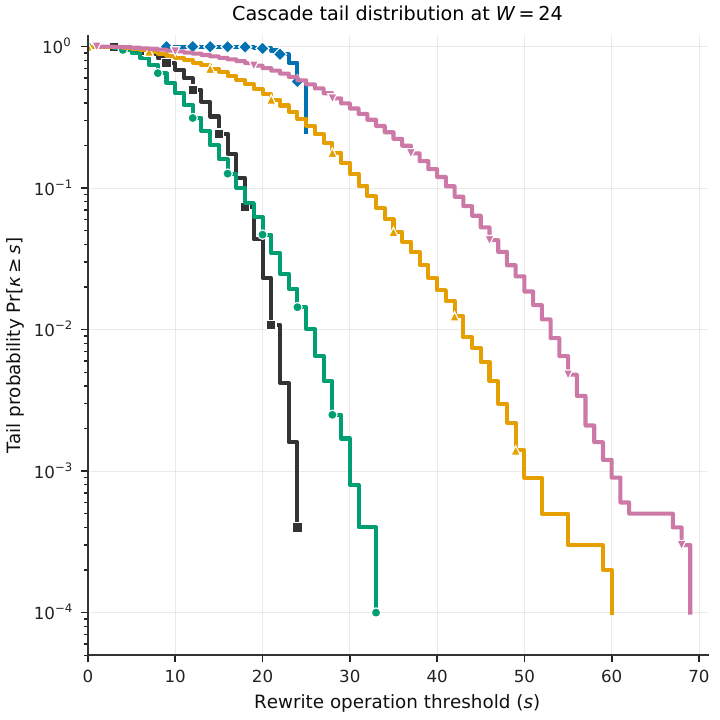}
        \caption{Cascade-tail distribution at \(W=24\).}
        \label{fig:exp2_sub_tail}
    \end{subfigure}
    \hfill
    \begin{subfigure}[t]{0.495\textwidth}
        \centering
        \includegraphics[width=\linewidth]{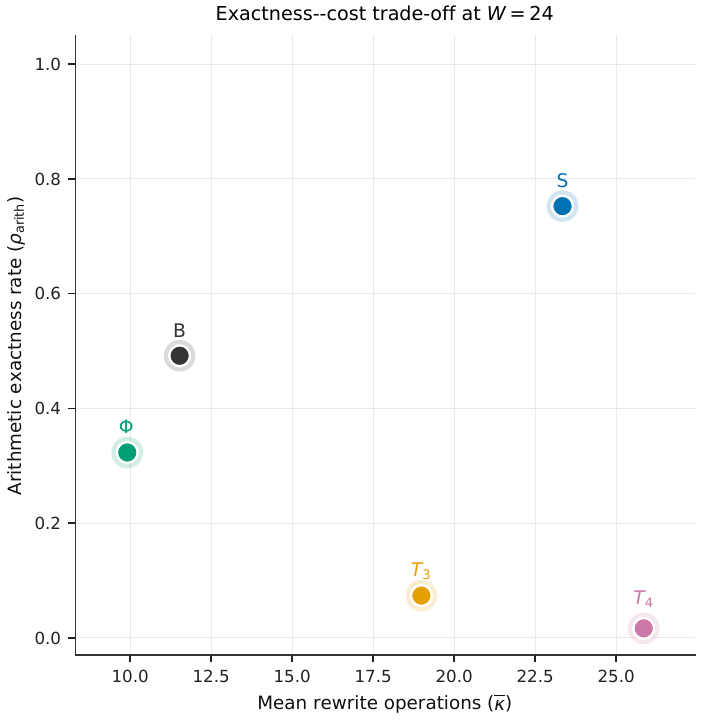}
        \caption{Exactness--cost distribution at \(W=24\).}
        \label{fig:exp2_sub_tradeoff}
    \end{subfigure}

    \vspace{0.45em}

    \includegraphics[width=0.98\textwidth]{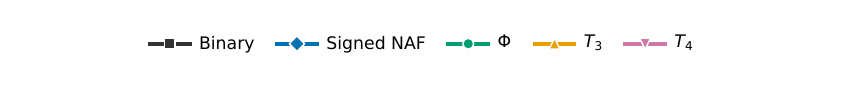}

    \caption{Arithmetic canonicalization after digit-wise addition.
    \hyperref[fig:exp2_sub_mean]{Panel~(a)} reports the mean native canonicalization cost \(\bar{\kappa}\),
    \hyperref[fig:exp2_sub_p99]{panel~(b)} reports the \(99\)th-percentile native cost \(p_{99}(\kappa)\),
    \hyperref[fig:exp2_sub_tail]{panel~(c)} shows the empirical cost-tail distribution
    \(\Pr[\kappa \geq s]\) at \(W=24\), and
    \hyperref[fig:exp2_sub_tradeoff]{panel~(d)} plots arithmetic exactness against
    mean native canonicalization cost at \(W=24\).
    For \(\Phi\), \(T_3\), and \(T_4\), \(\kappa\) counts local rewrite operations;
    for Binary it counts radix-\(2\) carry-resolution steps; for Signed NAF it
    counts bounded-window recoding iterations.}
    \label{fig:exp2_canonicalization_block}
\end{figure*}

\begin{figure*}[htbp]
    \centering

    \begin{subfigure}[t]{0.495\textwidth}
        \centering
        \includegraphics[width=\linewidth]{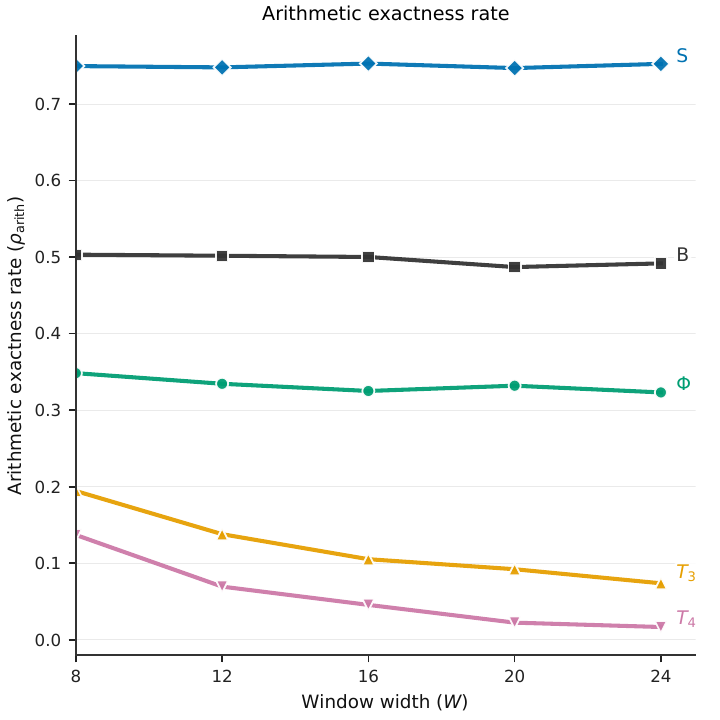}
        \caption{Arithmetic exactness rate versus window width.}
        \label{fig:exp2_sub_arithmetic_exactness}
    \end{subfigure}
    \hfill
    \begin{subfigure}[t]{0.495\textwidth}
        \centering
        \includegraphics[width=\linewidth]{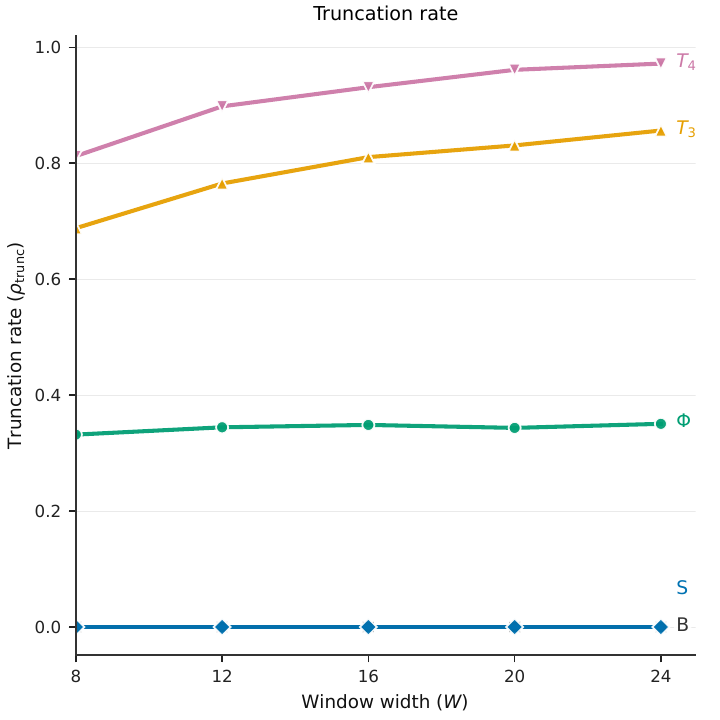}
        \caption{Truncation rate versus window width.}
        \label{fig:exp2_sub_truncation}
    \end{subfigure}

    \vspace{0.45em}

    \includegraphics[width=0.98\textwidth]{figures/fig_exp2_legend_strip.pdf}

    \caption{Arithmetic exactness and truncation behavior across window widths.
    \hyperref[fig:exp2_sub_arithmetic_exactness]{Panel~(a)} reports the arithmetic exactness rate
    \(\rho_{\mathrm{arith}}\), and
    \hyperref[fig:exp2_sub_truncation]{panel~(b)} reports the truncation rate
    \(\rho_{\mathrm{trunc}}\), both as functions of the window width \(W\).}
    \label{fig:exp2_exactness_truncation_block}
\end{figure*}

\hyperref[fig:exp2_canonicalization_block]{Figure~\ref*{fig:exp2_canonicalization_block}}
reports the native canonicalization cost and cost-tail behavior.
\hyperref[tab:exp2_summary]{Table~\ref*{tab:exp2_summary}} gives the numerical
summary at \(W=24\), and
\hyperref[fig:exp2_exactness_truncation_block]{Figure~\ref*{fig:exp2_exactness_truncation_block}}
reports exactness and truncation across window widths.

\begin{table*}[!htbp]
    \centering
    \caption{Arithmetic canonicalization summary at the maximum evaluated window
    width \(W=24\).}
    \label{tab:exp2_summary}
    \small
    \renewcommand{\arraystretch}{1.18}
    \setlength{\tabcolsep}{5pt}

    \begin{tabularx}{\textwidth}{@{}l *{7}{>{\centering\arraybackslash}X}@{}}
        \toprule
        System
        & Exactness
        & Success
        & Overflow
        & Trunc.
        & Mean cost
        & \(p_{99}\)
        & Max cost \\
        \midrule
        Binary
        & 0.4918 & 1.000 & 0.5082 & 0.0000 & 11.53 & 21 & 24 \\
        Signed NAF
        & 0.7525 & 1.000 & 0.2475 & 0.0000 & 23.35 & 25 & 25 \\
        \(\Phi\)
        & 0.3232 & 1.000 & 0.5010 & 0.3505 & 9.91 & 25 & 33 \\
        \(T_3\)
        & 0.0738 & 1.000 & 0.4998 & 0.8565 & 18.99 & 42 & 60 \\
        \(T_4\)
        & 0.0167 & 1.000 & 0.5001 & 0.9720 & 25.85 & 52 & 69 \\
        \bottomrule
    \end{tabularx}

    \vspace{1.5em}
    
    \begin{minipage}{0.96\textwidth}
        \footnotesize
        \emph{Note.} Exactness denotes \(\rho_{\mathrm{arith}}\), and success
        denotes \(p_{\mathrm{norm}}\). Overflow records discarded writes above
        the most significant boundary \(i>L\), while truncation records discarded
        writes below the least significant boundary \(i<-R\). Mean cost is
        \(\bar{\kappa}\), \(p_{99}\) is the \(99\)th percentile of \(\kappa\),
        and max cost is the largest observed \(\kappa\). The unit of \(\kappa\)
        is system-native: radix-\(2\) carry-resolution steps for Binary, NAF recoding
        iterations for Signed NAF, and local rewrite applications for \(\Phi\), \(T_3\),
        and \(T_4\). Thus the local-rewrite systems are directly comparable to one
        another, while Binary and Signed NAF serve as deterministic baseline procedures
        with separately defined cost units.
        \vspace{1em}
    \end{minipage}
\end{table*}

\hyperref[tab:exp2_summary]{Table~\ref*{tab:exp2_summary}} separates
admissibility restoration from arithmetic exactness. All systems return
structurally admissible outputs at \(W=24\), while exactness differs sharply
across representations: Signed NAF is highest (\(0.7525\)), Binary is
intermediate (\(0.4918\)), and the non-integer bases are lower, with \(\Phi\)
best within that family (\(0.3232\)).

For the non-integer systems, lower-boundary truncation is a major additional
loss mechanism. Binary and Signed NAF have zero truncation in this benchmark,
whereas \(\Phi\), \(T_3\), and \(T_4\) exhibit increasing truncation rates as the
rewrite rules become longer and more boundary-sensitive. At \(W=24\), the
truncation rates are \(0.3505\), \(0.8565\), and \(0.9720\), respectively.
Overflow remains a separate upper-boundary loss mechanism across systems, with
rates near \(0.5\) for Binary, \(\Phi\), \(T_3\), and \(T_4\) in
\hyperref[tab:exp2_summary]{Table~\ref*{tab:exp2_summary}}.

The cost measurements show a different ordering. In native operation count,
\(\Phi\) has the lowest mean canonicalization cost at \(W=24\) (\(9.91\)),
below Binary (\(11.53\)), Signed NAF (\(23.35\)), \(T_3\) (\(18.99\)), and
\(T_4\) (\(25.85\)). The tail behavior follows the same structural ordering
among the non-integer bases: \(\Phi\) has a much shorter tail than \(T_3\) and
\(T_4\), whose cascades extend to substantially larger rewrite thresholds.

At \(W=24\), the arithmetic benchmark therefore orders the non-integer bases as
\(\Phi\), \(T_3\), and \(T_4\) in both arithmetic exactness and mean native
canonicalization cost, while Signed NAF remains the highest-exactness baseline
under its global recoding convention.

\subsection{Experiment 3: Single-Digit Corruption Benchmark}
\label{subsec:exp3}

The third experiment evaluates the response to a single digit-level corruption
of an admissible stored representation. For each system and window width, an
admissible state was sampled, one digit position was corrupted, and the
deterministic bounded-window repair procedure was applied. The recorded
quantities include structural detectability, structural exactness, semantic
survival, numerical error, and native repair cost.

\begin{figure*}[p]
    \centering

    \begin{subfigure}[t]{0.495\textwidth}
        \centering
        \includegraphics[width=\linewidth]{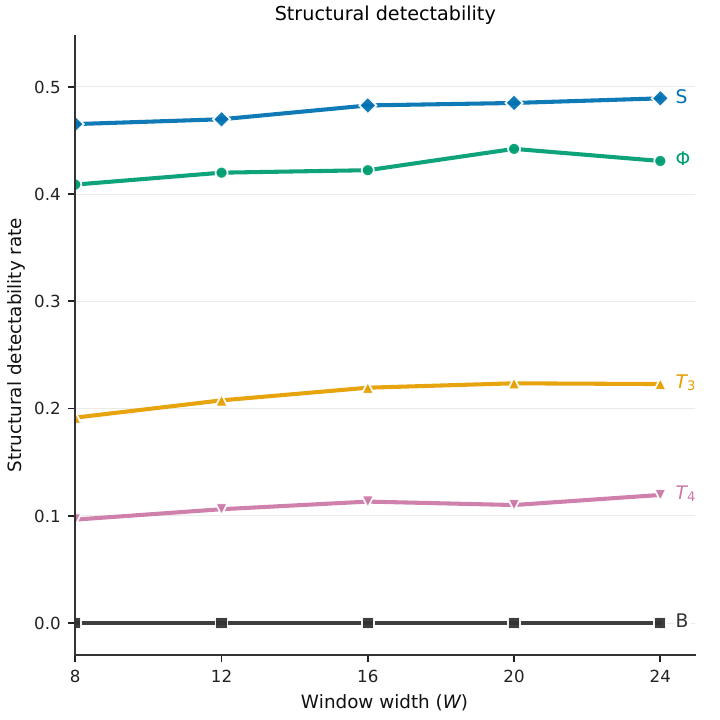}
        \caption{Structural detectability rate.}
        \label{fig:exp3_sub_detectability}
    \end{subfigure}
    \hfill
    \begin{subfigure}[t]{0.495\textwidth}
        \centering
        \includegraphics[width=\linewidth]{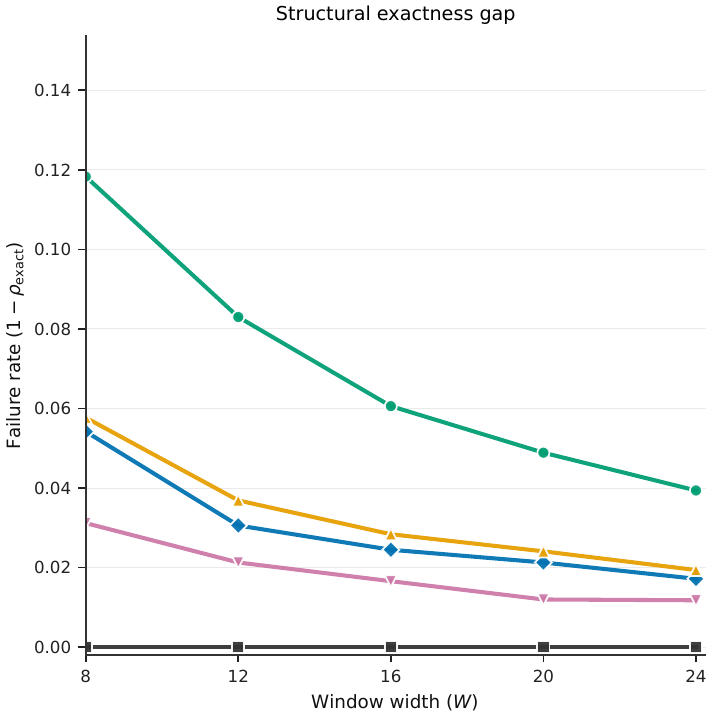}
        \caption{Structural exactness gap.}
        \label{fig:exp3_sub_exactness_gap}
    \end{subfigure}

    \vspace{1em}

    \begin{subfigure}[t]{0.495\textwidth}
        \centering
        \includegraphics[width=\linewidth]{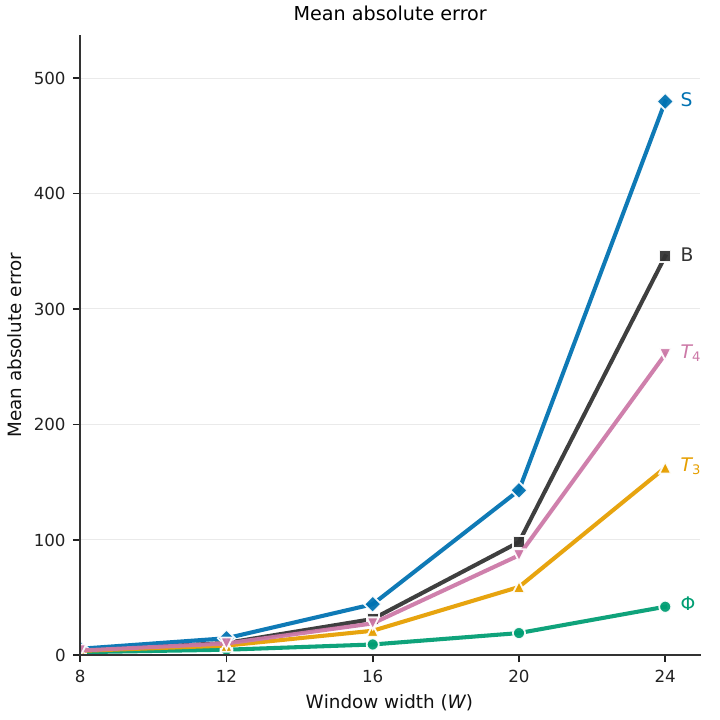}
        \caption{Mean absolute error.}
        \label{fig:exp3_sub_mean_abs_error}
    \end{subfigure}
    \hfill
    \begin{subfigure}[t]{0.495\textwidth}
        \centering
        \includegraphics[width=\linewidth]{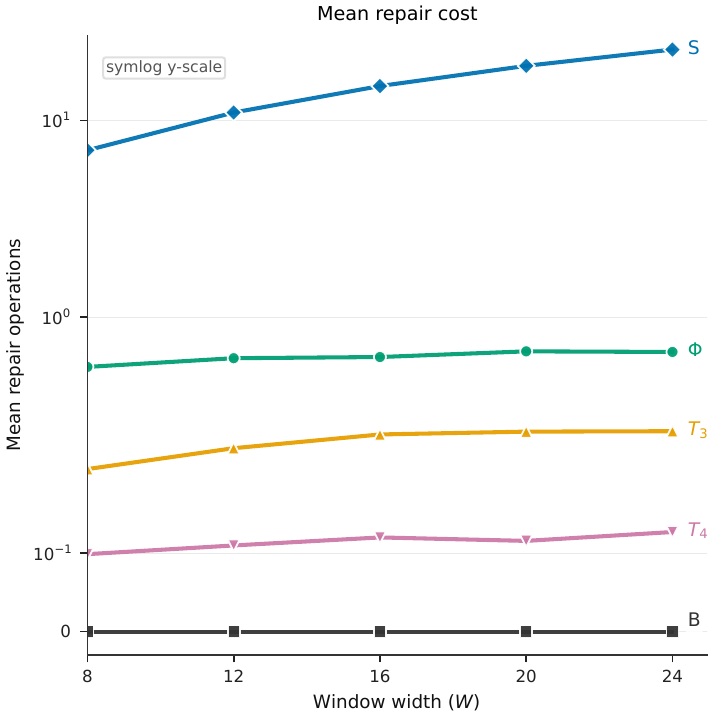}
        \caption{Mean repair cost.}
        \label{fig:exp3_sub_mean_repair}
    \end{subfigure}

    \vspace{0.45em}

    \includegraphics[width=0.98\textwidth]{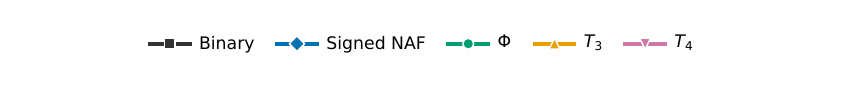}

    \caption{Single-digit corruption benchmark across window widths.
    \hyperref[fig:exp3_sub_detectability]{Panel~(a)} reports the structural
    detectability rate. \hyperref[fig:exp3_sub_exactness_gap]{Panel~(b)} reports the
    structural exactness gap \(1-\rho_{\mathrm{exact}}\), where
    \(\rho_{\mathrm{exact}}\) is the structural exactness rate.
    \hyperref[fig:exp3_sub_mean_abs_error]{Panel~(c)} reports the mean absolute
    error after bounded-window canonicalization.
    \hyperref[fig:exp3_sub_mean_repair]{Panel~(d)} reports the mean number of
    native repair operations; this panel uses a symmetric logarithmic \(y\)-axis,
    as marked in the plot.}
    \label{fig:exp3_summary_block}
\end{figure*}

\begin{figure*}[htbp]
    \centering

    \includegraphics[width=0.96\textwidth]{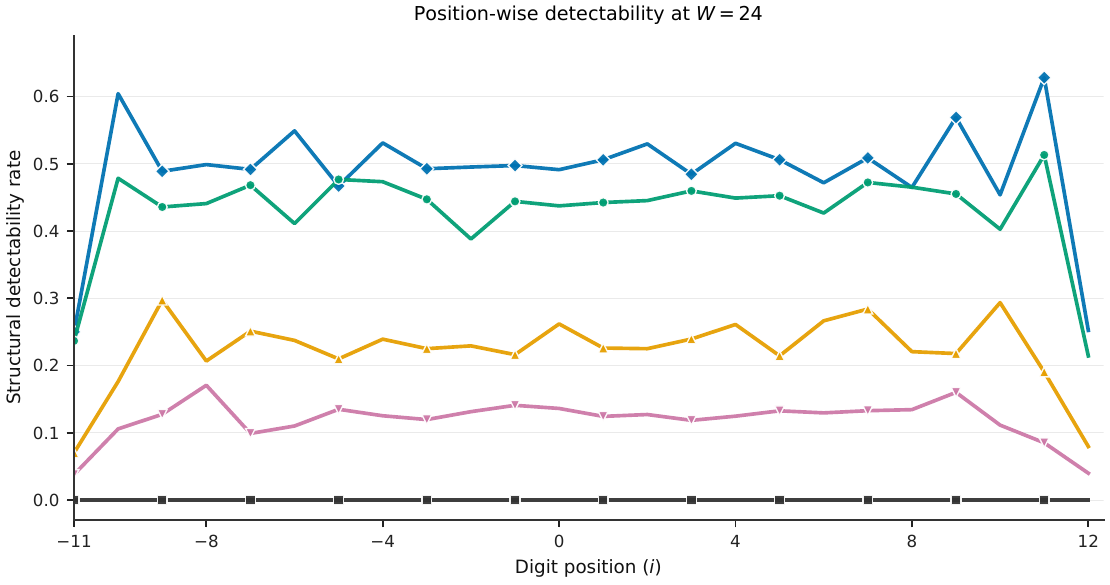}

    \vspace{0.45em}

    \includegraphics[width=0.98\textwidth]{figures/fig_exp3_legend_strip.pdf}

    \caption{Position-wise structural detectability at \(W=24\).
    The horizontal axis is the signed digit position \(i\), and the vertical
    axis is the structural detectability rate after single-digit corruption and
    bounded-window canonicalization.}
    \label{fig:exp3_position_detectability}
\end{figure*}

\hyperref[fig:exp3_summary_block]{Figure~\ref*{fig:exp3_summary_block}} reports
the main window-width-dependent quantities.
\hyperref[tab:exp3_summary]{Table~\ref*{tab:exp3_summary}} gives the numerical
summary at \(W=24\), and
\hyperref[fig:exp3_position_detectability]{Figure~\ref*{fig:exp3_position_detectability}}
reports position-wise detectability at \(W=24\).

\begin{table*}[!htbp]
    \centering
    \caption{Single-digit corruption summary at the maximum evaluated window
    width \(W=24\).}
    \label{tab:exp3_summary}
    \scriptsize
    \renewcommand{\arraystretch}{1.18}
    \setlength{\tabcolsep}{3.5pt}

    \begin{tabularx}{\textwidth}{@{}l *{9}{>{\centering\arraybackslash}X}@{}}
        \toprule
        System
        & Detect. \(\delta_{\mathrm{det}}\)
        & Exact \(\rho_{\mathrm{exact}}\)
        & Sem. surv. \(\rho_{\mathrm{sem}}\)
        & Raw MAE
        & NMAE
        & Raw \(\Delta_{\max}\)
        & N\(\Delta_{\max}\)
        & Mean repair
        & \(p_{99}\) repair \\
        \midrule
        Binary
        & 0.0000 & 1.0000 & 0.0000 & 345.68 & 0.04220 & 4096.00 & 0.50000 & 0.00 & 0 \\
        Signed NAF
        & 0.4892 & 0.9828 & 0.0000 & 479.76 & 0.08785 & 8192.00 & 1.50000 & 23.03 & 25 \\
        \(\Phi\)
        & 0.4308 & 0.9606 & 0.0000 & 41.97 & 0.08056 & 520.77 & 0.99956 & 0.66 & 4 \\
        \(T_3\)
        & 0.2228 & 0.9806 & 0.0000 & 162.54 & 0.05896 & 2756.99 & 0.99998 & 0.26 & 2 \\
        \(T_4\)
        & 0.1194 & 0.9882 & 0.0000 & 260.72 & 0.05141 & 5067.46 & 0.99927 & 0.13 & 1 \\
        \bottomrule
    \end{tabularx}

    \vspace{1.5em}

    \begin{minipage}{0.96\textwidth}
    \footnotesize
    \emph{Note.} Detect., Exact, and Sem. surv. denote
    \(\delta_{\mathrm{det}}\), \(\rho_{\mathrm{exact}}\), and
    \(\rho_{\mathrm{sem}}\), respectively. Exact is measured relative to the
    corrupted word after repair; Sem. surv. is measured relative to the original
    pre-fault word. Raw MAE and Raw \(\Delta_{\max}\) are unscaled residual
    errors, while NMAE and N\(\Delta_{\max}\) are normalized by
    \(V_{\max,X}^{L,R}\). Mean repair and \(p_{99}\) repair report
    \(\bar{\kappa}\) and the \(99\)th percentile of native repair cost.
    \(p_{\mathrm{norm}}=1.000\) for all systems at \(W=24\) and is omitted.
\end{minipage}
\end{table*}

The first result matches
\hyperref[thm:no-free-lunch]{Proposition~1}: semantic survival is \(0.0000\)
for every system and every evaluated window width. A genuine single-digit
corruption changes the represented value, so value-preserving repair normalizes
the observed corrupted word rather than reconstructing the pre-fault value.

The useful distinction is therefore structural, not semantic. Binary has zero
structural detectability because every bit flip remains a valid binary word.
The redundant languages expose a positive fraction of faults as syntactic
violations. At \(W=24\), Signed NAF detects the largest fraction
(\(0.4892\)), followed closely by \(\Phi\) (\(0.4308\)). The higher-order
multinacci systems detect fewer single-digit corruptions: \(0.2228\) for \(T_3\)
and \(0.1194\) for \(T_4\). Position-wise detectability follows the same
ordering, with lower endpoint detectability caused by reduced local context near
the window boundaries.

Structural exactness remains high once repair is attempted. At \(W=24\), the
exactness rates relative to the observed corrupted word are \(1.0000\) for
Binary, \(0.9828\) for Signed NAF, \(0.9606\) for \(\Phi\), \(0.9806\) for
\(T_3\), and \(0.9882\) for \(T_4\). Thus, the repair procedures usually perform
value-preserving re-admissibilization of the post-fault state.

The numerical error metrics add one important caution. On raw absolute scale,
\(\Phi\) has the smallest mean and worst-case errors at \(W=24\). After
normalization by the finite-window dynamic range, however, the ranking changes:
Binary has the lowest NMAE, and among the non-binary redundant systems \(T_4\)
and \(T_3\) have lower NMAE than \(\Phi\). This prevents an overstrong
interpretation of the raw MAE result.

That cost contrast separates Signed NAF from the local-rewrite systems. Signed
NAF requires a mean native repair cost of \(23.03\) operations at \(W=24\),
while the non-integer local-rewrite systems remain below one mean repair
operation: \(0.66\) for \(\Phi\), \(0.26\) for \(T_3\), and \(0.13\) for
\(T_4\). At the single-digit level, the observed pattern is therefore zero
semantic survival for all systems, positive structural detectability for the
redundant languages, and a large repair-cost separation between Signed NAF and
the local-rewrite non-integer bases.

\subsection{Experiment 4: Local Burst Corruption Benchmark}
\label{subsec:exp4_local_burst_corruption}

The fourth experiment evaluates admissible stored codewords under local burst
corruption. In each trial, we sampled an admissible codeword, selected a burst
width \(b \in \{2,3,4,5\}\), chose a contiguous burst window, and applied a
nonzero binary corruption mask using the alphabet-aware single-digit rule of
\hyperref[def:single-digit-corruption]{Definition~5}. The corrupted word was
then repaired using the same bounded-window convention as in
\hyperref[subsec:exp3]{Experiment~3}.

\begin{figure*}[htbp]
    \centering

    \begin{subfigure}[t]{0.495\textwidth}
        \centering
        \includegraphics[width=\linewidth]{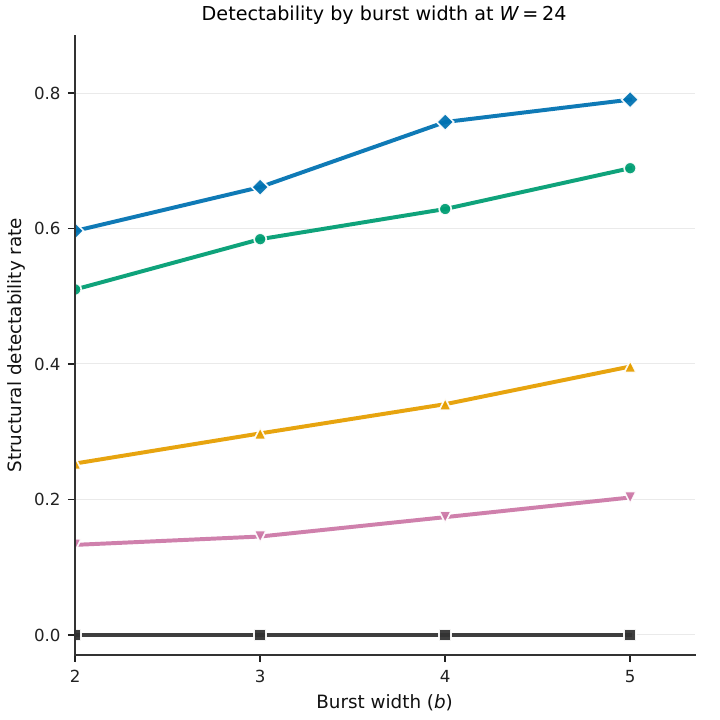}
        \caption{Structural detectability by burst width at \(W=24\).}
        \label{fig:exp4_sub_detectability_by_b}
    \end{subfigure}
    \hfill
    \begin{subfigure}[t]{0.495\textwidth}
        \centering
        \includegraphics[width=\linewidth]{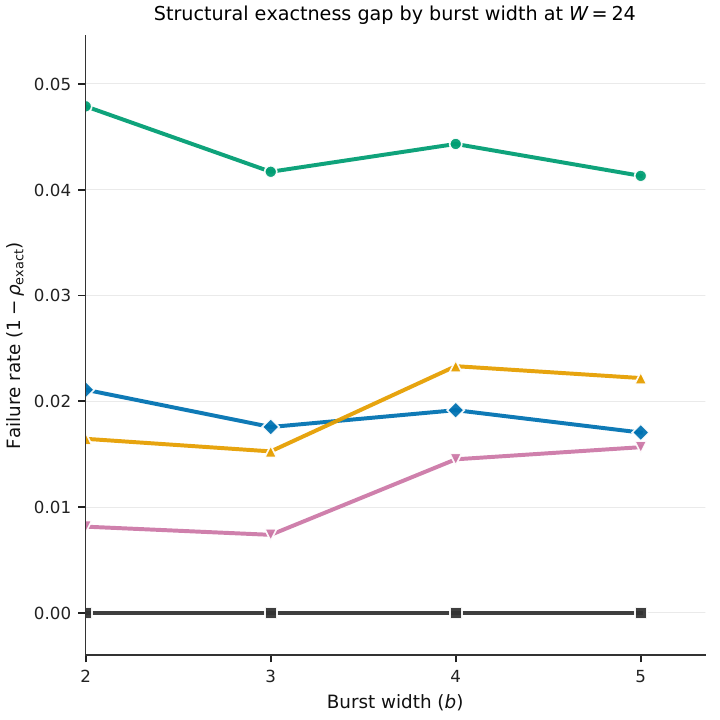}
        \caption{Structural exactness gap by burst width at \(W=24\).}
        \label{fig:exp4_sub_exactness_gap_by_b}
    \end{subfigure}

    \vspace{0.9em}

    \begin{subfigure}[t]{0.495\textwidth}
        \centering
        \includegraphics[width=\linewidth]{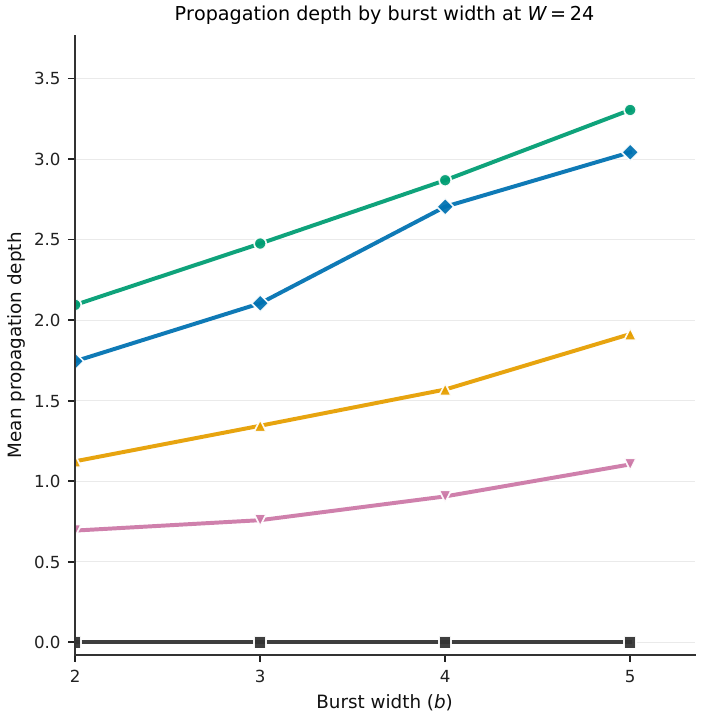}
        \caption{Mean propagation depth by burst width at \(W=24\).}
        \label{fig:exp4_sub_propagation_depth_by_b}
    \end{subfigure}
    \hfill
    \begin{subfigure}[t]{0.495\textwidth}
        \centering
        \includegraphics[width=\linewidth]{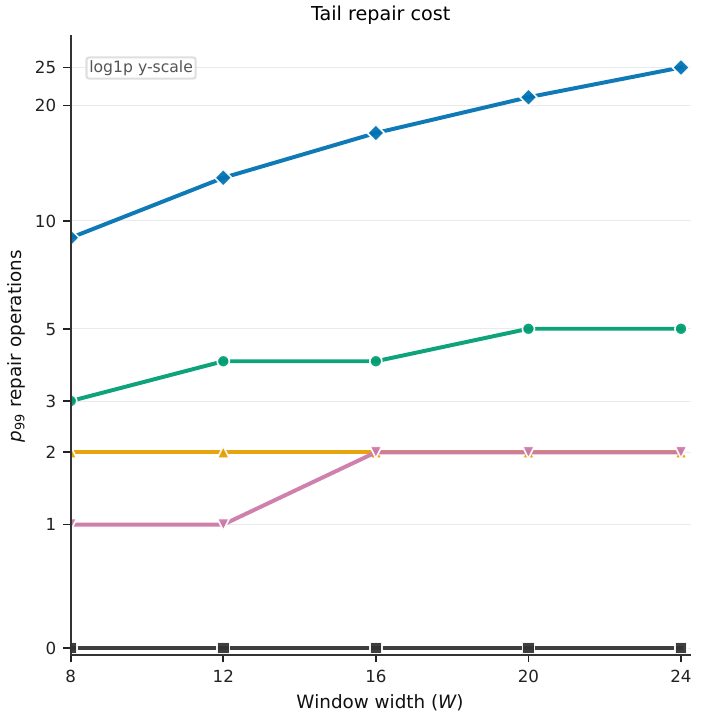}
        \caption{Tail repair cost as a function of \(W\).}
        \label{fig:exp4_sub_p99_repair_steps}
    \end{subfigure}

     \vspace{0.45em}

    \includegraphics[width=0.98\textwidth]{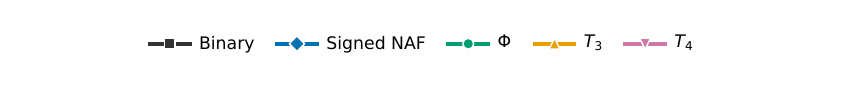}

    \caption{Local burst corruption results.
    \hyperref[fig:exp4_sub_detectability_by_b]{Panel~(a)} fixes the repair window
    width at \(W=24\) and plots the structural detectability rate as a function
    of the burst width \(b\).
    \hyperref[fig:exp4_sub_exactness_gap_by_b]{Panel~(b)} uses the same \(W=24\)
    slice and reports the structural exactness failure rate
    \(1-\rho_{\mathrm{exact}}\).
    \hyperref[fig:exp4_sub_propagation_depth_by_b]{Panel~(c)} reports the mean
    propagation depth under the same burst-width sweep.
    \hyperref[fig:exp4_sub_p99_repair_steps]{Panel~(d)} plots the 99th percentile
    of the native repair cost as a function of the repair window width \(W\); the
    vertical axis uses a log1p scale while retaining tick labels in native
    repair-operation units.}
    \label{fig:exp4_main_block}
\end{figure*}

\hyperref[tab:exp4_summary]{Table~\ref*{tab:exp4_summary}} reports the aggregate
metrics at \(W=24\), averaged over burst widths \(b \in \{2,3,4,5\}\).
\hyperref[fig:exp4_main_block]{Figure~\ref*{fig:exp4_main_block}} shows how
detectability, exactness gap, propagation depth, and tail repair cost vary with
burst width and window width.

\begin{table*}[htbp]
    \centering
    \caption{Local burst corruption summary at the maximum evaluated window
    width \(W=24\), aggregated over burst widths \(b \in \{2,3,4,5\}\).}
    \label{tab:exp4_summary}
    \small
    \renewcommand{\arraystretch}{1.18}
    \setlength{\tabcolsep}{5pt}

    \begin{tabularx}{\textwidth}{@{}l *{7}{>{\centering\arraybackslash}X}@{}}
        \toprule
        System
        & Detect. \(\delta_{\mathrm{det}}\)
        & Exact \(\rho_{\mathrm{exact}}\)
        & Semantic \(\rho_{\mathrm{sem}}\)
        & Raw MAE
        & NMAE
        & Mean cost
        & Mean depth \\
        \midrule
        Binary
        & 0.000 & 1.000 & 0.000 & 272.93 & 0.03332 & 0.000 & 0.000 \\
        Signed NAF
        & 0.702 & 0.981 & 0.023 & 396.55 & 0.07261 & 23.021 & 2.402 \\
        \(\Phi\)
        & 0.602 & 0.956 & 0.018 & 36.28 & 0.06964 & 1.045 & 2.682 \\
        \(T_3\)
        & 0.324 & 0.981 & 0.003 & 132.67 & 0.04812 & 0.395 & 1.498 \\
        \(T_4\)
        & 0.164 & 0.989 & 0.001 & 213.46 & 0.04209 & 0.178 & 0.865 \\
        \bottomrule
    \end{tabularx}

    \vspace{1.2em}

    \begin{minipage}{0.96\textwidth}
        \footnotesize
        \emph{Note.} Detect. is structural detectability
        \(\delta_{\mathrm{det}}\). Exact is value preservation relative to the corrupted
        word after repair; Semantic is survival of the original pre-fault value. NMAE is
        MAE normalized by \(V_{\max,X}^{L,R}\). Mean cost and depth are
        \(\bar{\kappa}\) and \(\bar{\pi}\), respectively. Structural correctness and
        normalization success are uniformly \(1.000\) and are omitted.
    \end{minipage}
\end{table*}

Burst corruption changes the picture in two ways. First, structural
detectability increases because a clustered perturbation is more likely to cross
a local admissibility constraint. At \(W=24\), the aggregate detectability rates
are \(0.702\) for Signed NAF, \(0.602\) for \(\Phi\), \(0.324\) for \(T_3\), and
\(0.164\) for \(T_4\), while Binary remains at \(0.000\). The same ordering is
visible across burst widths: Signed NAF is highest, \(\Phi\) follows, and the
higher-order multinacci bases occupy lower bands.

Second, unlike single-digit faults, burst faults can occasionally preserve the
original value. The semantic survival rates are small but nonzero for the
redundant systems: \(0.023\) for Signed NAF, \(0.018\) for \(\Phi\), \(0.003\)
for \(T_3\), and \(0.001\) for \(T_4\). These values are explained by
\hyperref[thm:algebraic-kernel]{Theorem~2}: under exact structural repair,
semantic survival requires the burst perturbation to lie in the algebraic kernel
of the evaluation map.

The error metrics again separate raw scale from normalized severity. On raw
absolute scale, \(\Phi\) has the smallest MAE at \(W=24\) (\(36.28\)). After
normalization by \(V_{\max,X}^{L,R}\), however, Binary has the lowest NMAE, and
among the non-binary redundant systems \(T_4\) and \(T_3\) are lower than
\(\Phi\).

The repair-cost comparison separates Signed NAF from the local-rewrite systems.
Signed NAF achieves the highest burst detectability but requires a mean native
repair cost of \(23.021\). The local-rewrite systems are much cheaper: \(1.045\)
for \(\Phi\), \(0.395\) for \(T_3\), and \(0.178\) for \(T_4\). Although
\(\Phi\) has the largest mean propagation depth among the local-rewrite systems,
its repair cost remains close to one native operation on average.

Relative to the single-digit benchmark, local bursts increase structural
detectability and introduce small nonzero semantic-survival rates for Signed NAF
and for the non-integer bases, while preserving the same qualitative repair-cost
separation between Signed NAF and the local-rewrite systems.

\subsection{Experiment 5: Exhaustive small-window verification}
\label{subsec:exp5_exhaustive}

To validate the single-digit impossibility result
\hyperref[thm:no-free-lunch]{Proposition~1} and the algebraic-kernel condition
\hyperref[thm:algebraic-kernel]{Theorem~2} without Monte Carlo sampling artifacts,
Experiment~5 performs an exhaustive finite-window enumeration. For small bounded
windows (\(W\le 16\)), we generated the complete admissible finite-window
codebook for each system. Every valid single-digit corruption and every
contiguous local burst mask with \(b\in\{2,3,4,5\}\) was then injected and
evaluated.

\begin{table*}[t]
    \centering
    
    \caption{Exhaustive verification of single-digit corruptions at \(W=16\).}
    \label{tab:exp5_single_exhaustive}
    \small
    \renewcommand{\arraystretch}{1.15}
    \setlength{\tabcolsep}{5pt}

    \begin{tabularx}{\textwidth}{@{}l *{5}{>{\centering\arraybackslash}X}@{}}
        \toprule
        System
        & Total valid injections
        & Detectability \(\delta_{\mathrm{det}}\)
        & Structural exactness \(\rho_{\mathrm{exact}}\)
        & Semantic survival count
        & Semantic survival \(\rho_{\mathrm{sem}}\) \\
        \midrule
        Binary     & 1,048,576 & 0.0000 & 1.0000 & 0 & 0.0000 \\
        Signed NAF & 2,796,192 & 0.4792 & 0.9740 & 0 & 0.0000 \\
        \(\Phi\)   & 41,344    & 0.4281 & 0.9375 & 0 & 0.0000 \\
        \(T_3\)    & 312,208   & 0.2158 & 0.9715 & 0 & 0.0000 \\
        \(T_4\)    & 634,368   & 0.1143 & 0.9842 & 0 & 0.0000 \\
        \bottomrule
    \end{tabularx}
\end{table*}

\begin{table*}[t]
    \centering
    \caption{Exhaustive verification of local burst corruptions at \(W=16\),
    aggregated over burst widths \(b \in \{2,3,4,5\}\).}
    \label{tab:exp5_burst_exhaustive}
    \small
    \renewcommand{\arraystretch}{1.15}
    \setlength{\tabcolsep}{5pt}

    \begin{tabularx}{\textwidth}{@{}l c *{4}{>{\centering\arraybackslash}X}@{}}
        \toprule
        System
        & Burst width \(b\)
        & Total valid injections
        & Detectability \(\delta_{\mathrm{det}}\)
        & Structural exactness \(\rho_{\mathrm{exact}}\)
        & Semantic survival \(\rho_{\mathrm{sem}}\) \\
        \midrule
        Binary     & \(2\)--\(5\) aggregated & 46,530,560  & 0.0000 & 1.0000 & 0.0000 \\
        Signed NAF & \(2\)--\(5\) aggregated & 386,923,068 & 0.8489 & 0.9657 & 0.0191 \\
        \(\Phi\)   & \(2\)--\(5\) aggregated & 1,834,640   & 0.6473 & 0.9244 & 0.0206 \\
        \(T_3\)    & \(2\)--\(5\) aggregated & 13,854,230  & 0.3500 & 0.9652 & 0.0044 \\
        \(T_4\)    & \(2\)--\(5\) aggregated & 28,150,080  & 0.1823 & 0.9805 & 0.0006 \\
        \bottomrule
    \end{tabularx}

    \vspace{1.5em}

    \begin{minipage}{0.96\textwidth}
        \footnotesize
        \emph{Note.} Total valid injections counts all admissible source words
        and all valid local burst masks of widths \(2\)--\(5\). Detectability
        denotes the fraction of injected words that are structurally invalid
        before repair. Structural exactness denotes value preservation relative
        to the corrupted word after bounded-window canonicalization. Semantic
        survival denotes value preservation relative to the original pre-fault
        word.
    \end{minipage}
\end{table*}

The exhaustive single-digit sweep confirms the limit described by
\hyperref[thm:no-free-lunch]{Proposition~1}. Across all valid injections for Binary,
Signed NAF, \(\Phi\), \(T_3\), and \(T_4\), the semantic survival count is zero,
as reported in
\hyperref[tab:exp5_single_exhaustive]{Table~\ref*{tab:exp5_single_exhaustive}}.
Thus, within the enumerated canonically injective finite-window codebooks, a
valid single-cell perturbation does not preserve the original represented value.
This finite-window enumeration is consistent with the zero semantic-survival
rates observed in \hyperref[subsec:exp3]{Experiment~3}.

The burst enumeration exhibits the complementary phenomenon described by
\hyperref[thm:algebraic-kernel]{Theorem~2}. Unlike single-digit faults,
localized burst corruptions can produce nonzero semantic survival when the
perturbation lies in the kernel of the evaluation map. As shown in
\hyperref[tab:exp5_burst_exhaustive]{Table~\ref*{tab:exp5_burst_exhaustive}},
Binary has zero burst semantic survival, while the redundant systems have small
positive survival rates: \(0.0191\) for Signed NAF, \(0.0206\) for \(\Phi\),
\(0.0044\) for \(T_3\), and \(0.0006\) for \(T_4\).

The mechanisms are different. For Signed NAF, the surviving bursts arise from
value-neutral signed-radix cancellations in the alphabet \(\{-1,0,1\}\). For the
multinacci systems, they arise from local algebraic rewrite identities, such as
the \(100 \leftrightarrow 011\) identity in the \(\Phi\)-system. The exhaustive
enumeration therefore confirms zero single-digit semantic survival, while burst
survival appears only through system-specific kernel mechanisms.

\subsection{Experiment 6: Algebraic Burst Injection and Kernel Survival}
\label{subsec:exp6_algebraic_burst}

Experiment~6 tests the constructive side of the algebraic-kernel condition
(\hyperref[thm:algebraic-kernel]{Theorem~2}). Instead of applying uniform random
burst noise, we inject perturbations that are algebraically value-neutral by
construction. In the golden-ratio system, canonical occurrences of \(100\) are
replaced by the forbidden but equivalent pattern \(011\). The corresponding
multinacci substitutions are
\[
    1000 \to 0111 \qquad \text{for } T_3,
    \qquad
    10000 \to 01111 \qquad \text{for } T_4 .
\]
Thus every injected perturbation satisfies
\[
    \operatorname{val}_{\beta}(\widetilde d-d^\star)=0 .
\]
The experiment therefore measures whether bounded-window canonicalization
preserves this algebraic equality after the corrupted word is repaired.
\vspace{-1em}

\begin{table}[htbp]
\centering
\caption{Experiment 6: targeted algebraic-kernel burst injections. ``Plac.'' is
the total number of source-pattern placements; ``Inj.'' is the number of injected
sampled words; ``Int./Bd.'' splits injections into interior and boundary-touching
placements; \(E/S\) reports structural exactness and semantic survival.}
\label{tab:exp6_algebraic_burst}
\small
\renewcommand{\arraystretch}{1.08}
\setlength{\tabcolsep}{4.8pt}
\begin{tabular*}{\linewidth}{@{\extracolsep{\fill}} llrrrrrr @{}}
\toprule
System & Rewrite & \(W\) & Plac. & Inj. & Int./Bd. & \(E/S\) & O/T \\
\midrule
\(\Phi\) & \(100\to011\)       & 8  & 10598 & 8377  & 5641/2736 & 1.000/1.000 & 0/0 \\
\(T_3\)  & \(1000\to0111\)     & 8  & 4092  & 4031  & 2491/1540 & 1.000/1.000 & 0/0 \\
\(T_4\)  & \(10000\to01111\)   & 8  & 1496  & 1496  & 795/701   & 1.000/1.000 & 0/0 \\
\midrule
\(\Phi\) & \(100\to011\)       & 12 & 17344 & 9648  & 7702/1946 & 1.000/1.000 & 0/0 \\
\(T_3\)  & \(1000\to0111\)     & 12 & 7512  & 6401  & 5146/1255 & 1.000/1.000 & 0/0 \\
\(T_4\)  & \(10000\to01111\)   & 12 & 3070  & 2996  & 2265/731  & 1.000/1.000 & 0/0 \\
\midrule
\(\Phi\) & \(100\to011\)       & 16 & 24160 & 9929  & 8524/1405 & 1.000/1.000 & 0/0 \\
\(T_3\)  & \(1000\to0111\)     & 16 & 10824 & 7807  & 6786/1021 & 1.000/1.000 & 0/0 \\
\(T_4\)  & \(10000\to01111\)   & 16 & 4464  & 4093  & 3461/632  & 1.000/1.000 & 0/0 \\
\midrule
\(\Phi\) & \(100\to011\)       & 20 & 31061 & 9986  & 8828/1158 & 1.000/1.000 & 0/0 \\
\(T_3\)  & \(1000\to0111\)     & 20 & 14124 & 8676  & 7777/899  & 1.000/1.000 & 0/0 \\
\(T_4\)  & \(10000\to01111\)   & 20 & 6113  & 5180  & 4569/611  & 1.000/1.000 & 0/0 \\
\midrule
\(\Phi\) & \(100\to011\)       & 24 & 37797 & 10000 & 9032/968  & 1.000/1.000 & 0/0 \\
\(T_3\)  & \(1000\to0111\)     & 24 & 17617 & 9272  & 8533/739  & 1.000/1.000 & 0/0 \\
\(T_4\)  & \(10000\to01111\)   & 24 & 7536  & 5919  & 5441/478  & 1.000/1.000 & 0/0 \\
\bottomrule
\end{tabular*}

\vspace{2mm}
\footnotesize
For all rows, structural detectability and structural correctness are also
\(1.000\). The mean and maximum semantic errors are \(0.0\) within the
\(10^{-9}\) tolerance, and no halt failures are observed.
\end{table}

Table~\ref{tab:exp6_algebraic_burst} reports the all-placement protocol. Each row
uses \(10^4\) sampled admissible words. An injection is made only when the sampled
word contains at least one canonical source pattern. The number of available
placements grows with \(W\), especially for \(\Phi\), where the shorter pattern
\(100\) appears most frequently. Longer multinacci patterns are rarer, so \(T_4\)
has fewer injections at every width. Boundary-touching placements remain present,
but their relative share decreases as the window grows.

This trend is stable across all tested widths. While the \(\Phi\)-system reaches 
near-complete injection coverage at moderate \(W\), \(T_3\) and \(T_4\) increase 
more gradually due to their longer source patterns. Despite these differing 
pattern frequencies, repair outcomes are identical across all three non-integer 
bases: every targeted algebraic burst is detected, exactly normalized, and 
semantically survives.

The boundary audit reflects this same pattern. Boundary-touching injections 
occur across all systems and widths without producing overflow or truncation. 
Thus, perfect survival extends beyond strictly interior placements. For the 
tested substitutions, the injected perturbation lies in the algebraic kernel, 
and bounded-window normalization preserves its value throughout the sampled range.

\subsection{Experiment 7: Guard-Digit Truncation Analysis}
\label{subsec:exp7_guard_digits}

To assess boundary canonicalization costs and compare accumulated lower-boundary
loss with the one-step tail scale of Lemma~2, Experiment~7 analyzes addition
under an extended fractional boundary with \(g \in\{0,2,4,8,12,16\}\)
lower-order guard digits. Raw digit-wise sums were canonicalized within
\(I_{L,R+g}\) and then projected back to the original \(W=16\) storage window.

\begin{figure*}[htbp]
    \centering

    \begin{subfigure}[t]{0.495\textwidth}
        \centering
        \includegraphics[width=\linewidth]{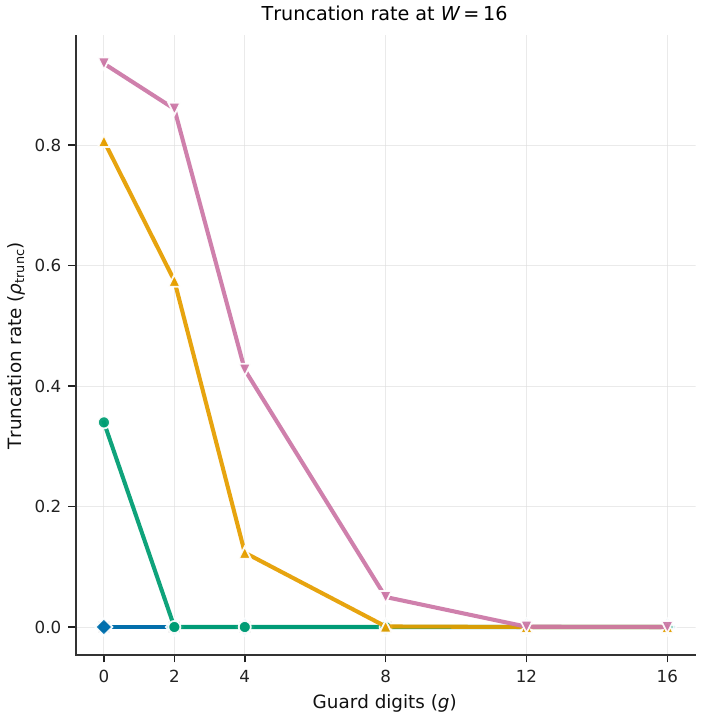}
        \caption{Truncation rate vs. guard digits.}
        \label{fig:exp7_truncation}
    \end{subfigure}
    \hfill
    \begin{subfigure}[t]{0.495\textwidth}
        \centering
        \includegraphics[width=\linewidth]{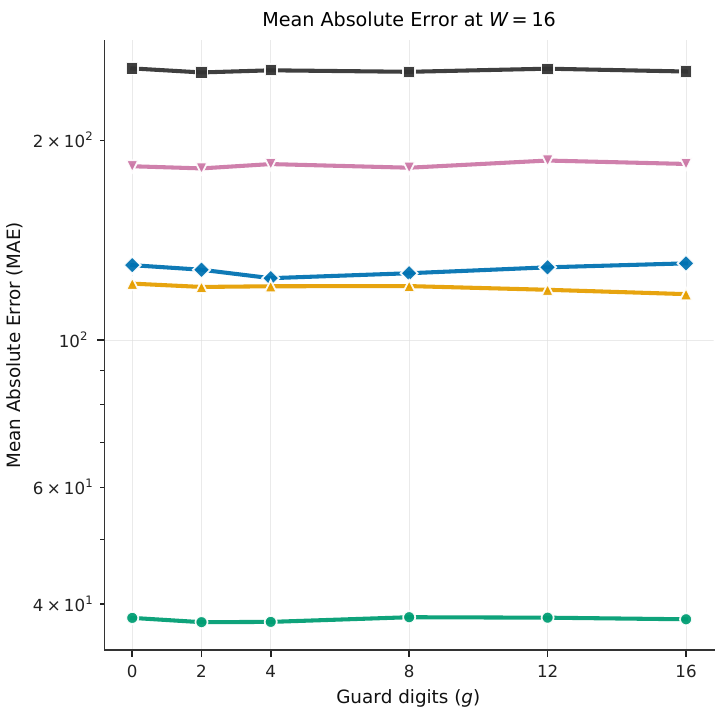}
        \caption{Mean absolute error vs. guard digits.}
        \label{fig:exp7_mae}
    \end{subfigure}

    \vspace{0.45em}

    \includegraphics[width=0.98\textwidth]{figures/fig_exp7_legend_strip.pdf}

    \caption{Impact of boundary guard digits (\(g\)) on finite-window
    canonicalization at \(W=16\). Panel~(a) shows the rapid truncation rate decrease
    as guard digits are added. Panel~(b) reports the corresponding MAE, which
    varies more modestly as it includes effects beyond lower-boundary truncation.}
    \label{fig:exp7_guard_digits}
\end{figure*}

Results are summarized in
\hyperref[tab:exp7_guard_digit_summary]{Table~\ref*{tab:exp7_guard_digit_summary}}
and visualized in \hyperref[fig:exp7_guard_digits]{Figure~\ref*{fig:exp7_guard_digits}}.
At \(g=0\), non-integer bases exhibit substantial truncation: \(0.3397\) for
\(\Phi\), \(0.8070\) for \(T_3\), and \(0.9355\) for \(T_4\). This decays rapidly
as guard digits are added. Truncation is eliminated by \(g=2\) for \(\Phi\),
while higher-order systems require wider buffers (\(T_3\) reaches \(0.0009\)
at \(g=8\); \(T_4\) hits \(0.0000\) at \(g=16\)).

Suppressing truncation improves arithmetic exactness. For example, \(\Phi\)
rises from \(0.3331\) to \(0.3726\), while \(T_4\) rises from \(0.0386\) to
\(0.2040\). The global MAE changes more modestly because, unlike the
lower-boundary bound audit
(\hyperref[tab:exp7_guard_digit_bound_audit]{Table~\ref*{tab:exp7_guard_digit_bound_audit}}),
it includes all finite-window arithmetic effects, including upper-boundary
overflow.

\begin{table*}[!htbp]
    \centering
    \caption{Guard-digit truncation and repair-cost summary at \(W=16\).}
    \label{tab:exp7_guard_digit_summary}
    \small
    \renewcommand{\arraystretch}{1.18}
    \setlength{\tabcolsep}{3.5pt}

    \begin{tabularx}{\textwidth}{@{}l *{8}{>{\centering\arraybackslash}X}@{}}
        \toprule
        System
        & \begin{tabular}{@{}c@{}}Trunc.\\ \(g=0\)\end{tabular}
        & \begin{tabular}{@{}c@{}}Trunc.\\ \(g=16\)\end{tabular}
        & \begin{tabular}{@{}c@{}}Exact\\ \(g=0\)\end{tabular}
        & \begin{tabular}{@{}c@{}}Exact\\ \(g=16\)\end{tabular}
        & \begin{tabular}{@{}c@{}}MAE\\ \(g=0\)\end{tabular}
        & \begin{tabular}{@{}c@{}}MAE\\ \(g=16\)\end{tabular}
        & \begin{tabular}{@{}c@{}}\(p_{99}\)\\ \(g=0\)\end{tabular}
        & \begin{tabular}{@{}c@{}}\(p_{99}\)\\ \(g=16\)\end{tabular} \\
        \midrule
        \(\Phi\) & 0.3397 & 0.0000 & 0.3331 & 0.3726 & 38.12  & 37.93  & 17 & 19 \\
        \(T_3\)  & 0.8070 & 0.0000 & 0.1036 & 0.2828 & 121.69 & 117.32 & 26 & 31 \\
        \(T_4\)  & 0.9355 & 0.0000 & 0.0386 & 0.2040 & 182.91 & 184.37 & 29 & 43 \\
        \bottomrule
    \end{tabularx}
\end{table*}

\begin{table*}[htbp]
    \centering
    \caption{Guard-digit tail-scale audit at \(W=16\). The observed value is the
    maximum accumulated lower-boundary error in trials without upper-boundary
    overflow. The comparison bound is the one-step tail scale from
    Lemma~2 with \(D_{\max}=2\).}
    \label{tab:exp7_guard_digit_bound_audit}
    \small
    \renewcommand{\arraystretch}{1.15}
    \setlength{\tabcolsep}{7pt}

    \begin{tabular*}{\textwidth}{@{\extracolsep{\fill}} lrrrrr @{}}
        \toprule
        System & \(g\) & \(N_{\mathrm{audit}}\) & Obs. max & Bound & Ratio \\
        \midrule
        \(\Phi\) & 0  & 4986 & \(3.44{\times}10^{-2}\)  & \(1.11{\times}10^{-1}\) & 0.309 \\
        \(\Phi\) & 16 & 5010 & \(1.42{\times}10^{-14}\) & \(5.05{\times}10^{-5}\) & \(2.81{\times}10^{-10}\) \\
        \(T_3\)  & 0  & 4949 & \(3.57{\times}10^{-2}\)  & \(3.35{\times}10^{-2}\) & 1.067 \\
        \(T_3\)  & 16 & 5130 & \(5.68{\times}10^{-14}\) & \(1.95{\times}10^{-6}\) & \(2.91{\times}10^{-8}\) \\
        \(T_4\)  & 0  & 5021 & \(2.96{\times}10^{-2}\)  & \(2.18{\times}10^{-2}\) & 1.358 \\
        \(T_4\)  & 16 & 4981 & \(1.14{\times}10^{-13}\) & \(6.00{\times}10^{-7}\) & \(1.89{\times}10^{-7}\) \\
        \bottomrule
    \end{tabular*}

    \vspace{0.5em}
    \begin{minipage}{0.96\textwidth}
        \footnotesize
        \textit{Note:} \(N_{\mathrm{audit}}\) excludes upper-boundary overflow.
        Ratios compare accumulated lower-boundary loss with the one-step tail
        scale of Lemma~2. Values above \(1\) indicate accumulated rewrite effects
        beyond a single isolated tail truncation.
    \end{minipage}
\end{table*}

The additional fractional workspace also avoids unbounded canonicalization tails.
The \(99\)th-percentile native repair cost \(p_{99}(\kappa)\) grows moderately:
\(\Phi\) increases from \(17\) to \(19\) operations, \(T_3\) from \(26\) to \(31\),
and \(T_4\) from \(29\) to \(43\). Thus, guard digits effectively reduce truncation
while keeping tail canonicalization costs finite and stable.

%% file: sections/5_discussion.tex
\section{Discussion}
\label{sec:discussion}

The experiments reveal a consistent pattern across the evaluated finite-window
numeration systems. \textbf{Intrinsic \(\beta\)-redundancy improves structural
fault visibility when it becomes a visible constraint on the stored language}:
faults can be exposed by admissibility violations, and addition intermediates can
be returned to canonical form through local rewrite identities. At the same time, this
robustness is selective. Structural detectability, semantic survival, arithmetic
exactness, and normalization cost respond to different features of the
representation system and therefore should not be reduced to a single
fault-tolerance score.

The primary conclusion is that finite-window behavior is governed by three
interacting mechanisms. The first is the density of the admissible language,
which controls how likely a perturbation is to leave the codebook. The second is
the algebraic kernel of the rewrite identities, which determines when a burst can
change the syntax without changing the represented value. The third is the
bounded hardware window itself: propagation can lose information both at the
most significant boundary through overflow and at the least significant boundary
through truncation. These mechanisms explain why the same system can be
structurally strong in one experiment, semantically weak in another, and
arithmetic-limited by boundary effects in a third.

\subsection{Clean codebooks as explanatory variables}
\label{subsec:discussion-exp1}

\hyperref[subsec:exp1]{Experiment~1} establishes the representational geometry
that later experiments probe under corruption and arithmetic. The clean-codebook
measurements in \hyperref[tab:clean_benchmark]{Table~\ref*{tab:clean_benchmark}}
are not robustness results by themselves; they describe how much structure is
already present before any fault is injected.

Within the multinacci family, this comparison is direct because the systems share
a binary alphabet and differ mainly in the forbidden-run constraint. The
\(\Phi\)-codebook forbids \(11\), \(T_3\) forbids \(111\), and \(T_4\) forbids
\(1111\). Increasing the order therefore makes the admissible language denser
and raises capacity, while weakening the local syntactic screen that exposes
faults. This gives the capacity--detectability trade-off used later to interpret
the single-digit and burst-corruption curves. Signed NAF provides a complementary
alphabet-driven baseline: its sparsity is measured inside a ternary ambient
alphabet and therefore represents a different redundancy mechanism.

Thus, the clean benchmark provides two explanatory axes rather than one universal
ranking: \textbf{constrained-language sparsity} for the non-integer bases and
\textbf{alphabet-driven canonicality} for Signed NAF. The later experiments show
which of these structural resources actually becomes visible under each fault
model.

\subsection{Arithmetic canonicalization and the two boundary losses}
\label{subsec:discussion-exp2}

\hyperref[subsec:exp2]{Experiment~2} shows that successful canonicalization is
not the same as exact finite-window arithmetic. In the reported benchmark, the
procedures return admissible outputs, but arithmetic exactness can still fail when
value is discarded at a boundary. This is a two-sided issue. Overflow loses value
above the most significant index \(L\), whereas truncation loses value below the
least significant index \(-R\).

This distinction is important because overflow is not specific to non-integer
bases. In \hyperref[tab:exp2_summary]{Table~\ref*{tab:exp2_summary}}, Binary has
zero truncation but still has only intermediate arithmetic exactness, precisely
because overflow is frequent. Therefore, truncation explains only part of the
arithmetic story. For the non-integer bases, lower-boundary truncation adds a
second loss mechanism: temporary-digit resolution can push part of the value
toward the fractional side of the register, and higher-order multinacci rewrites
increase this sensitivity. The truncation panel in
\hyperref[fig:exp2_sub_truncation]{Figure~\ref*{fig:exp2_sub_truncation}} is
therefore highly informative for ordering \(\Phi\), \(T_3\), and \(T_4\), but it
does not replace the need to account for upper-boundary overflow.

The conclusion from this experiment is that finite-window arithmetic is not
limited only by the existence of local rewrite rules. It also depends on how the
finite register treats both boundaries. Lower guard digits can address the
truncation component, but overflow requires a separate integer-side policy, such
as a wider most-significant window, carry-out reporting, or explicit overflow
detection.

\subsection{Single-digit corruption: local visibility without recovery}
\label{subsec:discussion-exp3}

\hyperref[subsec:exp3]{Experiment~3} tests whether a one-position fault becomes
visible to the admissible language. A single-digit fault is detected only if it
moves the word outside the canonical codebook. If the corrupted word remains
admissible, the intrinsic language has no independent information with which to
reject it.

The mechanisms behind the detectability rates in
\hyperref[tab:exp3_summary]{Table~\ref*{tab:exp3_summary}} differ across
systems. Binary has no structural detection because the full binary window is
admissible. Signed NAF detects many faults through a restrictive canonical subset
inside a ternary alphabet. Within the multinacci systems, the explanation is
local-language sparsity: creating a forbidden \(11\) is more likely than creating
\(111\), and creating \(111\) is more likely than creating \(1111\).

The zero semantic-survival result supplies the theoretical boundary of this
effect. It is the empirical form of
\hyperref[thm:no-free-lunch]{Proposition~1}: after one digit has changed,
admissibility checking can expose some faults, but value-preserving repair of
the observed state does not identify the original codeword without external
information.

\subsection{Local bursts: kernel survival and mechanism separation}
\label{subsec:discussion-exp4}

\hyperref[subsec:exp4_local_burst_corruption]{Experiment~4} changes the fault
model from an isolated digit to a clustered disturbance. This is where semantic
survival first becomes possible, because a multi-digit error vector can have zero
value even though its syntax changes. For the multinacci systems, this is exactly
the mechanism described by \hyperref[thm:algebraic-kernel]{Theorem~2}: a burst
survives semantically only when it aligns with a value-preserving identity of the
base, such as \(100\leftrightarrow011\) for \(\Phi\).

The Signed NAF entries in
\hyperref[tab:exp4_summary]{Table~\ref*{tab:exp4_summary}} and
\hyperref[tab:exp5_burst_exhaustive]{Table~\ref*{tab:exp5_burst_exhaustive}}
have a different interpretation. Their semantic-survival events arise from
value-neutral signed-radix cancellations in the alphabet \(\{-1,0,1\}\). This is
an alphabet-driven radix-\(2\) mechanism, not the multinacci rewrite mechanism
realized by identities such as \(100\leftrightarrow011\). Thus, Signed NAF should
be treated as a baseline with its own cancellation structure, whereas the
non-integer bases exhibit semantic survival through multinacci algebraic rewrite
identities.

Once this signed-radix mechanism is separated from the intrinsic non-integer
mechanism, the multinacci ordering has the expected meaning. The \(\Phi\)
identity is shortest, whereas the \(T_3\) and \(T_4\) identities require longer
aligned patterns. Random bursts are therefore less likely to be semantically
neutral as the multinacci order increases. The burst experiment consequently
supports the algebraic-kernel condition, while also showing why sampled semantic
survival should be interpreted together with exhaustive enumeration.

\subsection{Exhaustive verification as a consistency check}
\label{subsec:discussion-exp5}

\hyperref[subsec:exp5_exhaustive]{Experiment~5} is useful because it removes
Monte Carlo ambiguity from the sampled corruption benchmarks. It confirms that
zero single-digit semantic survival is not a sampling accident: across all
enumerated one-cell corruptions, the original value is never recovered under
value-preserving repair of the corrupted state.

The burst exhaustive results reinforce the algebraic-kernel condition in a
different way. Binary retains zero burst semantic survival, while Signed NAF and
the non-integer bases show small positive rates. The mechanisms, however, are
not the same. Signed NAF survivals are signed-radix cancellations inside the
alphabet \(\{-1,0,1\}\), whereas \(\Phi\), \(T_3\), and \(T_4\) survivals correspond
to multinacci rewrite identities. In this role, the exhaustive experiment prevents
a misleading reading of sampled finite-window coincidences as a single universal
fault-tolerance mechanism.

\subsection{Algebraic burst injection: forced kernel alignment}
\label{subsec:discussion-exp6}

\hyperref[subsec:exp6_algebraic_burst]{Experiment~6} directly tests the mechanism
behind rare burst survivals. Instead of sampling arbitrary local disturbances,
it injects the native value-preserving substitutions of each non-integer base:
\(100\to011\) for \(\Phi\), \(1000\to0111\) for \(T_3\), and
\(10000\to01111\) for \(T_4\). In all three cases, the perturbation is chosen
from the algebraic kernel of the evaluation map.

The resulting pattern is clean. As the window grows, source-pattern placements
become more frequent, especially for \(\Phi\), whose kernel pattern is shortest.
The longer \(T_3\) and \(T_4\) patterns appear less often, so their injection
coverage grows more gradually. Once an injection is made, however, the outcome is
the same across the non-integer bases: structural detectability, structural
exactness, and semantic survival all remain \(1.000\), with zero observed
semantic error.

The boundary audit strengthens this interpretation. Boundary-touching placements
occur in every system and at every tested width, but no overflow or truncation is
observed in this experiment. Thus, the observed perfect survival is not confined
to strictly interior placements. For the targeted substitutions tested here,
bounded-window normalization preserves the value of the algebraic-kernel
perturbation throughout the sampled range.

\subsection{Guard digits: controlling lower-boundary truncation}
\label{subsec:discussion-exp7}

\hyperref[subsec:exp7_guard_digits]{Experiment~7} revisits the lower-boundary
part of the arithmetic problem. The truncation curves in
\hyperref[fig:exp7_truncation]{Figure~\ref*{fig:exp7_truncation}} show that
adding least-significant guard digits rapidly suppresses truncation for \(\Phi\),
\(T_3\), and \(T_4\). In
\hyperref[tab:exp7_guard_digit_summary]{Table~\ref*{tab:exp7_guard_digit_summary}},
all three systems reach zero observed truncation by \(g=16\).

The bound audit in
\hyperref[tab:exp7_guard_digit_bound_audit]{Table~\ref*{tab:exp7_guard_digit_bound_audit}}
adds a sharper check against the one-step tail scale in \hyperref[lem:guard-digits]{Lemma~2}. At \(g=16\), the accumulated lower-boundary losses are far
below the one-step tail scale for all three non-integer bases. At \(g=0\),
\(\Phi\) remains within this scale, while \(T_3\) and \(T_4\) slightly exceed the
simple \(D_{\max}=2\) one-step estimate. This does not contradict the guard-digit
bound: the audited quantity is accumulated finite-window loss after a rewrite
cascade, not a single isolated tail truncation. The comparison therefore supports
the interpretation that guard digits suppress lower-boundary cascade loss.

Guard digits isolate the lower-boundary component of arithmetic loss. Arithmetic
exactness improves from \(0.3331\) to \(0.3726\) for \(\Phi\), and from
\(0.0386\) to \(0.2040\) for \(T_4\) in the reported table. The remaining gap is
expected because the upper boundary is unchanged: guard digits extend the
fractional side of the window, but they do not prevent carry-out above \(L\).
Thus, Experiment~7 shows that \textbf{truncation-driven loss is tunable}; it
does not by itself restore full finite-window arithmetic exactness.

This distinction also explains why the error curves change more slowly than the
truncation indicators. Guard digits remove near-boundary fractional losses, but
they do not redesign the whole finite-window value distribution or resolve
upper-boundary carry-out. A complete arithmetic architecture would need both
least-significant guard space and a most-significant overflow policy.

\subsection{Comparison with classical error-control metrics}
\label{subsec:discussion-classical-ecc}

The experimental metrics clarify how intrinsic \(\beta\)-redundancy differs from
classical error-control redundancy. Classical parity checks, Hamming-type codes,
and CRCs add check structure external to the represented arithmetic value
\cite{Hamming1950,PetersonBrown1961,LinCostello2004}. In a linear-code
formulation, a received word \(y=x+e\) is tested by a syndrome
\[
    s = Hy^\top = He^\top,
\]
which reveals properties of the error vector \(e\). The syndrome is not the
arithmetic value itself; it is auxiliary information stored or computed for
protection.

The \(\beta\)-codebooks studied here are different. Their positive signal is not
an external syndrome but a violation of the admissible language. The word is
rejected when a forbidden local pattern such as \(11\), \(111\), or \(1111\)
appears. This is why structural detectability in
\hyperref[fig:exp3_sub_detectability]{Figure~\ref*{fig:exp3_sub_detectability}}
and
\hyperref[fig:exp4_sub_detectability_by_b]{Figure~\ref*{fig:exp4_sub_detectability_by_b}}
is the natural analogue of a detection metric, whereas semantic recovery is much
stronger and usually unavailable. A Hamming-style code is designed so that some
syndromes identify correctable errors. The intrinsic \(\beta\)-language can only
say that the observed word is not a valid canonical representative; by itself,
it does not identify the unique pre-fault word after a generic single-digit
change.

Signed NAF sits between these viewpoints. Its high detectability in
\hyperref[tab:exp3_summary]{Table~\ref*{tab:exp3_summary}} and
\hyperref[tab:exp4_summary]{Table~\ref*{tab:exp4_summary}} comes from a
restrictive canonical subset inside a larger ternary alphabet, not from a
non-integer base. The experiments therefore separate three mechanisms:
\textbf{external check redundancy}, \textbf{alphabet-driven redundancy}, and
\textbf{intrinsic constrained-language redundancy}. Under the selected protocol,
\(\Phi\) achieves slightly lower single-digit detectability than Signed NAF but
requires much lower local repair cost. Signed NAF remains stronger in arithmetic
exactness because its global recoding dynamics and boundary behavior differ.

\subsection{Future directions: local syndromes, reversible rewriting, and hybrid protection}
\label{subsec:discussion-future-directions}

The finite-window results point to several extensions. One direction is to treat
forbidden-pattern checks as local admissibility syndromes. For the no-\(1^m\)
language \(C_m^{L,R}\), one may define
\[
    s_i(d)
    =
    \mathbf{1}\{d_i d_{i+1}\cdots d_{i+m-1}=1^m\}.
\]
The detectability rates in
\hyperref[fig:exp3_sub_detectability]{Figure~\ref*{fig:exp3_sub_detectability}}
and
\hyperref[fig:exp4_sub_detectability_by_b]{Figure~\ref*{fig:exp4_sub_detectability_by_b}}
can then be interpreted as activation probabilities of these local checks under
the chosen corruption model. This would enable studies of syndrome density,
false negatives, and capacity--detectability trade-offs across larger families of
constrained numeration systems, including Parry and Pisot bases.

A second direction is hybrid protection. Intrinsic admissibility can expose some
local faults, but cannot supply the external syndrome information needed to
invert arbitrary single-digit corruptions. Future architectures could combine
lightweight classical checks, such as parity or CRC-style constraints, with
intrinsic multinacci admissibility checks. External redundancy would localize or
classify errors, while the constrained language would provide low-cost structural
filtering and value-preserving canonicalization when the perturbation matches a
native rewrite identity.

A third direction is reversible and quantum-compatible rewriting. Formally, a
finite codebook can be associated with the computational-basis subspace
\[
    \mathcal{H}_{C_m}
    =
    \operatorname{span}\{|d\rangle : d\in C_m^{L,R}\},
\]
and forbidden-run checks can be represented by diagonal projectors
\[
    \Pi_i^{(m)}
    =
    |1^m\rangle\!\langle 1^m|_{i,\ldots,i+m-1}.
\]
This gives a limited analogy with syndrome extraction in stabilizer quantum
error correction, where local or semi-local measurements determine whether a
state has left the codespace without measuring the encoded logical state
\cite{Gottesman1997,NielsenChuang2010}. The analogy should not be overextended:
the bounded-window normalization rule \(011\to100\) is directed and many-to-one,
whereas a quantum circuit must be unitary unless it measures and records
ancillary information. A reversible implementation would therefore require
embedding the rewrite relation, for example by retaining history registers, using
ancilla-assisted syndrome extraction, or restricting the operation to explicitly
paired value-equivalent configurations.

Finally, the present study leaves open implementation-level and model-level
extensions. The experiments measure scheduler-dependent rewrite activity,
propagation depth, overflow, and truncation; they do not synthesize circuits. A
natural next step is to construct finite-state or block-prefix canonicalizers for
the tested rule sets and compare area, depth, switching activity, and boundary
behavior. Similarly, the arithmetic benchmark isolates addition and upward carry
propagation. Subtraction, signed intermediate states, borrow cascades, mixed
workloads, and physically informed multi-bit upset models may change the balance
between detectability, exactness, and repair cost. Extending the same
finite-window methodology to these cases would test whether the observed
capacity--detectability and kernel-survival patterns persist beyond the additive
and logically contiguous corruption models studied here.

%% file: sections/6_conclusion.tex
\clearpage
\section{Conclusion}

This study gives a finite-window answer to a precise digital robustness
question: intrinsic \(\beta\)-redundancy provides structural fault visibility
and value-preserving re-admissibilization of observed non-canonical states,
while true semantic survival is restricted to algebraic-kernel-aligned
perturbations. This separates notions often grouped under the broad label of
``robustness'': ambient semantic non-uniqueness, canonical admissibility of
stored words, normalization of corrupted states, recovery of the original value,
and arithmetic exactness.

The main theoretical boundary is single-digit recovery. Within a canonically
injective finite codebook, a genuine single-digit corruption cannot be
semantically recovered using only the internal structure of the representation.
Exact structural repair can preserve the value of the observed corrupted
configuration, but it cannot reconstruct the original value without external
information. This clarifies the difference between redundancy in the ambient
representation space and error correction inside a finite storage format.

Intrinsic redundancy nevertheless has a concrete operational role. Local
algebraic identities can make certain multi-digit perturbations semantically
neutral. In the \(\varphi\)-system this role is represented by
\(100 \leftrightarrow 011\); in higher-order multinacci systems analogous
identities involve longer rewrite neighborhoods. The experiments show that such
semantic survival is rare under random bursts, but when it occurs it is governed
by the algebraic kernel of the evaluation map rather than by generic fault
correction.

The finite-window experiments expose the trade-offs behind this behavior.
Stricter languages such as \(\Phi\) provide stronger local syntactic screening
and lower local rewrite cost, while denser languages such as \(T_3\) and \(T_4\)
admit more stored words but expose fewer local faults and are more sensitive to
lower-boundary truncation. Binary provides the nonredundant baseline, whereas
Signed NAF provides a strong alphabet-driven redundancy baseline with different
recoding dynamics and boundary behavior.

The resulting interpretation is modest but sharp. Intrinsic
\(\beta\)-redundancy does not turn a finite canonical codebook into a classical
error-correcting code. It does provide a native constrained-language mechanism
for detecting structural violations, re-admissibilizing observed faulty or
arithmetic intermediate states, and preserving semantic value for specific
algebraic burst patterns. This is a narrower claim than generic fault-tolerant
arithmetic, but it is more precise, measurable, and architecturally meaningful.

%% file: sections/appendix.tex
\clearpage
\appendix

\section[Ambient State-Space Graph of phi-Representations]
{Ambient State-Space Graph of \texorpdfstring{$\varphi$}{phi}-Representations}
\phantomsection
\label{app:redundancy-graph}

\begin{figure}[htbp]
    \centering
    \vspace*{\fill}

    \makebox[\textwidth][c]{%
        \begin{adjustbox}{
            max width=\textwidth,
            max totalheight=\dimexpr\textheight-7\baselineskip\relax,
            center
        }
            \input{figures/fig_phi_redundancy_flow_B}
        \end{adjustbox}%
    }
    \vspace{5px}
    \caption{Alluvial-style ambient semantic flow diagram for finite $\varphi$-representations up to length $4$. Each vertical layer corresponds to all binary strings of fixed length $n$, while each node represents a semantic state labelled by its exact value $a+b\varphi$. Highlighted nodes indicate semantic collisions in the ambient representation space, where multiple distinct strings evaluate to the same value. These collisions should not be interpreted as multiple admissible representatives inside the no-$11$ codebook. The visible merges are generated by the local equivalence $011 \leftrightarrow 100$, induced by $\varphi^2=\varphi+1$.}
    \label{fig:phi-redundancy-flow}

    \vspace*{\fill}
\end{figure}
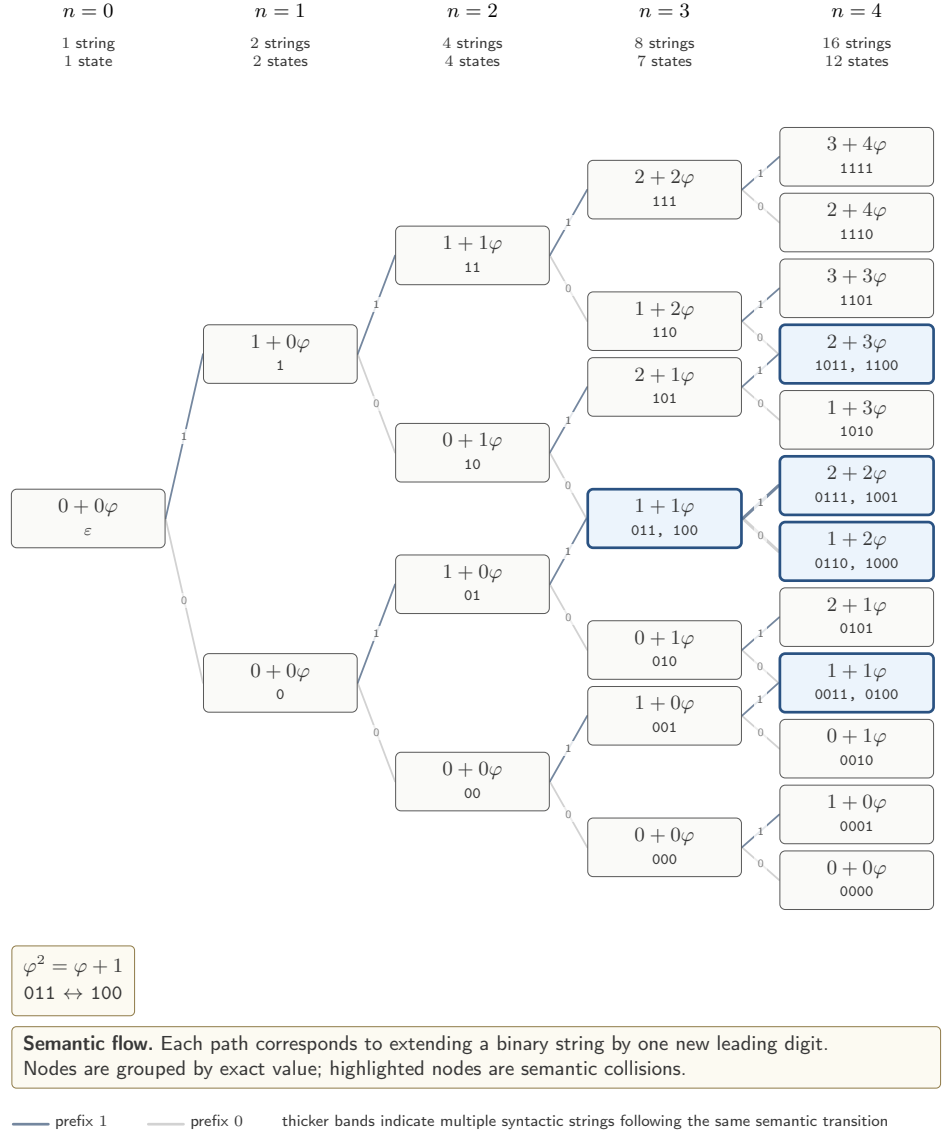

To make the ambient redundancy structure of finite $\varphi$-representations explicit, we constructed the layered state-space graph in Fig.~\ref{fig:phi-redundancy-flow}. 
It is obtained by enumerating all binary strings of a fixed length, not only admissible no-$11$ strings, and merging strings with the same value in base $\varphi$. 
Thus, it is neither a prefix tree nor the stored codebook, but an \emph{ambient semantic quotient graph}: distinct syntactic strings may collapse into one semantic node.

Each node is labeled by the exact value $a+b\varphi$ and the strings at that layer evaluating to it. 
Edges are one-bit extensions by $0$ or $1$. 
Thick borders mark \emph{semantic collisions}, where multiple ambient strings encode the same value. 
These collisions visualize redundancy in the ambient space, not multiplicity inside $C_\varphi^{L,R}$.

The merge structure is generated by
\[
\varphi^2=\varphi+1,
\]
which induces
\[
011 \leftrightarrow 100.
\]
Thus, any substring $011$ can be replaced by $100$ without changing value, and longer collision classes are built recursively from this local equivalence. 
The first nontrivial examples are
\[
011 = 100,\qquad
0011 = 0100,\qquad
0110 = 1000.
\]

This viewpoint is useful for repair because the number of distinct \emph{semantic} states grows more slowly than the number of \emph{syntactic} strings. 
Let $c_n$ denote the number of distinct ambient semantic nodes at layer $n$. Then
\[
c_0=1,\qquad c_1=2,\qquad c_n=c_{n-1}+c_{n-2}+1 \quad (n\ge 2),
\]
and equivalently
\[
c_n = F_{n+3}-1,
\]
where $F_k$ is the $k$th Fibonacci number. 
The exact combinatorial derivation and rigorous enumeration of bounded $\varphi$-representations are established by Dekking and van Loon~\cite{DekkingVanLoon2023}. 
By contrast, there are $2^n$ binary strings at layer $n$. 
Thus the ambient semantic state space is substantially smaller than the ambient syntactic one.

This gap is not multiple admissible representatives of the same value. 
The no-$11$ codebook selects canonical representatives, whereas the graph shows pre-canonical collisions. 
For repair, it visualizes how non-canonical or corrupted strings may share values with other ambient strings and be mapped by normalization toward the admissible language.

Because the full graph is too large for a standard page, Fig.~\ref{fig:phi-redundancy-flow} shows only the first layers, up to length $4$, with continuation indicated schematically. 
Even there, the graph shows the essential phenomenon: local algebraic equivalence induces systematic semantic merging in the ambient representation space.

\clearpage
\clearpage

\section{A Logarithmic-Depth Perspective on Canonicalization}
\phantomsection
\label{app:parallel-canonicalization}

\noindent\textbf{Caution on the scope of Proposition~B.1.} \hyperref[prop:parallel-finite-state]{Proposition~B.1} is a metatheorem about an
implementation pathway: if a canonicalization procedure is already expressed as
a fixed number of deterministic finite-state transducer passes with bounded
output delay, then those passes can be parallelized by a prefix-composition
network. The proposition is not a proof that the particular bounded-window
MSB-first scheduler used in the experiments, based on
Eqs.~\eqref{eq:phi-011-to-100}--\eqref{eq:phi-two-carry-rule}, has itself been
verified as such an FST realization. Establishing that equivalence would require
an explicit transducer construction, a proof that its output agrees with the
selected canonical normal form under the stated boundary policy, and a separate
implementation-level analysis. This verification task is left open in the
present paper and is not claimed as solved here.

The bounded-window experiments in this work use a strict serial rewrite order:
the most significant visible defect is resolved first, and the word is rescanned
until no forbidden pattern or extended digit remains. This convention makes the
rewrite count and propagation metrics reproducible, but it should not be
interpreted as an intrinsic lower bound on implementation latency. The same
normalization relation may admit a substantially more parallel realization when
a suitable finite-state implementation is available.

\paragraph{Local rules can still induce global propagation.}
The locality of an individual rewrite does not imply that the entire repair has constant latency. Consider the admissible family
\[
u_m=(01)^m\,\underline{0}\,010,
\]
where the underlined digit is corrupted from $0$ to $1$. The faulty word is
\[
\widetilde{u}_m=(01)^m\,\underline{1}\,010.
\]
Repeated application of $011\to100$ yields the cascade
\[
(01)^m1 0^1 10
\longrightarrow
(01)^{m-1}1 0^3 10
\longrightarrow\cdots\longrightarrow
1 0^{2m+1}10.
\]
Thus a single-digit corruption can trigger $m=\Theta(W)$ dependent rewrites in a word of width $W=2m+4$. Consequently, an exact explicit repair cannot in general be bounded by the number of corrupted digits alone. In particular, a sequential local implementation may have worst-case latency $\Theta(W)$ even when only one digit is corrupted.

This phenomenon is analogous to carry propagation in an ordinary ripple-carry adder: each local operation is simple, but the dependency chain may span the full word. The appropriate analogue of carry lookahead is therefore not to execute all rewrites eagerly, but to summarize how a block transforms its incoming boundary state and then compose these summaries in parallel.

\paragraph{Finite-state block summaries.}
Suppose that one canonicalization pass is realized by a deterministic finite-state transducer
\[
\mathcal{T}=(Q,E,\delta,\lambda,q_{\mathrm{in}}),
\]
where $Q$ is a fixed finite state set, $E$ is the working alphabet, $\delta$ is the state-transition map, and $\lambda$ produces the output associated with a transition. For an input word
\[
x=x_0x_1\cdots x_{W-1}
\]
written in the order in which the pass scans it, each symbol induces a state transformation
\[
f_i:Q\to Q,
\qquad
f_i(q):=\delta(q,x_i).
\]
The state immediately before position $i$ is determined by the prefix composition
\[
q_i=(f_{i-1}\circ f_{i-2}\circ\cdots\circ f_0)(q_{\mathrm{in}}).
\]
More generally, a contiguous block $B=x_a\cdots x_b$ can be represented by the summary function
\[
F_B:=f_b\circ f_{b-1}\circ\cdots\circ f_a.
\]
For adjacent blocks $B_1$ and $B_2$ scanned in that order,
\[
F_{B_1B_2}=F_{B_2}\circ F_{B_1}.
\]
Function composition is associative, so the block summaries form an associative prefix problem. Since $Q$ is fixed independently of $W$, every function $Q\to Q$ has a constant-size representation, and the composition of two summaries has constant cost.

\paragraph{Proposition B.1 (Parallel evaluation of finite-state passes).}\phantomsection\label{prop:parallel-finite-state}
Assume that canonicalization is expressed as a fixed number of deterministic finite-state passes, each with a fixed finite state set and bounded output delay. Then a word of width $W$ can be canonicalized by a combinational network of depth
\[
O(\log W)
\]
and size
\[
O(W).
\]
The total sequential work remains $O(W)$.

\emph{Proof sketch.}
For one pass, all symbol transformations $f_i$ are generated independently in constant depth. A work-efficient parallel-prefix network computes all prefix compositions
\[
f_{i-1}\circ\cdots\circ f_0
\]
in $O(\log W)$ depth and $O(W)$ total size. Applying each resulting prefix transformation to $q_{\mathrm{in}}$ gives every state $q_i$ in parallel, after which the output symbols are produced locally by $\lambda$ in constant additional depth. A right-to-left pass is handled by reversing the indexing. Composing a fixed number of such passes changes only the constant factors, so the asymptotic depth remains $O(\log W)$ and the size remains $O(W)$.

The proposition concerns \emph{parallel depth}, not running time on a single sequential processor. On a conventional serial machine, evaluating or materializing all $W$ output digits still requires linear work. The logarithmic bound describes the critical path of a parallel circuit or a sufficiently parallel machine.

\paragraph{Relation to known Fibonacci and $\varphi$ arithmetic.}
Finite-state normalization is available for broad classes of Pisot numeration systems, including the golden-ratio base~\cite{Frougny1992,Frougny2002}. A closely related explicit precedent is Zeckendorf arithmetic. Ahlbach et al. give linear-time addition and subtraction algorithms for Fibonacci representations and show that these operations can be implemented by combinational logic networks of linear size and logarithmic depth~\cite{Ahlbach2013Zeckendorf}. Their setting shares the no-adjacent-ones canonical language and the Fibonacci carry identity underlying $011\to100$, although it is not identical to the two-sided finite-window $\varphi$ model used here.

A different route is to enlarge the digit alphabet. For the golden mean, constant-time parallel addition is possible on the redundant signed alphabet $\{-1,0,1\}$~\cite{Frougny2011Parallel}. This result does not remove the distinction relevant to the present work: producing a redundant intermediate representation in constant depth is not the same task as returning the strict binary greedy representative in $\{0,1\}$ with no occurrence of $11$.

\paragraph{Interpretation of the experimental cost metrics.}
The metrics $\kappa$ and $\pi$ introduced in Section~\ref{subsec:correctness-criteria-metrics} quantify, respectively, the amount of local rewrite activity and the spatial extent of propagation under the deterministic serial convention. They should therefore be interpreted as scheduler-dependent structural
indicators of normalization effort, not as implementation-independent latency
bounds: two implementations may realize the same canonical map while having
different critical-path depth, area, and switching behavior.

In particular, the experiments do not estimate VLSI energy, switching activity, fan-out, or gate area. 
The propagation metrics reported here are structural metrics of the selected rewrite convention, not measurements of a synthesized circuit.

An event-driven serial implementation could maintain a queue of positions whose local neighbourhood may contain a defect, avoiding repeated full rescans and reducing the bookkeeping cost to $O(W+\kappa)$. A finite-state parallel implementation instead replaces the carry chain by block summaries and a prefix-composition network, potentially reducing the critical path to $O(\log W)$ while retaining $O(W)$ total work. Overflow and truncation can be represented by additional boundary states or by guard digits, so the finite-window policy can be incorporated without changing the asymptotic depth.

\paragraph{Scope of this appendix.}
The discussion above establishes an implementation pathway rather than presenting a synthesized circuit for the complete rule set in Eqs.~\eqref{eq:phi-011-to-100}--\eqref{eq:phi-two-carry-rule}. A complete hardware construction would require an explicit transducer state table, a proof that its output agrees with the selected canonical normal form, and an implementation-level evaluation of gate count, fan-out, and boundary handling. The present experiments intentionally retain the serial priority rule because it gives a transparent and reproducible definition of rewrite cost. The logarithmic-depth perspective shows, however, that the observed carry-cascade length should not automatically be identified with the minimum achievable parallel latency.
 
\section{Admissibility Conditions for Higher-Order Multinacci Bases}
\phantomsection
\label{app:multinacci-admissibility}

While Section 2.1 explicitly formalizes the structural admissibility constraint for the golden-ratio base ($\beta=\varphi$), the experimental protocol in \hyperref[subsec:comparison-systems-experimental-protocol]{Section~\ref*{subsec:comparison-systems-experimental-protocol}} also evaluates the tribonacci ($T_3$) and tetranacci ($T_4$) systems. To ensure syntactic rigor across all evaluated comparison systems, we formally state their respective admissibility conditions here.

For an integer $m \ge 2$, the $m$-step multinacci base $\beta_m$ is defined as the unique real root $\beta_m > 1$ of the polynomial equation
\[
x^m = x^{m-1} + x^{m-2} + \dots + x + 1.
\]
The golden ratio corresponds to $m=2$, tribonacci to $m=3$, and tetranacci to $m=4$. 

By Parry's theorem~\cite{Parry1960}, the admissibility of a digit string in base $\beta_m$ is strictly governed by the quasi-greedy expansion of $1$. For any $m \ge 2$, the algebraic relation defining the base implies that the finite greedy expansion of $1$ is exactly $m$ digits long:
\[
1 = \sum_{i=1}^{m} \beta_m^{-i} = 0.\underbrace{11\dots1}_{m \text{ times}}.
\]
Consequently, the infinite quasi-greedy expansion of $1$, which serves as the universal upper bound for lexicographically valid greedy tails, is the periodic sequence
\[
(\underbrace{11\dots1}_{m-1 \text{ times}} 0)^\infty.
\]
Restricting any proper suffix from lexicographically exceeding this boundary sequence structurally forbids the occurrence of $m$ consecutive ones~\cite{Frougny2002}. This provides a direct, mathematically rigorous generalization of the admissibility constraint to higher-order systems.

\paragraph{Tribonacci admissibility ($T_3$).}
For the tribonacci base $\beta_3$ ($m=3$), the forbidden pattern is the block $111$. A finite binary string $d \in \{0,1\}^{(\mathbb{Z})}$ is structurally admissible if and only if it contains no three consecutive ones, meaning:
\[
d_i + d_{i-1} + d_{i-2} \le 2 \qquad \text{for all } i \in \mathbb{Z}.
\]

\paragraph{Tetranacci admissibility ($T_4$).}
For the tetranacci base $\beta_4$ ($m=4$), the forbidden pattern is the block $1111$. A finite binary string $d \in \{0,1\}^{(\mathbb{Z})}$ is structurally admissible if and only if:
\[
d_i + d_{i-1} + d_{i-2} + d_{i-3} \le 3 \qquad \text{for all } i \in \mathbb{Z}.
\]

\paragraph{Context-sensitive canonicalization and burst overlap resolution.}
In the finite-window computational model, the respective valid codebooks $C_{T_3}^{L,R}$ and $C_{T_4}^{L,R}$ are defined strictly as the subsets of $\{0,1\}^{I_{L,R}}$ that satisfy these local inequalities at every valid index boundary. Whenever arithmetic operations or transient digital corruptions introduce a forbidden sequence, the state must be re-admissibilized to return to the canonical codebook.

Crucially, this canonicalization is governed by a context-sensitive grammar where the presence of a prefix zero is mandatory to initiate a value-preserving carry. The corresponding directed rewrite rules must be explicitly defined with this left-sided context:
\[
0111 \longrightarrow 1000 \qquad (\text{for } T_3)
\]
and
\[
01111 \longrightarrow 10000 \qquad (\text{for } T_4).
\]
The inclusion of the leading zero in the search pattern guarantees that the local substitution strictly preserves the semantic value and appropriately allocates the carry to an empty higher-order position.

When localized burst errors or raw arithmetic states create overlapping
forbidden patterns---such as the corrupted string \(1111\) in the \(T_3\)
system---the algebraic identities alone do not determine which local occurrence
should be rewritten first. We therefore use the bounded-window scheduler defined
in \hyperref[alg:bounded-msb-scheduler]{Section~\ref*{subsec:corruption-model-canonicalization}}
to make the experimental procedure deterministic.

Under this scheduler, the currently visible defect with the largest anchor index
is processed first. If several defects have the same anchor, the fixed tie rule
specified in the scheduler is used. Thus the MSB-first rule is a reproducibility
convention for resolving overlap ambiguity in finite-window experiments, not a
mathematical assertion that lower-order-first or parallel schedules could not
also normalize the same abstract value under different implementation choices.
The reported propagation depth and normalization cost for \(T_3\) and \(T_4\)
should therefore be read as measurements of this specified scheduler.

General finite-state normalization results for Pisot numeration systems support
the existence of suitable normalization mechanisms~\cite{Frougny1992,Frougny2002},
but they are not used here as a confluence or strong-termination proof for this
specific bounded-window priority schedule and boundary convention. Termination
of the reported experimental procedure is instead bounded operationally by the
finite rewrite budget \(K_{\max}(W)\) in
\hyperref[alg:bounded-msb-scheduler]{the bounded-window scheduler}. Trials that
exhaust this budget are recorded with halt flag \(H=0\) and do not contribute to
normalization success. If a burst
corruption creates a forbidden pattern at the left boundary of the window, so
that the required prefix zero falls outside the tracked window \(I_{L,R}\), the
procedure records the resulting leftward carry as overflow and applies the
prescribed boundary policy rather than claiming exact value-preserving
normalization at the boundary.

\paragraph{Temporary digit resolution for arithmetic canonicalization.}

In arithmetic trials, digit-wise addition initially produces digits only in
\(\{0,1,2\}\). During subsequent carry propagation, however, several carries may
accumulate at the same position, so the scheduler must also define how to handle
temporary digits \(q\ge 3\).

For an \(m\)-step multinacci base \(\beta_m\), the defining identity
\[
\beta_m^m=\beta_m^{m-1}+\beta_m^{m-2}+\cdots+\beta_m+1
\]
implies the carry identity
\[
2\beta_m^k=\beta_m^{k+1}+\beta_m^{k-m}.
\]
Indeed, multiplying
\[
2\beta_m^m=\beta_m^{m+1}+1
\]
by \(\beta_m^{k-m}\) yields the stated relation. We therefore use the following
general temporary-digit resolution schema:
\[
d_k=q\ge 2
\quad\Longrightarrow\quad
\begin{cases}
d_k := q-2,\\
d_{k+1} := d_{k+1}+1,\\
d_{k-m} := d_{k-m}+1.
\end{cases}
\tag{C.1}\label{eq:multinacci-general-temp-rule}
\]
This rule is value-preserving whenever the required write positions are inside
the tracked window, since
\[
q\beta_m^k
=
(q-2)\beta_m^k+\beta_m^{k+1}+\beta_m^{k-m}.
\]
For \(q=2\), the digit at position \(k\) is eliminated. For \(q>2\), one
application reduces it by two, and repeated applications are allowed until the
digit belongs to \(\{0,1\}\), unless the finite rewrite budget is exhausted.

Consequently, temporary digits in the tribonacci system are resolved by
\[
d_k=q\ge 2
\quad\Longrightarrow\quad
\begin{cases}
d_k := q-2,\\
d_{k+1} := d_{k+1}+1,\\
d_{k-3} := d_{k-3}+1,
\end{cases}
\]
using
\[
2\beta_3^k=\beta_3^{k+1}+\beta_3^{k-3}.
\]
For the tetranacci system, the corresponding rule is
\[
d_k=q\ge 2
\quad\Longrightarrow\quad
\begin{cases}
d_k := q-2,\\
d_{k+1} := d_{k+1}+1,\\
d_{k-4} := d_{k-4}+1,
\end{cases}
\]
using
\[
2\beta_4^k=\beta_4^{k+1}+\beta_4^{k-4}.
\]

The same bounded-window convention used for the \(\varphi\)-system applies:
writes to indices \(i>L\) are recorded as overflow and discarded, while writes
to indices \(i<-R\) are recorded as truncation and discarded. The evaluation
strategy is deterministic: at each step, the procedure resolves the most
significant visible defect, either a temporary digit \(q\ge 2\) or a forbidden
block of \(m\) consecutive ones. A temporary digit \(q\ge 3\) is therefore not
outside the definition of the algorithm; it is resolved by repeated applications
of \eqref{eq:multinacci-general-temp-rule}. The procedure halts successfully
when no temporary digit or forbidden block remains.

\subsection{Canonical Injectivity of the Comparison Codebooks}
\label{app:canonical-injectivity-comparison}

This subsection makes explicit the finite-window injectivity assumption used in
the single-digit recovery limit. The proposition below verifies this assumption
for every canonical codebook used in the comparison experiments.

\paragraph{Proposition C.1 (Finite-window canonical injectivity).}
\phantomsection\label{prop:comparison-codebook-injectivity}
For every finite window \(I_{L,R}\), the value map is injective on each canonical
codebook used in the experiments: standard binary, signed NAF, \(\Phi\), \(T_3\),
and \(T_4\).

\begin{proof}
For standard binary, multiplying by \(2^R\) reduces the claim to the usual
uniqueness of finite base-\(2\) integer expansions. For \(\Phi\), the claim is
exactly
\hyperref[lem:phi-codebook-injective]{Lemma 1}.

For signed NAF, multiplying by \(2^R\) reduces the finite-window representation
to a finite signed binary expansion of an integer with digits in
\(\{-1,0,1\}\) and no adjacent nonzero digits. Let \(n\) be the represented
integer and let \(d_0\) be the least significant digit. If \(n\) is even, then
\(d_0=0\). If \(n\) is odd, then \(d_0\in\{-1,1\}\), and the no-adjacent-nonzero
constraint forces the next digit to be \(0\). Hence \(n-d_0\) must be divisible
by \(4\), so \(d_0\) is uniquely determined by \(n \bmod 4\). Removing this digit
and dividing by \(2\) gives uniqueness by induction.

For \(T_3\) and \(T_4\), we prove a slightly more general statement for an
arbitrary \(m\)-step multinacci base. Let \(\beta_m\) be the real root
\[
\beta_m^m
=
\beta_m^{m-1}+\beta_m^{m-2}+\cdots+\beta_m+1,
\qquad m\ge 2.
\]
The canonical binary language forbids runs of \(m\) consecutive ones. After
multiplying by \(\beta_m^R\), it is enough to consider nonnegative indices.

We first prove the lower-order domination bound needed for injectivity. For
\(n\ge 0\), define
\[
M_n
:=
\max
\left\{
\sum_{j=0}^{n-1} u_j\beta_m^j
:
u\in\{0,1\}^{\{0,\ldots,n-1\}}
\text{ contains no block }1^m
\right\}.
\]
We claim that
\[
M_n<\beta_m^n
\qquad\text{for all } n\ge 0.
\]

For \(n=0\), this is immediate since \(M_0=0\). We next handle
\(1\le n<m\). Since \(\beta_m\in(1,2)\), the defining identity is equivalent to
\[
\beta_m^m(2-\beta_m)=1.
\]
Hence, for \(1\le n<m\),
\[
\beta_m^n(2-\beta_m)<\beta_m^m(2-\beta_m)=1.
\]
Equivalently,
\[
\beta_m^{n+1}-2\beta_m^n+1>0,
\]
and therefore
\[
\beta_m^n
>
1+\beta_m+\cdots+\beta_m^{n-1}.
\]
Thus \(M_n<\beta_m^n\) for all \(1\le n<m\), since in these shorter windows
even the all-one word has value strictly below \(\beta_m^n\).

For \(n=m\), the all-one word of length \(m\) is forbidden. The largest
possible contribution is therefore strictly below
\[
1+\beta_m+\cdots+\beta_m^{m-1}
=
\beta_m^m,
\]
so \(M_m<\beta_m^m\).

Now let \(n>m\), and assume the claim has been proved for all smaller lengths.
Consider any admissible word of length \(n\), and let \(r\) be the length of its
terminal run of ones at the most significant end. Since \(m\) consecutive ones
are forbidden, \(0\le r\le m-1\).

If \(r=0\), then the most significant digit is zero, and the word has value at
most \(M_{n-1}<\beta_m^{n-1}<\beta_m^n\).

If \(1\le r\le m-1\), then the \(r\) most significant positions contribute
\[
\beta_m^{n-1}+\beta_m^{n-2}+\cdots+\beta_m^{n-r}.
\]
The next lower position, if present, must be zero; otherwise the terminal run
would have length \(r+1\). The remaining lower part has length \(n-r-1\), so by
the induction hypothesis its value is strictly less than
\(\beta_m^{n-r-1}\). Hence the whole word has value strictly less than
\[
\beta_m^{n-1}+\beta_m^{n-2}+\cdots+\beta_m^{n-r}
+
\beta_m^{n-r-1}.
\]
If \(r\le m-2\), this is strictly smaller than
\[
\beta_m^{n-1}+\beta_m^{n-2}+\cdots+\beta_m^{n-m}
=
\beta_m^n,
\]
because at least one positive term from the defining multinacci sum is missing.
If \(r=m-1\), the displayed upper bound equals \(\beta_m^n\), but the inequality
is still strict because the lower part is strictly less than
\(\beta_m^{n-m}\). Thus \(M_n<\beta_m^n\) in all cases.

We can now prove injectivity. Let
\[
a,b\in\{0,1\}^{\{0,\ldots,N\}}
\]
be two admissible words with no block \(1^m\), and suppose
\[
\sum_{j=0}^{N} a_j\beta_m^j
=
\sum_{j=0}^{N} b_j\beta_m^j.
\]
If \(a\neq b\), let \(k\) be the largest index at which they differ. Interchanging
\(a\) and \(b\), if necessary, assume \(a_k=1\) and \(b_k=0\). Then
\[
0
=
\sum_{j=0}^{N}(a_j-b_j)\beta_m^j
=
\beta_m^k+\sum_{j=0}^{k-1}(a_j-b_j)\beta_m^j.
\]
The negative lower-order contribution is maximized when the lower digits of
\(b\) have the largest admissible value below \(k\), which is at most \(M_k\).
By the bound just proved,
\[
M_k<\beta_m^k.
\]
Therefore
\[
\sum_{j=0}^{N}(a_j-b_j)\beta_m^j
\ge
\beta_m^k-M_k
>
0,
\]
contradicting equality. Hence \(a=b\). This proves finite-window canonical
injectivity for the \(m\)-step multinacci codebook, and in particular for
\(T_3\) and \(T_4\).
\end{proof}

\clearpage
\section{Matched-Range Robustness Check}
\phantomsection
\label{app:matched_range}

The main experiments compare systems at equal window width \(W\), which fixes
the storage budget but not the numerical dynamic range. As a robustness check, we
also run a matched-\(V_{\max}\) comparison. For each system \(X\), we select a
balanced window whose
\[
    V^{L,R}_{\max,X}
    =
    \max_{d\in C_X^{L,R}} \operatorname{val}^{L,R}_X(d)
\]
is closest to the reference value \(V_{\max}=75.9787\), obtained from \(\Phi\) at
\(W=16\).

\vspace{-1.5em}

\begin{table}[htbp]
    \centering
    \caption{Matched-range windows. The reference target is
    \(V_{\max}=75.9787\), obtained from \(\Phi\) at \(W=16\).}
    \label{tab:matched_range_windows}
    \small
    \renewcommand{\arraystretch}{1.12}
    \setlength{\tabcolsep}{5pt}
    \begin{tabular}{lrrrrr}
        \toprule
        System & \(W\) & \(L\) & \(R\) & \(V_{\max}\) & Rel. err. \\
        \midrule
        Binary     & 11 & 5 & 5 & 63.969 & 0.158 \\
        Signed NAF & 12 & 6 & 5 & 85.313 & 0.123 \\
        \(\Phi\)   & 16 & 8 & 7 & 75.979 & 0.000 \\
        \(T_3\)    & 13 & 6 & 6 & 71.189 & 0.063 \\
        \(T_4\)    & 12 & 6 & 5 & 98.831 & 0.301 \\
        \bottomrule
    \end{tabular}
\end{table}

\vspace{-1.5em}

The same native admissibility, corruption, and repair rules are then applied to
these matched windows. The burst check uses width \(b=3\), matching the shortest
non-integer kernel identity \(100\leftrightarrow011\).

\vspace{-1.5em}

\begin{table*}[htbp]
    \centering
    \caption{Matched-range robustness check. Each row uses \(10^4\) trials.
    Burst columns use width \(b=3\). NMAE is normalized by the matched-window
    \(V_{\max}\).}
    \label{tab:matched_range_robustness}
    \small
    \renewcommand{\arraystretch}{1.14}
    \setlength{\tabcolsep}{4.2pt}
    \begin{tabular*}{\textwidth}{@{\extracolsep{\fill}} lrrrrrrrr @{}}
        \toprule
        System
        & \(W\)
        & \begin{tabular}{@{}c@{}}Single\\ det.\end{tabular}
        & \begin{tabular}{@{}c@{}}Single\\ sem.\end{tabular}
        & \begin{tabular}{@{}c@{}}Single\\ NMAE\end{tabular}
        & \begin{tabular}{@{}c@{}}Burst\\ det.\end{tabular}
        & \begin{tabular}{@{}c@{}}Burst\\ sem.\end{tabular}
        & \begin{tabular}{@{}c@{}}Burst\\ NMAE\end{tabular}
        & \begin{tabular}{@{}c@{}}\(p_{99}\)\\ \(\kappa\)\end{tabular} \\
        \midrule
        Binary     & 11 & 0.000 & 0.000 & 0.0877 & 0.000 & 0.000 & 0.0839 & 0  \\
        Signed NAF & 12 & 0.473 & 0.000 & 0.1747 & 0.664 & 0.027 & 0.1518 & 13 \\
        \(\Phi\)   & 16 & 0.426 & 0.000 & 0.0998 & 0.568 & 0.025 & 0.0936 & 4  \\
        \(T_3\)    & 13 & 0.212 & 0.000 & 0.0933 & 0.293 & 0.000 & 0.0789 & 2  \\
        \(T_4\)    & 12 & 0.111 & 0.000 & 0.0888 & 0.147 & 0.000 & 0.0809 & 1  \\
        \bottomrule
    \end{tabular*}
\end{table*}

The matched-range check preserves the main qualitative trends. Single-digit
semantic survival remains zero for all systems, Binary remains structurally
undetectable, and multinacci detectability decreases from \(\Phi\) to \(T_3\) to
\(T_4\). Under \(b=3\) bursts, \(\Phi\) retains nonzero multinacci kernel
survival, while \(T_3\) and \(T_4\) do not because their native kernel identities
require longer aligned patterns. The nonzero Signed NAF burst value is retained
only as an alphabet-driven baseline and should not be interpreted as the same
multinacci kernel mechanism. Thus, the equal-\(W\) conclusions are not merely
artifacts of dynamic-range mismatch.

\clearpage
\section{Full Clean Representation Benchmark}
\phantomsection
\label{app:full-clean-benchmark}

The main text reports the \(W=24\) clean representation benchmark in
\hyperref[tab:clean_benchmark]{Table~\ref*{tab:clean_benchmark}}. For
completeness, \hyperref[tab:clean_benchmark_full]{Table~\ref*{tab:clean_benchmark_full}}
gives the full sweep over \(W \in \{8,12,16,20,24\}\), including codebook size,
ambient cardinality, sparsity, capacity, normalized deficit, and average nonzero
density for each system.

\begin{table*}[!htbp]
\centering
\caption{Representational metrics and structural properties of bounded
codebooks across varying window widths \(W\).}
\label{tab:clean_benchmark_full}

\begingroup
\scriptsize
\renewcommand{\arraystretch}{1.18}
\setlength{\tabcolsep}{0pt}

\begin{tabular*}{\textwidth}{@{\extracolsep{\fill}}lrrrrrr@{}}
\toprule
\textbf{System}
& \textbf{Codebook}
& \textbf{Ambient}
& \textbf{Sparsity}
& \textbf{Capacity}
& \textbf{Norm.}
& \textbf{Density} \\
& \textbf{\((|C|)\)}
& \textbf{\((|\Omega|)\)}
& \textbf{\((s)\)}
& \textbf{\((C_X)\)}
& \textbf{deficit \((D_X)\)}
& \textbf{\((\bar{\omega})\)} \\
\midrule

\multicolumn{7}{@{}l}{\textit{Window Width \(W = 8\), \(L = 4\), \(R = 3\)}} \\
Binary (\hyperref[sys:standard-binary]{System B}) & 256 & 256 & 1.0000 & 1.0000 & 0.0000 & 0.4997 \\
Signed NAF (\hyperref[sys:signed-digit-radix-2]{System S}) & 341 & 6,561 & 0.0520 & 1.0517 & 0.3364 & 0.3611 \\
\(\Phi\)-System (\hyperref[sys:golden-ratio-base]{System \(\Phi\)}) & 55 & 256 & 0.2148 & 0.7227 & 0.2773 & 0.2945 \\
Tribonacci (\hyperref[sys:tribonacci-base]{System \(T_3\)}) & 149 & 256 & 0.5820 & 0.9024 & 0.0976 & 0.3997 \\
Tetranacci (\hyperref[sys:tetranacci-base]{System \(T_4\)}) & 208 & 256 & 0.8125 & 0.9626 & 0.0374 & 0.4519 \\

\midrule
\multicolumn{7}{@{}l}{\textit{Window Width \(W = 12\), \(L = 6\), \(R = 5\)}} \\
Binary (\hyperref[sys:standard-binary]{System B}) & 4,096 & 4,096 & 1.0000 & 1.0000 & 0.0000 & 0.5002 \\
Signed NAF (\hyperref[sys:signed-digit-radix-2]{System S}) & 5,461 & 531,441 & 0.0103 & 1.0346 & 0.3473 & 0.3517 \\
\(\Phi\)-System (\hyperref[sys:golden-ratio-base]{System \(\Phi\)}) & 377 & 4,096 & 0.0920 & 0.7132 & 0.2868 & 0.2881 \\
Tribonacci (\hyperref[sys:tribonacci-base]{System \(T_3\)}) & 1,705 & 4,096 & 0.4163 & 0.8946 & 0.1054 & 0.3937 \\
Tetranacci (\hyperref[sys:tetranacci-base]{System \(T_4\)}) & 2,872 & 4,096 & 0.7012 & 0.9573 & 0.0427 & 0.4474 \\

\midrule
\multicolumn{7}{@{}l}{\textit{Window Width \(W = 16\), \(L = 8\), \(R = 7\)}} \\
Binary (\hyperref[sys:standard-binary]{System B}) & 65,536 & 65,536 & 1.0000 & 1.0000 & 0.0000 & 0.4977 \\
Signed NAF (\hyperref[sys:signed-digit-radix-2]{System S}) & 87,381 & 43,046,721 & 0.0020 & 1.0259 & 0.3527 & 0.3466 \\
\(\Phi\)-System (\hyperref[sys:golden-ratio-base]{System \(\Phi\)}) & 2,584 & 65,536 & 0.0394 & 0.7085 & 0.2915 & 0.2856 \\
Tribonacci (\hyperref[sys:tribonacci-base]{System \(T_3\)}) & 19,513 & 65,536 & 0.2977 & 0.8908 & 0.1092 & 0.3923 \\
Tetranacci (\hyperref[sys:tetranacci-base]{System \(T_4\)}) & 39,648 & 65,536 & 0.6050 & 0.9547 & 0.0453 & 0.4427 \\

\midrule
\multicolumn{7}{@{}l}{\textit{Window Width \(W = 20\), \(L = 10\), \(R = 9\)}} \\
Binary (\hyperref[sys:standard-binary]{System B}) & 1,048,576 & 1,048,576 & 1.0000 & 1.0000 & 0.0000 & 0.5005 \\
Signed NAF (\hyperref[sys:signed-digit-radix-2]{System S}) & 1,398,101 & 3,486,784,401 & 0.0004 & 1.0208 & 0.3560 & 0.3439 \\
\(\Phi\)-System (\hyperref[sys:golden-ratio-base]{System \(\Phi\)}) & 17,711 & 1,048,576 & 0.0169 & 0.7056 & 0.2944 & 0.2831 \\
Tribonacci (\hyperref[sys:tribonacci-base]{System \(T_3\)}) & 223,317 & 1,048,576 & 0.2130 & 0.8884 & 0.1116 & 0.3880 \\
Tetranacci (\hyperref[sys:tetranacci-base]{System \(T_4\)}) & 547,337 & 1,048,576 & 0.5220 & 0.9531 & 0.0469 & 0.4415 \\

\midrule
\multicolumn{7}{@{}l}{\textit{Window Width \(W = 24\), \(L = 12\), \(R = 11\)}} \\
Binary (\hyperref[sys:standard-binary]{System B}) & 16,777,216 & 16,777,216 & 1.0000 & 1.0000 & 0.0000 & 0.5003 \\
Signed NAF (\hyperref[sys:signed-digit-radix-2]{System S}) & 22,369,621 & 282,429,536,481 & \(7.92{\times}10^{-5}\) & 1.0173 & 0.3582 & 0.3427 \\
\(\Phi\)-System (\hyperref[sys:golden-ratio-base]{System \(\Phi\)}) & 121,393 & 16,777,216 & 0.0072 & 0.7037 & 0.2963 & 0.2823 \\
Tribonacci (\hyperref[sys:tribonacci-base]{System \(T_3\)}) & 2,555,757 & 16,777,216 & 0.1523 & 0.8869 & 0.1131 & 0.3883 \\
Tetranacci (\hyperref[sys:tetranacci-base]{System \(T_4\)}) & 7,555,935 & 16,777,216 & 0.4504 & 0.9520 & 0.0480 & 0.4407 \\

\bottomrule
\end{tabular*}

\endgroup
\end{table*}

For completeness, \(\rho_{\mathrm{rt}}\) was recorded for all systems and window
widths in Experiment~1. It was \(1.0000\) in every case, since the benchmark
uses valid finite-window codewords without corruption, repair, or arithmetic
canonicalization.

\clearpage

%% file: figures/fig_phi_redundancy_flow_B.tex
\begingroup%
\definecolor{PhiNodeFill}{RGB}{250,250,248}%
\definecolor{PhiNodeDraw}{RGB}{90,90,90}%
\definecolor{PhiCollisionFill}{RGB}{236,244,252}%
\definecolor{PhiCollisionDraw}{RGB}{42,82,130}%
\definecolor{PhiFlowZero}{RGB}{180,180,180}%
\definecolor{PhiFlowOne}{RGB}{82,105,135}%
\definecolor{PhiText}{RGB}{55,55,55}%
\definecolor{PhiNoteFill}{RGB}{252,250,242}%
\definecolor{PhiNoteDraw}{RGB}{145,125,75}%

\def\PhiStateNode#1#2{%
  \begin{tabular}{@{}c@{}}
    $#1$\\[-1pt]
    {\scriptsize \texttt{#2}}
  \end{tabular}%
}%

\def\FlowZero#1#2#3{%
  \draw[flowzero,line width=#1] (#2.east) -- node[pos=0.5, fill=white, text=PhiText, inner sep=1.5pt] {\tiny 0} (#3.west);
}%
\def\FlowOne#1#2#3{%
  \draw[flowone,line width=#1] (#2.east) -- node[pos=0.5, fill=white, text=PhiText, inner sep=1.5pt] {\tiny 1} (#3.west);
}%

\noindent\begin{tikzpicture}[
    x=1cm,
    y=0.9cm, 
    font=\sffamily\footnotesize,
    state/.style={
        rectangle,
        rounded corners=2.4pt,
        draw=PhiNodeDraw,
        fill=PhiNodeFill,
        line width=0.45pt,
        align=center,
        minimum width=2.4cm,
        minimum height=0.85cm,
        inner xsep=4pt,
        inner ysep=2.5pt,
        text=PhiText
    },
    collision/.style={
        state,
        draw=PhiCollisionDraw,
        fill=PhiCollisionFill,
        line width=1.2pt
    },
    layerlabel/.style={
        font=\sffamily\small\bfseries,
        text=black,
        align=center,
        anchor=center
    },
    countlabel/.style={
        font=\sffamily\scriptsize,
        text=PhiText,
        align=center,
        anchor=center
    },
    flowzero/.style={
        draw=PhiFlowZero,
        opacity=0.6,
        line cap=round
    },
    flowone/.style={
        draw=PhiFlowOne,
        opacity=0.8,
        line cap=round
    },
    note/.style={
        rectangle,
        rounded corners=2pt,
        draw=PhiNoteDraw,
        fill=PhiNoteFill,
        line width=0.45pt,
        align=left,
        inner xsep=5pt,
        inner ysep=4pt,
        text=PhiText
    }
]

\useasboundingbox (-1.2, -10.5) rectangle (13.2, 9.4);

\def\xA{0.0}
\def\xB{3.0}
\def\xC{6.0}
\def\xD{9.0}
\def\xE{12.0}

\node[layerlabel] at (\xA, 8.8) {$n=0$};
\node[countlabel] at (\xA, 8.1) {$1$ string\\$1$ state};

\node[layerlabel] at (\xB, 8.8) {$n=1$};
\node[countlabel] at (\xB, 8.1) {$2$ strings\\$2$ states};

\node[layerlabel] at (\xC, 8.8) {$n=2$};
\node[countlabel] at (\xC, 8.1) {$4$ strings\\$4$ states};

\node[layerlabel] at (\xD, 8.8) {$n=3$};
\node[countlabel] at (\xD, 8.1) {$8$ strings\\$7$ states};

\node[layerlabel] at (\xE, 8.8) {$n=4$};
\node[countlabel] at (\xE, 8.1) {$16$ strings\\$12$ states};

\node[state] (L0_0) at (\xA, 0.00) {\PhiStateNode{0+0\varphi}{$\varepsilon$}};

\node[state] (L1_1) at (\xB, 2.85) {\PhiStateNode{1+0\varphi}{1}};
\node[state] (L1_0) at (\xB, -2.85) {\PhiStateNode{0+0\varphi}{0}};

\node[state] (L2_3) at (\xC, 4.56) {\PhiStateNode{1+1\varphi}{11}};
\node[state] (L2_2) at (\xC, 1.14) {\PhiStateNode{0+1\varphi}{10}};
\node[state] (L2_1) at (\xC, -1.14) {\PhiStateNode{1+0\varphi}{01}};
\node[state] (L2_0) at (\xC, -4.56) {\PhiStateNode{0+0\varphi}{00}};

\node[state] (L3_6) at (\xD, 5.70) {\PhiStateNode{2+2\varphi}{111}};
\node[state] (L3_5) at (\xD, 3.42) {\PhiStateNode{1+2\varphi}{110}};
\node[state] (L3_4) at (\xD, 2.28) {\PhiStateNode{2+1\varphi}{101}};
\node[collision] (L3_3) at (\xD, 0.00) {\PhiStateNode{1+1\varphi}{011, 100}};
\node[state] (L3_2) at (\xD, -2.28) {\PhiStateNode{0+1\varphi}{010}};
\node[state] (L3_1) at (\xD, -3.42) {\PhiStateNode{1+0\varphi}{001}};
\node[state] (L3_0) at (\xD, -5.70) {\PhiStateNode{0+0\varphi}{000}};

\node[state] (L4_11) at (\xE, 6.27) {\PhiStateNode{3+4\varphi}{1111}};
\node[state] (L4_10) at (\xE, 5.13) {\PhiStateNode{2+4\varphi}{1110}};
\node[state] (L4_9) at (\xE, 3.99) {\PhiStateNode{3+3\varphi}{1101}};
\node[collision] (L4_8) at (\xE, 2.85) {\PhiStateNode{2+3\varphi}{1011, 1100}};
\node[state] (L4_7) at (\xE, 1.71) {\PhiStateNode{1+3\varphi}{1010}};
\node[collision] (L4_6) at (\xE, 0.57) {\PhiStateNode{2+2\varphi}{0111, 1001}};
\node[collision] (L4_5) at (\xE, -0.57) {\PhiStateNode{1+2\varphi}{0110, 1000}};
\node[state] (L4_4) at (\xE, -1.71) {\PhiStateNode{2+1\varphi}{0101}};
\node[collision] (L4_3) at (\xE, -2.85) {\PhiStateNode{1+1\varphi}{0011, 0100}};
\node[state] (L4_2) at (\xE, -3.99) {\PhiStateNode{0+1\varphi}{0010}};
\node[state] (L4_1) at (\xE, -5.13) {\PhiStateNode{1+0\varphi}{0001}};
\node[state] (L4_0) at (\xE, -6.27) {\PhiStateNode{0+0\varphi}{0000}};

\begin{pgfonlayer}{background}
\FlowZero{0.8pt}{L0_0}{L1_0} \FlowOne{0.8pt}{L0_0}{L1_1}
\FlowZero{0.8pt}{L1_0}{L2_0} \FlowOne{0.8pt}{L1_0}{L2_1}
\FlowZero{0.8pt}{L1_1}{L2_2} \FlowOne{0.8pt}{L1_1}{L2_3}
\FlowZero{0.8pt}{L2_0}{L3_0} \FlowOne{0.8pt}{L2_0}{L3_1}
\FlowZero{0.8pt}{L2_1}{L3_2} \FlowOne{0.8pt}{L2_1}{L3_3}
\FlowZero{0.8pt}{L2_2}{L3_3} \FlowOne{0.8pt}{L2_2}{L3_4}
\FlowZero{0.8pt}{L2_3}{L3_5} \FlowOne{0.8pt}{L2_3}{L3_6}
\FlowZero{0.8pt}{L3_0}{L4_0} \FlowOne{0.8pt}{L3_0}{L4_1}
\FlowZero{0.8pt}{L3_1}{L4_2} \FlowOne{0.8pt}{L3_1}{L4_3}
\FlowZero{0.8pt}{L3_2}{L4_3} \FlowOne{0.8pt}{L3_2}{L4_4}
\FlowZero{1.4pt}{L3_3}{L4_5} \FlowOne{1.4pt}{L3_3}{L4_6}
\FlowZero{0.8pt}{L3_4}{L4_7} \FlowOne{0.8pt}{L3_4}{L4_8}
\FlowZero{0.8pt}{L3_5}{L4_8} \FlowOne{0.8pt}{L3_5}{L4_9}
\FlowZero{0.8pt}{L3_6}{L4_10} \FlowOne{0.8pt}{L3_6}{L4_11}
\end{pgfonlayer}

\node[note, anchor=north west] at (-1.2, -7.4) {%
  \begin{tabular}{@{}c@{}}
    $\varphi^2=\varphi+1$\\[1pt]
    \texttt{011} $\leftrightarrow$ \texttt{100}
  \end{tabular}%
};

\node[note, anchor=north west, text width=14.15cm] at (-1.2, -8.8) {%
  \textbf{Semantic flow.} Each path corresponds to extending a binary string by one new leading digit.\\
  Nodes are grouped by exact value; highlighted nodes are semantic collisions.
};

\node[anchor=north west, font=\sffamily\scriptsize, text=PhiText, inner xsep=0pt] at (-1.2, -10.2) {%
  \tikz{\draw[flowone,line width=1.2pt] (0,0)--(0.55,0);} prefix $1$
  \qquad
  \tikz{\draw[flowzero,line width=1.2pt] (0,0)--(0.55,0);} prefix $0$
  \qquad
  thicker bands indicate multiple syntactic strings following the same semantic transition
};

\end{tikzpicture}%
\endgroup%